\documentclass[11pt]{article}

\usepackage{changepage}

\DeclareMathAlphabet\mathbfcal{OMS}{cmsy}{b}{n}
\usepackage{xfrac}
\usepackage{nicefrac}
\usepackage{empheq,fancybox}  
\usepackage{amsmath}	
\usepackage{bm}		    
\usepackage{bbm}
\usepackage{amssymb}	
\usepackage{makeidx}	
\usepackage{color} 		
\usepackage{graphicx} 	
\usepackage{enumitem}
\usepackage{multirow}	
\usepackage{times} 		
\usepackage{tensor} 	
\usepackage{hyperref} 	
\usepackage{cite} 		
\usepackage{slashed}
\usepackage[all]{hypcap}
\usepackage[title]{appendix}
\usepackage{dsfont}
\usepackage{notoccite}
\usepackage[font=small]{caption}
\hypersetup{
    colorlinks=true,	
    linkcolor=red,		
    citecolor=blue,		
    filecolor=magenta,	
    urlcolor=black	    
}
\newcommand{\per}{\ . \ }
\newcommand{\com}{\ , \ }

\newcommand{\xvec}{{\bm x}}

\newcommand{\kvec}{{\bm k}}

\newcommand{\khat}{{\hat{\kvec}}}
\newcommand{\gvec}{{\bm \gamma}}

\newcommand{\del}[1]{\nabla_{\! \! #1}}

\newcommand{\calU}{\mathcal{U}}
\newcommand{\cs}{c_s}
\newcommand{\eref}[1]{Eq.~(\ref{#1})}
\newcommand{\erefs}[2]{Eqs.~(\ref{#1})~and~(\ref{#2})}
\newcommand{\fref}[1]{Fig.~\ref{#1}}
\newcommand{\sref}[1]{Sec.~\ref{#1}}
\newcommand{\aref}[1]{Appendix~\ref{#1}}

\newcommand{\pref}[1]{(\ref{#1})}

\newcommand{\half}{{\textstyle\frac{1}{2}}}
\newcommand{\const}{\, \mathrm{const.}}

\newcommand{\Mpl}{M_{\rm Pl}}

\usepackage{accents}
\newcommand{\ubar}[1]{\underaccent{\bar}{#1}}
\newcommand{\ThreeHalf}{\nicefrac{3}{2}}
\newcommand{\Half}{\nicefrac{1}{2}}
\newcommand{\psiThreeHalf}{\psi_{\ThreeHalf,\kvec}}
\newcommand{\psiHalf}{\psi_{\Half,\kvec}}
\newcommand{\nn}{\nonumber \\}
\numberwithin{equation}{section}
\begin{document}

\title{Catastrophic Production of Slow Gravitinos}
\author{Edward W.\ Kolb$^1$, \ Andrew J.\ Long$^2$, \ and \ Evan McDonough$^1$}
\date{%
    $^1$Kavli Institute for Cosmological Physics and Enrico Fermi Institute\\  The University of Chicago, Chicago, IL 60637\\%
    $^2$Department of Physics and Astronomy, Rice University, Houston, TX 77005\\[2ex]%
    \today
}
\maketitle

\begin{abstract} 
We study gravitational particle production of the massive spin-$3/2$ Rarita-Schwinger field, and its close relative, the gravitino, in FRW cosmological spacetimes. For masses lighter than the value of the Hubble expansion rate after inflation, $m_{3/2} \lesssim H$, we find catastrophic gravitational particle production, wherein the number of gravitationally produced particles is divergent, caused by a transient vanishing of the helicity-1/2 gravitino sound speed. In contrast with the conventional gravitino problem, the spectrum of produced particles is dominated by those with momentum at the UV cutoff.  This suggests a breakdown of effective field theory, which might be cured by new degrees of freedom that emerge in the UV. We study the UV completion of the Rarita-Schwinger field, namely ${\cal N}=1$, $d=4$, supergravity. We reproduce known results for models with a single superfield and models with an arbitrary number of chiral superfields,  find a simple geometric expression for the sound speed in the latter case, and extend this to include nilpotent constrained superfields and orthogonal constrained superfields.  We find supergravity models where the catastrophe is cured and models where it persists. Insofar as quantizing the gravitino is tantamount to quantizing gravity, as is the case in any UV completion of supergravity, the models exhibiting catastrophic production are prime examples of 4-dimensional effective field theories that become inconsistent when gravity is quantized, suggesting a possible link to the Swampland program.  We propose the {\it Gravitino Swampland Conjecture}, which is consistent with and indeed follows from the KKLT and Large Volume scenarios for moduli stabilization in string theory.

\end{abstract}
\begingroup
\setlength{\parskip}{0.0cm}
\hypersetup{linkcolor=black}
\newpage
\tableofcontents
\endgroup
\newpage

\section{Introduction}\label{sec:intro}

Gravitational particle production (GPP) is ubiquitous in quantum field theories in curved spacetime. In a cosmological context GPP was first noted by Erwin Schrodinger in 1939 \cite{Schrodinger:1939:PVE}. In a modern context, gravitational particle production has risen to prominence as a mechanism to produced the observed dark matter density, beginning with early work on heavy scalar fields \cite{Chung:1998zb,Chung:1998ua,Chung:1998rq,Kolb:1998ki, Kuzmin:1998uv, Chung:2001cb,Chung:2004nh}, and later extended to spin-1/2 \cite{Chung:2011ck,Ema:2019yrd, Herring:2020cah}, spin-1 \cite{Chung:2011ck, Ema:2019yrd, Herring:2020cah,Kolb:2020fwh}, spin-3/2 \cite{Kallosh:1999jj, Kallosh:2000ve,Giudice:1999yt,Giudice:1999am,BasteroGil:2000je}, spin-2 \cite{Albornoz:2017yup}, and spin-$s$ ($s>2$) fields \cite{Alexander:2020gmv}. A gravitational origin of dark matter is particularly compelling given the dearth of evidence for the interactions between dark matter and the Standard Model of particle physics upon which conventional WIMP models are premised.

More generally, early universe inflationary cosmology has risen to the fore of particle physics, in particular, under the banner of Cosmological Collider Physics \cite{Chen:2009we,Arkani-Hamed:2015bza,Chen:2015lza}, and the Cosmological Bootstrap Program \cite{Arkani-Hamed:2018kmz,Baumann:2019oyu,Baumann:2020dch},  with the hope that one may use the cosmic microwave background non-Gaussianity as a particle detector \cite{Arkani-Hamed:2015bza,Lee:2016vti, Alexander:2019vtb, Alexander:2019vtb,Wang:2020uic}. Complementary to this, reheating as Cosmological Heavy Ion Collider \cite{McDonough:2020tqq}, namely, reheating as a playground in which to study thermalization in quantum field theory, has seen a recent resurgence \cite{Harigaya:2013vwa,Harigaya:2014waa,Mukaida:2015ria,Ema:2016oxl,Harigaya:2019tzu, Brandenberger:2019njw,McDonough:2020tqq}, as has reheating as a testing ground for Higgs physics \cite{Litsa:2020rsm}. With the rise of cosmology as a particle physics laboratory, it is natural to continue the study of gravitational particle production and  of inflation as a particle factory; the B-factory to the Cosmological Collider's LHC. Many very recent works have proceeded along these lines.

Parallel to the developments in particle cosmology have been many developments in string theory and supergravity. Gravitational production of spin-3/2 particles was examined in detail two decades ago by several groups \cite{Kallosh:1999jj,Kallosh:2000ve,Giudice:1999yt,Giudice:1999am,BasteroGil:2000je}. Since then, relatively little attention has been paid to this question (with exception of \cite{Addazi:2016bus,Addazi:2017kbx,Hasegawa:2017hgd, Hasegawa:2017nks}). Meanwhile, string theory saw the advent of stabilized flux compactifications \cite{Giddings:2001yu, Kachru:2003aw,Balasubramanian:2005zx}  and the string landscape \cite{Bousso:2000xa,Susskind:2003kw}. These string theory setups have very recently made contact with models of supergravity, with the anti-D3 brane of KKLT encapsulated by a nilpotent constrained superfield \cite{Kallosh:2014wsa,Bergshoeff:2015jxa,Vercnocke:2016fbt,GarciadelMoral:2017vnz, Cribiori:2019hod}. With all these developments, it is well past due to re-examine the gravitational production of gravitinos in the early universe.

In this work we study in detail the dynamics of massive spin-3/2 fields in cosmological spacetimes. The massive spin-3/2 carries within it a helicity-1/2 state and a helicity-3/2 state, which enjoy decoupled equations of motion. The former has a sound speed $\cs^2$ which in general differs from $1$, and can vanish if $m_{3/2} \lesssim H$ (here $m_{3/2}$ is the mass of the spin-3/2 field). We compute the gravitational particle production of these fields in a background cosmology generated by an oscillating scalar field, and find that for $m_{3/2} \lesssim H $ there is a production of particles with arbitrarily high-$k$.  That the spectrum is dominated by modes in the deep UV (including at the cutoff of the effective field theory) is in stark contrast with the conventional gravitino problem \cite{KHLOPOV1984265,ELLIS1984181}. We thus refer to this phenomenon as ``catastrophic particle production.'' 

One might hope that supergravity cures this instability. We demonstrate this is {\it not} always the case: there are a plethora of supergravity models that exhibit this instability and a plethora that do not. Indeed, the catastrophic production we study in detail here was first observed in the context of a supergravity model by Hasegawa et al.\ \cite{Hasegawa:2017hgd}. We demonstrate that a generic result in supergravity is a sound speed $\cs^2$ that differs from $1$; this was shown long ago \cite{BasteroGil:2000je,Kallosh:2000ve}, and has been referred to in the literature as a ``slow gravitino'' \cite{Benakli:2014bpa,Benakli:2015mbb,Kahn:2015mla}.  Whether the sound speed ever vanishes, $\cs^2=0$, is dependent on details of the model, such as the precise form of the superpotential and K\"{a}hler potential. We derive a simple expression for the gravitino sound speed in supergravity models with an arbitrary number of chiral superfields, and find a simple geometric interpretation of slow gravitinos in terms of an angle in field space. We find for the sound speed of the helicity-1/2 state of gravitinos
\begin{align}
\cs^2 = 1 -  \frac{ 4 |\vec{\dot{\Phi}}|^2 |\vec{F}|^2  }{(|\vec{\dot{\Phi}}|^2 +|\vec{F}|^2)^2} \left( 1- \cos^2 (\theta) \right) \com
\end{align}
where $\theta$ is the angle between the field-space vector of the cosmological evolution, $\vec{\dot{\Phi}}$, and the $F$-term vector $\vec{F}$. This matches Bastero-Gil and Mazumdar \cite{BasteroGil:2000je}, and provides a simple geometric interpretation:   The sound speed vanishes whenever $|\vec{\dot{\Phi}}|=|\vec{F}|$ and $\cos(\theta)=0$. We show this applies also in models of constrained nilpotent superfields, and with a small modification, to models of orthogonal constrained superfields. In all these cases, we find models and parameter choices which exhibit catastrophic particle production, and models which do not. We give an example of a toy model which interpolates between these two regimes. Note that the additional fermionic degrees of freedom in supergravity, which allow the helicity-1/2 gravitino to decay, and the identification of the gravitino as a linear combination of spin-1/2 fermions that may change with time, does not ameliorate the breakdown of effective field theory that defines catastrophic production. This is in contrast with the conventional gravitino problem (for example \cite{Eberl:2020fml}), where the EFT is valid at all times (unlike the catastrophe), and where the changing identity of the gravitino can significantly amelioriate the problem \cite{Nilles:2001ry,Nilles:2001fg} (see also \cite{Dalianis:2017okk} in the context of orthogonal constrained superfield supergravity models).

In summary, we find that quantization of the gravitino leads to an instability in certain quantum field theories, such as the Rarita-Schwinger model, which, at the classical level, appear to be perfectly well behaved. In the context of supergravity, at energy scales above the supersymmetry breaking scale there is no distinction between a graviton and a gravitino, the two being related by a linear supersymmetry transformation. Thus, the instabilities found here are no less severe than any that emerge from quantizing gravity. The study of apparently consistent field theories which become inconsistent when embedded in quantum gravity, and hence cannot be completed into quantum gravity in the ultraviolet, is known as the Swampland Program \cite{Vafa:2005ui} (for a review, see \cite{Palti:2019pca,Brennan:2017rbf}). With this in mind, we propose the {\it Gravitino Swampland Conjecture}. The conjecture is given in \sref{sec:GSC}  and elaborated upon in a separate article \cite{Kolb:2021nob}. 

The structure of this paper is as follows: In \sref{sec:model} we review the Rarita-Schwinger theory of a massive spin-3/2 field in flat space. In \sref{sec:Spin_32} we generalize this to an FRW space, and derive the equations of motion for a massive spin-3/2 field with a general time-dependent mass. In \sref{sec:GPP} we numerically compute the gravitational particle production of a constant-mass spin-3/2 field, and find catastrophic particle production of helicity-1/2 particles in the case that $m_{3/2} \lesssim H_e$, where $H_e$ is the Hubble parameter at the end of inflation. In \sref{sec:SUGRA} we consider a variety of supergravity models, and observe cases where there are, and cases where there are not, catastrophic particle production. Motivated by this, in \sref{sec:GSC} we propose the Gravitino Swampland Conjecture. We conclude in \sref{sec:conc} with a discussion of directions for future work.  There are three appendices.  Appendix \ref{app:framefields} is a brief discussion of frame fields used to promote the Minkowski-space field theory to curved space.  Appendix \ref{app:Projection} discusses how to extract the helicity components from the vector-spinor $\psi_\mu$.  A final Appendix \ref{app:SUGRA} reviews the salient features of supergravity we employ to study GPP. 

\textit{Conventions.}
We adopt the Landau-Lifshitz timelike conventions \cite{LL} for the signature of the metric ($\mathrm{sign}[\eta_{00}]=+1$ where $\eta_{\mu\nu}$ is the Minkowski metric), the Riemann curvature tensor  ($\tensor{R}{^\rho_{\sigma\mu\nu}} = + \partial_\mu \Gamma^\rho_{\nu\sigma} \cdots$), and the sign of the Einstein tensor $G_{\mu\nu}=+8\pi G_N T_{\mu\nu}$. To translate these conventions to other conventions, see the introductory material in Misner, Thorne, and Wheeler \cite{MTW}. Our sign conventions correspond to $(-,+,+)$ in their table.

\section{Rarita--Schwinger in Minkowski spacetime\label{sec:model}}

Two useful references for this section are Giudice, Riotto, and Tkachev \cite{Giudice:1999yt} and Kallosh, Kofman, Linde, and Van Proeyen \cite{Kallosh:1999jj}.  In the discussion we hew closer to the Giudice et al.\ treatment.

The relativistic quantum theory of spin-3/2 particles was constructed in 1941 by Rarita and  Schwinger \cite{Rarita:1941mf}.  The Rarita-Schwinger field $\psi_\mu(x)$ is a ``vector-spinor'' resulting from the direct product of the $(\half,\half)$ vector representation of the Lorentz group and the $(\half,0)\oplus(0,\half)$ representation of the Lorentz group for a Dirac spinor, resulting in the reducible representation 
\begin{align}
(\half,\half)\otimes\left[(\half,0) \oplus (0,\half)\right] = (\half,1) \oplus (\half,0) \oplus (1,\half) \oplus (0,\half) \per
\end{align}
The $(\half,1) \oplus (1,\half)$ reducible representation, corresponding to a spin-3/2 Rarita-Schwinger field, can be further decomposed into irreducible representations corresponding to the  $\pm \tfrac{3}{2}$ and $\pm\half$ helicity states. One may additionally impose a reality, i.e., Majorana, condition on $\psi_\mu(x)$. This choice finds a natural home in supergravity, since the number of degrees of freedom of a massless Majorana $\psi_\mu$ then matches that of the metric. In what follows we treat $\psi_\mu(x)$ as Dirac and impose a Majorana condition only when explicitly stated.

The properties of the Rarita-Schwinger field of mass $m$ in Minkowski space are derived from the action 
\begin{align}\label{eq:RS_flat}
S[\psi_\mu,\bar{\psi}_\mu] = & \int \! d^4\!x\left[ \frac{i}{2} \bar{\psi}_\mu \gamma^{\mu\rho\nu} \left(\partial_\rho\psi_\nu\right)  - \frac{i}{2} \left(\partial_\rho\bar{\psi}_\mu\right) \gamma^{\mu\rho\nu} \psi_\nu +  2 m\bar{\psi}_\mu \Sigma^{\mu\nu} \psi_\nu  \right] 
\end{align}
where $\Sigma^{\mu\nu} \equiv \tfrac{1}{2} \gamma^{\mu\nu} =  \tfrac{1}{2} \gamma^{[\mu} \gamma^{\nu]} = \tfrac{1}{4} (\gamma^\mu \gamma^\nu - \gamma^\nu \gamma^\mu) = \tfrac{1}{4}[\gamma^\mu,\gamma^\nu]$ and $\gamma^{\mu\rho\nu} = \gamma^{[\mu}\gamma^\rho\gamma^{\nu]} = \tfrac{1}{2} \left(\gamma^\mu\gamma^\rho\gamma^\nu - \gamma^\nu\gamma^\rho\gamma^\mu\right)$.
There are several equivalent forms for the action.  The first two are
\begin{align}\label{eq:sf}
S[\psi_\mu,\bar{\psi}_\mu] & = \int \! d^4\!x\left[ i \bar{\psi}_\mu \gamma^{\mu\rho\nu} \left(\partial_\rho\psi_\nu\right) +  2 m\bar{\psi}_\mu \Sigma^{\mu\nu} \psi_\nu \right]
= \int \! d^4\!x\left[ - i (\partial_\rho\bar{\psi}_\mu) \gamma^{\mu\rho\nu} \psi_\nu +  2 m\bar{\psi}_\mu \Sigma^{\mu\nu} \psi_\nu \right] \com
\end{align}
where in both expressions we have discarded a surface term resulting from integration by parts which will not contribute to the equation of motion.  In a manner reminiscent of the case for vectors, the $\mu=0$ component of $\psi_\mu$ is not dynamical, i.e., $\partial_0\psi_0$ does not appear in the action.  The third form of the action makes use of the identity 
\begin{align}\label{eq:randomidentity}
\gamma^{\mu\rho\nu} = - i\epsilon^{\mu\rho\nu\sigma}\gamma^5\gamma_\sigma\com
\end{align} 
where $\epsilon^{\mu\rho\nu\sigma}$ is the Levi-Civita symbol with convention $\epsilon^{0123}=+1$.  Using this, the action can be written in the modern form 
\begin{subequations}
\begin{align}
S[\psi_\mu,\bar{\psi}_\mu] = & \int \! d^4\!x  \left[ \epsilon^{\mu\sigma\rho\nu} \bar{\psi}_\mu  \gamma^5\gamma_\sigma(\partial_\rho\psi_\nu) + 2m\bar{\psi}_\mu \Sigma^{\mu\nu} \psi_\nu\right] \\
= & \int \! d^4\!x\left[ - \epsilon^{\mu\sigma\rho\nu}(\partial_\rho \bar{\psi}_\mu)  \gamma^5\gamma_\sigma\psi_\nu + 2m\bar{\psi}_\mu \Sigma^{\mu\nu} \psi_\nu\right] \per
\end{align}
\end{subequations}
Writing the action in this form makes transparent that $\psi_0$ is not a dynamical degree of freedom since the term proportional to $\partial_0\psi_0$ vanishes by action of the Levi-Civita symbol.  

Variation of the action \pref{eq:sf} with respect to $\psi_\mu$ and $\bar{\psi}_\mu$ yield the field equations
\begin{subequations}
\begin{align}
i\gamma^{\mu\rho\nu}(\partial_\rho\psi_\nu) + 2m\Sigma^{\mu\nu}\psi_\nu & = \epsilon^{\mu\sigma\rho\nu}\gamma^5\gamma_\sigma(\partial_\rho\psi_\nu) + 2m\Sigma^{\mu\nu}\psi_\nu = 0 \label{eq:fss} \\
- i (\partial_\rho\bar{\psi}_\mu)\gamma^{\mu\rho\nu}+  2 m\bar{\psi}_\mu \Sigma^{\mu\nu} & = - \epsilon^{\mu\sigma\rho\nu} (\partial_\rho \bar{\psi}_\mu)  \gamma^5\gamma_\sigma + 2m\bar{\psi}_\mu \Sigma^{\mu\nu} = 0 \per
\end{align}
\end{subequations}
The next step is to find the two constraint equations. They are found from operations on \eref{eq:fss}:
\begin{subequations}
\begin{align}
& \gamma_\mu \cdot \left[ i\gamma^{\mu\rho\nu}(\partial_\rho\psi_\nu) + 2m\Sigma^{\mu\nu}\psi_\nu \right] =  2i\left(\gamma^\mu\gamma^\nu\partial_\mu\psi_\nu-\eta^{\mu\nu}\partial_\mu\psi_\nu\right) +3m\gamma^\mu\psi_\mu =0 \\
& \partial_\mu \cdot \left[ i\gamma^{\mu\rho\nu}(\partial_\rho\psi_\nu) + 2m\Sigma^{\mu\nu}\psi_\nu \right] = m \left(\gamma^\mu\gamma^\nu\partial_\mu\psi_\nu - \eta^{\mu\nu}\partial_\mu\psi_\nu\right) = 0 \per
\end{align}
\end{subequations}
For $m \neq 0$ the constraint equations can be expressed as 
\begin{align}\label{eq:constraint}
	\gamma^\mu \psi_\mu = 0 
	\qquad \text{and} \qquad 
	\eta^{\mu\nu} \partial_\mu \psi_\nu = 0 
	\per
\end{align}
Enforcing these two constraints in \eref{eq:fss}, after tedious manipulation of $\gamma$-matrices, leads to
\begin{align}\label{eq:psi_mu_Dirac_eqn}
i \gamma^\mu\partial_\mu\psi_\nu-m\psi_\nu=0 \per
\end{align}
Thus, in Minkowski space the field equation for the spin-$3/2$ field $\psi_\nu$ is simply four copies of the Dirac equation labeled by the spacetime index $\nu$. The difference is that the original four components of $\psi_\nu$ are reduced to two physical components because of the constraint equations \pref{eq:constraint}.  

We will remove the nondynamical degrees of freedom after an expansion of the field into Fourier modes:
\begin{align}\label{eq:psi_Fourier}
\psi_\mu(t,\xvec) = \int \! \! \frac{d^3 \kvec}{(2\pi)^3} \ \psi_{\mu,\kvec}(t) \ e^{i\kvec\cdot\xvec} \per
\end{align}
Consider a single Fourier mode for which we choose $\kvec$ along the third axis; $\kvec=(0,0,k_z)$, i.e., $k_z \equiv k^3$. From \erefs{eq:psi_mu_Dirac_eqn}{eq:psi_Fourier}, the equation of motion for that Fourier mode is 
\begin{align}
i\gamma^0\partial_0\psi_{\mu,\kvec} - \gamma^3k_z\psi_{\mu,\kvec}-m\psi_{\mu,\kvec} = 0 \per
\end{align}
The first constraint from \eref{eq:constraint}, $\gamma^\mu\psi_\mu = 0$, can be used to express $\psi_{0,\kvec} = -\gamma^0\gamma^i\psi_{i,\kvec}$. Using $\partial_0\psi_{0,\kvec}$ from the $0$-th component of the equation of motion, we can solve the second constraint equation for $\psi_{3,\kvec}$.  The constraint equations imply for the Fourier modes
\begin{equation}
\begin{split}
\gamma^0\psi_{0,\kvec} & = -\gamma^i\psi_{i,\kvec} \\
\psi_{3,\kvec} & = \left(\dfrac{k_z}{m} + \gamma^3\right) \left(\gamma^1\psi_{1,\kvec} + \gamma^2 \psi_{2,\kvec} \right) \per
\end{split}
\end{equation}

The four spinor fields $\psi_0\ldots\psi_3$ separately obey the Dirac equation with canonically normalized kinetic terms.  However, they are not independent as there are two constraint equations. In \aref{app:Projection} we discuss the construction of two orthogonal combinations of $\psi_0\ldots\psi_3$ that are helicity eigenstates, have correctly normalized kinetic terms, and satisfy the Dirac equation.\footnote{The derivation in \aref{app:Projection} departs from Giudice et al.\ \cite{Giudice:1999yt}, but the final results are the same.}  The two states are \pref{eq:projthreehalf}--\pref{eq:projonehalf}:
\begin{align}\label{eq:threeone}
\psi_{{\ThreeHalf,\kvec}} & = \dfrac{1}{\sqrt{2}}\left( \psi_{1,\kvec}+\gamma^1\gamma^2\psi_{2,\kvec} \right) \nn
\psi_{{\Half,\kvec}} & = \dfrac{\sqrt{6}}{2}\left( \psi_{1,\kvec}-\gamma^1\gamma^2\psi_{2,\kvec} \right)  \per
\end{align}
The fields $\psi_{{\ThreeHalf,\kvec}}$ and $\psi_{{\Half,\kvec}}$ are constructed from helicity projections, have canonical kinetic terms, are orthogonal, and obey the Dirac equation (see \aref{app:Projection}).

Using \eref{eq:threeone} and the constraint equations we can also express the four fields $\psi_{\mu,\kvec}$ in terms of the two independent fields $\psi_{\Half,\kvec}$ and $\psi_{\ThreeHalf,\kvec}$: 
\begin{align}\label{eq:psio3}
\gamma^0\psi_{0,\kvec} & = - \frac{2}{\sqrt{6}}\frac{k_z}{m} \gamma^3\gamma^1\psi_{\Half,\kvec} \nn
\gamma^1\psi_{1,\kvec} & = \frac{1}{\sqrt{6}}\gamma^1\psi_{\Half,\kvec} + \frac{1}{\sqrt{2}}\gamma^1\psi_{\ThreeHalf,\kvec} \nn
\gamma^2\psi_{2,\kvec} & = \frac{1}{\sqrt{6}} \gamma^1\psi_{\Half,\kvec} - \frac{1}{\sqrt{2}}\gamma^1\psi_{\ThreeHalf,\kvec} \nn
\gamma^3\psi_{3,\kvec} & = \frac{2}{\sqrt{6}}\frac{k_z}{m}\gamma^3\gamma^1 \psi_{\Half,\kvec} - \frac{2}{\sqrt{6}}\gamma^1 \psi_{\Half,\kvec} \per
\end{align}

The equations of motion for $\psi_{\Half,\kvec}$ and $\psi_{\ThreeHalf,\kvec}$ can be found from the equations of motion for $\psi_{\mu,\kvec}$:
\begin{align}\label{eq:minz}
\left(i\gamma^0\partial_0-k_z\gamma^3-m\right)\psi_{\Half,\kvec} & = 0 \nn
\left(i\gamma^0\partial_0-k_z\gamma^3-m\right)\psi_{\ThreeHalf,\kvec} & = 0 \per
\end{align}
So the $\psi_{\Half,\kvec}$ and $\psi_{\ThreeHalf,\kvec}$ are, as advertised, independent states which satisfy the same equation of motion as an ordinary Dirac field, and the quantization can proceed as in the Dirac fermion case.  In deriving \eref{eq:minz} we have assumed $\kvec = k_z\hat{z}$. The obvious rotationally-invariant mode equation would be to replace $k_z\gamma^3$ in \eref{eq:minz} by $\kvec\cdot\gvec$.

\section{Rarita-Schwinger in a Friedmann--Robertson--Walker background \label{sec:Spin_32}}

The quantum field theory in curved spacetime is discussed in many excellent reviews and textbooks; see for instance Freedman and Van Proeyen \cite{Freedman:2012zz}.  As a specific application, the phenomenon of gravitational particle production applied to a massive gravitino during inflation is discussed in early work~\cite{Kallosh:1999jj,Giudice:1999yt,Giudice:1999am}, and see also more recent articles \cite{Addazi:2016bus,Addazi:2017kbx, Hasegawa:2017hgd}.

\subsection{Action and field equations}

In a curved spacetime the Rarita-Schwinger action from \eref{eq:RS_flat} generalizes to
\begin{align}\label{eq:32_action_1a} 
	S[\Psi_\mu(x), \bar{\Psi}_\mu(x), \tensor{e}{^\beta_\nu}(x)] 
	& = \int \! d^4 x \, \sqrt{-g} \, \biggl[  \frac{i}{2} \bar{\Psi}_\mu  \ubar{\gamma}^{\mu\rho\nu} (\del{\rho} \Psi_\nu) - \frac{i}{2} (\del{\rho} \bar{\Psi}_\mu) \ubar{\gamma}^{\mu\rho\nu} \Psi_\nu  + 2 m \bar{\Psi}_\mu \ubar{\Sigma}^{\mu\nu} \Psi_\nu \biggr] 	\per
\end{align}
Varying the action with respect to $\bar{\Psi}_\mu$ and $\Psi_\mu$ yields the field equations
\begin{align}\label{eq:32_field_eqn_1}
	i  \ubar{\gamma}^{\mu\rho\sigma}  \del{\rho} \Psi_\sigma + 2 m \ubar{\Sigma}^{\mu\sigma} \Psi_\sigma & = 0 \qquad \text{and} \qquad
	i \del{\rho} \bar{\Psi}_\mu  \ubar{\gamma}^{\mu\rho\sigma} - 2 m \bar{\Psi}_\mu \ubar{\Sigma}^{\mu\sigma}  = 0 
	\per
\end{align}
Here we work in the frame-field formalism; see \aref{app:framefields} for a brief review.  The frame field is denoted by $\tensor{e}{^\alpha_\mu}(x)$ where the first index is raised/lowered by the Minkowski metric $\eta_{\alpha\beta}$ and the second index by the spacetime metric $g_{\mu\nu}$.  
Gamma matrices become local objects, $\ubar{\gamma}^\mu(x) = \tensor{e}{_\alpha^\mu}(x) \, \gamma^\alpha$, while $\gamma^\alpha$ obeys the Clifford algebra in Minkowski space.  
The Dirac conjugate of $\Psi_\mu$ is given by $\bar{\Psi}_\mu = \Psi_\mu^\dagger \gamma^0$.  
The covariant derivatives are given by $\del{\rho} \Psi_\nu = \partial_\rho \Psi_\nu + \Gamma_\rho \Psi_\nu$ and $\del{\rho} \bar{\Psi}_\mu = \partial_\rho \bar{\Psi}_\mu - \bar{\Psi}_\mu \Gamma_\rho$ where $\Gamma_\mu(x)$ is the spin connection, and see \aref{app:framefields} for an explicit expression.  We assume that $\Psi_\mu$ is minimally coupled to gravity, but we suppose that it may also interact with other fields, and in this background the mass parameter may develop a time-dependence $m = m(\eta)$, where $\eta$ is conformal time defined by ${\rm d}\eta = {\rm d}t/a$.  
For example, in \sref{sec:SUGRA} we will show that this can happen when the spin-3/2 field is embedded into theories of supergravity as the gravitino.  

Specifying to the FRW background, the frame-field components and spin connection are given by 
\begin{align}\label{eq:vierbein_FRW}
	\tensor{e}{^0_0} = a(\eta) 
	\ , \qquad 
	\tensor{e}{^i_j} = a(\eta) \, \delta^i_j 
	\ , \qquad 
	\tensor{e}{^0_i} = \tensor{e}{^i_0} = 0  
	\ , \qquad 
	\Gamma_\mu = a(\eta) \, H(\eta) \, \eta_{\mu\nu} \, \Sigma^{\nu0} 
	\com
\end{align}
where $H \equiv \partial_\eta a/a^2$ is the Hubble expansion rate, which satisfies the Friedmann equation $H = \sqrt{(8\pi G/3) \rho}$ for energy density $\rho$.  
The action \pref{eq:32_action_1a} becomes 
\begin{equation}\label{eq:32_action_2}
\begin{split}
	S[\psi_\mu(\eta,\xvec), \bar{\psi}_\mu(\eta,\xvec)] = \int_{-\infty}^{\infty} \! d \eta \int \! d^3 \xvec \, \biggl[ 
& i \bar{\psi}_\mu  \gamma^{\mu\rho\nu}  \partial_\rho \psi_\nu 
+ 2 a m \, \bar{\psi}_\mu \Sigma^{\mu\nu} \psi_\nu \\ & 
	- i aH \, \bar{\psi}_\mu \bigl( \gamma^\mu \eta^{\nu0} - \gamma^\nu \eta^{\mu0} \bigr) \psi_\nu	\biggr] 
\end{split}
\end{equation}
where all indices are raised/lowered with the Minkowski metric $\eta_{\mu\nu}$.  
To ensure that that field's kinetic term is canonically normalized we have introduced $\psi_\mu(\eta,\xvec)$ according to 
\begin{align}\label{eq:32_comoving_field}
	\psi_\mu(\eta,\xvec) = a^{1/2}(\eta) \Psi_\mu(\eta,\xvec) \per
\end{align}
This transformation partially absorbs the factor of $\sqrt{-g} = a^4$; in the kinetic terms the remaining $a^3$ is canceled by the inverse frame field $\tensor{e}{_\alpha^\mu} \propto a^{-1}$, and in the mass term the factor of $a$ remains.  Of course if one sets $H=0$ and $a=1$, the Minkowski result \pref{eq:sf} is recovered.
In the FRW background the field equations \pref{eq:32_field_eqn_1} become
\begin{subequations}\label{eq:32_field_eqn_2}
\begin{align}
	i  \gamma^{\mu\rho\nu}\partial_\rho \psi_\nu - i aH \, \left( \gamma^\mu \eta^{\nu0} - \gamma^\nu \eta^{\mu0} \right) \psi_\nu
	+ 2 a m \,\Sigma^{\mu\nu} \psi_\nu & = 0 \\
	i \partial_\rho \bar{\psi}_\nu \gamma^{\mu\rho\nu}
	+ i aH \, \bar{\psi}_\nu \left( \gamma^\mu \eta^{\nu0} - \gamma^\nu \eta^{\mu0} \right) 
- 2 a m \, \bar{\psi}_\nu \Sigma^{\mu\nu} & = 0 \per
\end{align}
\end{subequations}
There are two ways in which this equation differs from the Rarita-Schwinger equation in flat space:  the mass is replaced by $m \to a(\eta) m$ and there is an additional Hubble-dependent term.  This is unlike the case of a Dirac spinor field for which the mass was the only source of conformal symmetry breaking.  

\subsection{Constraints}
 
The field $\psi_\mu$ can be viewed as a set of four Dirac or Majorana spinors labeled by $\mu = 0, 1, 2, 3$.  From a naive accounting, it looks like this field describes $16$ degrees of freedom (or $8$ if the Majorana condition is imposed).  However, the field equations \pref{eq:32_field_eqn_2} also contain constraints, which cut the degrees of freedom down by half.  

It's useful to first write the field equations \pref{eq:32_field_eqn_2} as 
\begin{subequations}\label{eq:32_field_eqn_3}
\begin{align}
	\mathcal{R}^\mu = 0
	\qquad \text{where} \qquad 
	\mathcal{R}^\mu = \mathcal{R}^{\mu\nu} \, \psi_\nu 
\end{align}
and where 
\begin{align}
\mathcal{R}^{\mu\nu} = i \gamma^{\mu\rho\nu} \partial_\rho 
	- \, i aH \, \left( \gamma^\mu \eta^{\nu0} - \gamma^\nu \eta^{\mu0} \right)
	+ 2 a m \, \Sigma^{\mu\nu} 
\end{align} 
is a linear differential operator.  
Written out explicitly, 
\begin{align}
\mathcal{R}^0 & = 2i \gamma^0 \Sigma^{ij} \partial_i \psi_j 
	+ i aH \gamma^i \psi_i 	+ a m \, \gamma^0 \gamma^i \psi_i \\
\mathcal{R}^i & = i \gamma^{i\rho\nu} \partial_\rho \psi_\nu
	- \bigl( i aH + am \gamma^0 \bigr) \, \gamma^i \psi_0 
	+ 2 a m \, \Sigma^{ij} \psi_j 
	\com
\end{align}
\end{subequations}
one can see that $\mathcal{R}^\mu = 0$ is equivalent to \eref{eq:32_field_eqn_2}.   The first constrain equation is 
\begin{align}\label{eq:constraint_1}
	\gamma_\mu \mathcal{R}^\mu
	= 0 = 
	2i \bigl[ 
	(\gamma^\rho \partial_\rho) (\gamma^\sigma \psi_\sigma) 
	- (\partial^\mu \psi_\mu) 
	\bigr] 
	- \, 4 i aH \, \psi_0 
	+ \, \left( i aH \gamma^0 + 3 am \right) (\gamma^\sigma \psi_\sigma) 
	\per
\end{align}
Note that $\gamma^\mu \mathcal{R}_\mu$ does not contain a term like $\partial_0 \psi_0$, which cancels between the two terms in square brackets.  
To extract the second constraint, consider the differential operator
\begin{align}\label{eq:32_D_mu}
	\mathcal{D}_\mu = \partial_\mu + \Gamma_\mu + \frac{i}{2} m \gamma_\mu = \partial_\mu + aH \, \eta_{\mu\nu} \Sigma^{\nu0} +\frac{i}{2} a m \eta_{\mu\nu} \gamma^\nu \per
\end{align}
Applying this operator to the field equation gives
\begin{align}\label{eq:32:DR}
	\mathcal{D}_\mu \mathcal{R}^\mu 
	= 0 =
	\mathcal{R}^0 
	- \frac{i}{2} a^2 \left[\frac{1}{3} R + H^2 - 3m^2 \right] (\gamma^i \psi_i) + \frac{i}{2} a^2 \left[ 3 (m^2 + H^2)  \right] (\gamma^0 \psi_0)  +a(\partial_\eta m)\, \gamma^0(\gamma^i\psi_i) 
	\per
\end{align}
In this expression, $R = - (6 \, \partial_\eta^2 a) / a^3$ is the Ricci scalar curvature, and the Friedmann equation enforces $R=-8\pi G(\rho-3p)$ where $p$ is the cosmological pressure.  The last term in \eref{eq:32:DR} arises in models that allow for a time-dependent mass, and in \sref{sec:SUGRA} we will encounter nonzero $\partial_\eta m$ in the context of supergravity where a Majorana $\psi_\mu$ is the gravitino.  Solving \eref{eq:32:DR} gives 
\begin{align}\label{eq:constraint_2}
	\gamma^0 \psi_0(\eta,\xvec) = C_c(\eta) \, \gamma^i \psi_i(\eta,\xvec)
\end{align}
where the time-dependent coefficient, $C_c(\eta)$, is given by (note appearance of $\partial_\eta m$)
\begin{align}\label{eq:32_Cc_def}
	C_c = \frac{\frac{1}{3} R + H^2 - 3 m^2}{3 (H^2 + m^2)} + \frac{ 2i \gamma^0 a\, \partial_\eta m}{3a^2(H^2+m^2)} 
	\per
\end{align}
In the Minkowski limit with static mass we have $C_c=-1$.

\subsection{Fourier decomposition and mode equations}
 
Since the field equation \pref{eq:32_field_eqn_2} is left invariant under spatial translations, ${\bm x}^\prime = {\bm x} + {\bm \epsilon}$, we anticipate that its solutions can be labeled by a 3-vector $\kvec$ and will be proportional $e^{i \kvec \cdot \xvec}$.  So we decompose the field $\Psi_\mu(x^0, \xvec)$ into mode functions $\Psi_{\mu,\kvec}(x^0)$ labeled by $3$-vectors $\kvec$, which are called comoving wavevectors.  We can write the Fourier transform of the field $\Psi_\mu$ and the comoving field $\psi_\mu$ as
\begin{align}\label{eq:32_mode_decomp}
\Psi_\mu(x^0,\xvec)  = \int \! \! \frac{d^3 \kvec}{(2\pi)^3} \, \Psi_{\mu,\kvec}(x^0) \, e^{i \kvec \cdot \xvec} \qquad \text{and} \qquad	\psi_\mu(\eta,\xvec)  = \int \! \! \frac{d^3 \kvec}{(2\pi)^3} \, \psi_{\mu,\kvec}(\eta) \, e^{i \kvec \cdot \xvec} 
\end{align}
where $\eta$ is the conformal time.  
  
After using \eref{eq:32_mode_decomp}, the field equation \pref{eq:32_field_eqn_3} becomes a set of mode equations 
\begin{subequations}\label{eq:32_field_eqn_3b}
\begin{align}
\mathcal{R}_\kvec^0 = 0 \qquad \text{and} \qquad \mathcal{R}_\kvec^i = 0 
\end{align}
where\footnote{For the sake of clarity and completeness, we note the following. The four-vectors $k^\mu$ and $x^\mu$ 4-vectors are naturally contravariant.  We choose $k$ to be a space-like 4-vector, so $k^\mu=(\omega,\kvec)$, and $x^\mu=(x^0,\xvec)$. This defines $\kvec$ and $\xvec$.  The 4-vector product $k\cdot x=\eta_{\mu\nu}k^\mu x^\nu=\eta_{00} k^0 x^0 + \eta_{ij}k^ix^j=\omega x^0-\kvec\cdot\xvec$.  We will take $\kvec$ in the $3$-direction ($z$-direction), so $\kvec\cdot\xvec=k_zx^3$. The Dirac matrices $\gamma^\mu$ are also contravariant vectors:  $\gamma^\mu = (\gamma^0,\bm{\gamma})$, so $k \cdot \gamma = \eta_{\mu\nu} k^\mu\gamma^\nu = \eta_{00} k^0 \gamma^0 + \eta_{ij}k^i\gamma^j = \eta_{00} k^0 \gamma^0 -\kvec\cdot\bm{\gamma}$.  When we take $\kvec$ in the $z$-direction, $k \cdot \gamma = \omega\gamma^0-k_z\gamma^3$.  Now $\psi_\mu = (\psi_0,\bm{\psi})$ is naturally a covariant 4-vector.  So $\gamma\cdot\psi=\gamma^0\psi_0+\gamma^i\psi_i = \gamma^0\psi_0+\bm{\gamma}\cdot\bm{\psi}$. We also note a subtle difference: $\kvec\cdot\xvec = k_zx^3$ and  $\kvec\cdot\bm{\gamma} = k_z\gamma^3$, while $\bm{\gamma}\cdot\bm{\psi} = \gamma^i\psi_i$. }
\begin{align}
	\mathcal{R}_\kvec^0 & = - \gamma^0 (\gvec \cdot \kvec) (\gvec \cdot \bm{\psi}_\kvec) - \gamma^0 (\kvec \cdot \bm{\psi}_\kvec) + \left( i aH + am \gamma^0 \right) (\gvec \cdot \bm{\psi}_\kvec) \\ 
	\mathcal{R}_\kvec^i & = - i \gamma^0 \left( \gamma^i \gamma^j - \eta^{ij} \right) \partial_0 \psi_{j,\kvec} + \left[\frac{1}{2} \left( \gamma^{ijl} - \gamma^{lji} \right) k_j + a m \, \left( \gamma^i \gamma^l - \eta^{il} \right) \right] \psi_{l,\kvec} \\ & \qquad - \left[ k^i + \gamma^i \left( \kvec \cdot \gvec + i aH \gamma^0 - am \right) \right] \gamma^0 \psi_{0,\kvec} \nonumber 	\per
\end{align}
\end{subequations}
The constraints in \erefs{eq:constraint_1}{eq:constraint_2} have Fourier representations of 
\begin{align}
	(\kvec \cdot \bm{\psi}_\kvec) & =  \left[- (\gvec \cdot \kvec) + i aH \gamma^0 + am \right] (\gvec \cdot {\bm{\psi}}_\kvec) \label{eq:constraint_1_Fourier} \\ 
	\gamma^0 \psi_{0,\kvec} & = C_c \, (\gvec \cdot \bm{\psi}_\kvec) \label{eq:constraint_2_Fourier}
\end{align}
where we've introduced $\bm{\psi}_\kvec$ through $\psi_{\mu,\kvec}=\{\psi_{0,\kvec}, \bm{\psi}_\kvec\}$.  
Recall that the time-dependent coefficient $C_c(\eta)$ is given by \eref{eq:32_Cc_def}.  

The two constraints in \erefs{eq:constraint_1_Fourier}{eq:constraint_2_Fourier} reduce the degrees of freedom by half.   Now we'd like to decompose the four spinor fields, $\psi_{\mu,\kvec}(\eta)$ into their constituent degrees of freedom.  
We will denote these by $\psiHalf(\eta)$ and $\psiThreeHalf(\eta)$ since they correspond to the helicity-1/2 and helicity-3/2 states.  This decomposition is easier to identify by work in the frame where $\kvec = (0, 0, k_z)$.  In this frame, the constraint from \eref{eq:constraint_1_Fourier} gives 
\begin{align}\label{eq:constraint_1_Fourierb}
\psi_{3,\kvec} 	= \left( C_d + \gamma^3 \right) \left( \gamma^1 \psi_{1,\kvec} + \gamma^2 \psi_{2,\kvec} \right) 	\com
\end{align}
where we have defined 
\begin{align}\label{eq:32_Cd_def}
C_d = \frac{k_z}{a^2 (H^2 + m^2)} \left( iaH \gamma^0 + am \right) \per
\end{align}
In the Minkowski limit $C_d=k_z/m$.  We decompose the fields $\psi_{\mu,\kvec}$ onto $\psiHalf$ and $\psiThreeHalf$ by writing [cf.\ \eref{eq:psio3}] with $C_d=k_k/m,\ C_c=-1$)
\begin{subequations}\label{eq:32_chi_decompose}
\begin{align}
\gamma^0\psi_{0,\kvec} & = \frac{2}{\sqrt{6}} \,  C_c\gamma^3 C_d \gamma^1 \, \psiHalf \\ 
\gamma^1\psi_{1,\kvec} & = \frac{1}{\sqrt{6}}\gamma^1 \, \psiHalf + \frac{1}{\sqrt{2}}\gamma^1 \, \psiThreeHalf \\ 
\gamma^2\psi_{2,\kvec} & =  \frac{1}{\sqrt{6}} \gamma^1 \, \psiHalf - \frac{1}{\sqrt{2}}\gamma^1 \, \psiThreeHalf  \\  
\gamma^3\psi_{3,\kvec} & =  \frac{2}{\sqrt{6}} \,  C_d \gamma^3\gamma^1\psiHalf - \frac{2}{\sqrt{6}} \, \gamma^1 \psiHalf
	\per
\end{align}
\end{subequations}
Equation (\ref{eq:32_chi_decompose}) implies [cf.\ \eref{eq:threeone}]
\begin{equation}\label{eq:invert}
\begin{split}
\psiHalf & = \sqrt{\frac{3}{2}}\left( \psi_{1,\kvec} - \gamma^1\gamma^2 \psi_{2,\kvec} \right) \\
\psiThreeHalf & = \sqrt{\frac{1}{2}} \left( \psi_{1,\kvec} + \gamma^1 \gamma^2\psi_{2,\kvec} \right) \per
\end{split}
\end{equation}
After this decomposition, the mode equations \pref{eq:32_field_eqn_3b} become
\begin{subequations}
\begin{align}
	\left[ i \gamma^0 \partial_\eta - k_z \gamma^3 - am \right] \psiThreeHalf & = 0 \\
	\left[ i \gamma^0 \partial_\eta - k_z \left( C_A + i C_B \gamma^0 \right) \gamma^3 - am 	\right] \psiHalf & = 0 
\end{align}
\end{subequations}
where we have defined
\begin{subequations}\label{eq:CA_CB_def}
\begin{align}
	C_A & = \frac{1}{3(H^2 + m^2)^2} \left[ (m^2 - H^2)\left( - \frac{1}{3} R - H^2 + 3 m^2 \right) -4Hm \frac{\partial_\eta\,m}{a} \right] \\
	C_B & = \frac{2 m}{3 (H^2 + m^2)^2}\left[H \left( - \frac{1}{3} R - H^2 + 3 m^2 \right) + (m^2-H^2)\frac{\partial_\eta\,m}{ma} \right] \per
\end{align}
\end{subequations}
Here we note for future use that 
\begin{align}\label{eq:ca2pcb2}
	\cs^2 \equiv C_A^2+C_B^2 = \frac{1}{9(H^2 + m^2)^2}  \left[ \left( - \frac{1}{3} R - H^2 + 3 m^2 \right)^2 +4\frac{(\partial_\eta m)^2}{a^2} \right] \per
\end{align}
Although we have taken $\kvec = (0,0,k_z)$ to derive these relations, we know that the system is rotationally invariant, and so we can extend to the Lorentz-covariant mode equations:
\begin{subequations}\label{eq:32_field_eqn_3d}
\begin{align}
\left[ i \gamma^0 \partial_\eta - \kvec \cdot \gvec - am \right] \psiThreeHalf 	& = 0 \\
\left[ 	i \gamma^0 \partial_\eta - \left( C_A + i C_B \gamma^0 \right) \kvec \cdot \gvec - am \right] \psiHalf 	& = 0 \per
\end{align}
\end{subequations}
In Minkowski space $C_A=1$ and $C_B=0$ and the mode equations reduce to \eref{eq:minz}.  

It is useful to compare \eref{eq:32_field_eqn_3d} with the mode equation for a Dirac spinor field.  We see that the helicity-3/2 mode function, $\psiThreeHalf$, satisfies precisely the same equation of motion as the Dirac spinor mode function, while the equation of motion for a helicity-1/2 mode function, $\psiHalf$, differs from the corresponding equation for a Dirac spinor if $C_A\neq1$ \textit{and} $C_B\neq0$.

\subsection{Helicity eigenspinors}
 
For each $\kvec$, the mode equations \pref{eq:32_field_eqn_3d} will admit be two solutions, which we can distinguish by writing $\psiThreeHalf^s(\eta)$ and $\psiHalf^s(\eta)$ where the new label $s$ takes values $s = \pm 3/2$ or $s = \pm 1/2$ as appropriate. In the Dirac representation of the gamma matrices it is convenient to parametrize the spinor wavefunctions as 
\begin{align}\label{eq:32_chiA_chiB}
	\psiThreeHalf^s(\eta) = \begin{pmatrix} \chi_{A,\ThreeHalf,\kvec}(\eta)  \\ (2s/3) \, \chi_{B,\ThreeHalf,\kvec}(\eta)\end{pmatrix} \otimes h_{\hat{\kvec}}^{2s/3}
	\qquad \text{and} \qquad 
	\psiHalf^s(\eta) = \begin{pmatrix} \chi_{A,\Half,\kvec}(\eta)  \\ (2s) \, \chi_{B,\Half,\kvec}(\eta)\end{pmatrix} \otimes h_{\hat{\kvec}}^{2s}
\end{align}
where $\chi_{A,\ThreeHalf,\kvec}(\eta)$, $\chi_{B,\ThreeHalf,\kvec}(\eta)$, $\chi_{A,\Half,\kvec}(\eta)$, and $\chi_{B,\Half,\kvec}(\eta)$ are complex-valued mode functions, and $h_\khat^\lambda$ is a 2-component complex column vector, called the helicity 2-spinor, which only depends upon $\khat = \kvec / |\kvec|$.  In particular, $h_\khat^\lambda$ is defined to be an eigenfunction of the helicity operator with eigenvalue $\lambda$:
\begin{align}\label{eq:32_h_def}
	\khat \cdot {\bm \sigma} \, h_\khat^\lambda & = \lambda \, h_\khat^\lambda \qquad \text{for} \qquad \lambda = \pm 1 	\per
\end{align}
Putting the Ans\"atze \pref{eq:32_chiA_chiB} into the mode equation \pref{eq:32_field_eqn_3d} yields 
\begin{subequations}\label{eq:32_mode_eqn_chiA_chiB}
\begin{align}\label{eq:32_mode_eqn_ThreeHalf}
	i \partial_\eta \begin{pmatrix} \chi_{A,\ThreeHalf,k}(\eta) \\ \chi_{B,\ThreeHalf,k}(\eta) \end{pmatrix} 
	& = \mathcal{A}_{\ThreeHalf} \begin{pmatrix} \chi_{A,\ThreeHalf,k}(\eta) \\ \chi_{B,\ThreeHalf,k}(\eta) \end{pmatrix} \\ 
	i \partial_\eta \begin{pmatrix} \chi_{A,\Half,k}(\eta) \\ \chi_{B,\Half,k}(\eta) \end{pmatrix} 
	& = \mathcal{A}_{\Half}\begin{pmatrix} \chi_{A,\Half,k}(\eta) \\ \chi_{B,\Half,k}(\eta) \end{pmatrix} \label{eq:32_mode_eqn_OneHalf}
\end{align}
\end{subequations}
where the 2-by-2 complex matrices $\mathcal{A}_{\ThreeHalf}$ and $\mathcal{A}_{\Half}$ are defined by 
\begin{align}
	\mathcal{A}_{\ThreeHalf}(\eta) = \begin{pmatrix} am &  k  \\  k  & -am \end{pmatrix} 
	\qquad \text{and} \qquad 
	\mathcal{A}_{\Half}(\eta) = \begin{pmatrix} am & (C_A + i C_B) \, k  \\ (C_A - i C_B) \, k & -am  \end{pmatrix} 
	\com
\end{align}
and $k \equiv |\kvec|$.  We will see below that the eigenvalues of $\mathcal{A}_{\ThreeHalf}$ and $\mathcal{A}_{\Half}$ are
\begin{subequations}
\begin{align}\label{eq:evaluesThreeHalf}
	\lambda_{\ThreeHalf\pm} 
	& = \pm\omega_{\ThreeHalf,k} 
	\qquad \text{where} \qquad 
	\omega_{\ThreeHalf,k} \equiv \sqrt{k^2+a^2m^2} \\ 
	\lambda_{\Half\pm} 
	& = \pm\omega_{\Half,k} 
	\qquad \text{where} \qquad 	
	\omega_{\Half,k} \equiv \sqrt{\cs^2 k^2+a^2m^2} \label{eq:evaluesHalf}
\end{align}
\end{subequations}
where $\cs^2 = C_A^2 + C_B^2$ is given by \eref{eq:ca2pcb2}.  
These relations let us interpret $\cs$ as the time-dependent sound speed of the helicity-1/2 modes, while the helicity-3/2 modes have unit sound speed.  
The coupled first-order mode equations \pref{eq:32_mode_eqn_chiA_chiB} can be combined to obtain second-order mode equations for the helicity-3/2 mode functions 
\begin{subequations}\label{eq:2nd_order_mode_eqns}
\begin{equation}
\begin{split}
	\partial_\eta^2 \chi_{A,\ThreeHalf,k} & = - \bigl( \omega_{\ThreeHalf,k}^2 + i a^2 Hm \bigr) \ \chi_{A,\ThreeHalf,k} \\ 
	\partial_\eta^2 \chi_{B,\ThreeHalf,k} & = - \bigl( \omega_{\ThreeHalf,k}^2 - i a^2 Hm \bigr) \ \chi_{B,\ThreeHalf,k} 
\end{split}
\end{equation}
and the helicity-1/2 mode functions
\begin{equation}
\begin{split}
	\partial_\eta^2 \chi_{A,\Half,k} & = - \bigl( \omega_{\Half,k}^2 + i a^2 Hm \bigr) \ \chi_{A,\Half,k} - \bigl( i k \partial_\eta C_A \bigr) \ \chi_{B,\Half,k} \\ 
	\partial_\eta^2 \chi_{B,\Half,k} & = - \bigl( \omega_{\Half,k}^2 - i a^2 Hm \bigr) \ \chi_{B,\Half,k} - \bigl( i k \partial_\eta C_A \bigr) \ \chi_{A,\Half,k} 
	\per
\end{split}
\end{equation}
\end{subequations}
Notice that the helicity-3/2 mode functions decouple, whereas the helicity-1/2 mode functions remain coupled through the $\partial_\eta C_A$ term.  
We will see below that the mode functions must satisfy a constraint \pref{eq:normalization}, which arises from quantizing the field, and one can easily verify that the mode equations \pref{eq:32_mode_eqn_chiA_chiB} respect this constraint.  

\subsection{Helicity-3/2 modes}

The helicity-3/2 mode equation \pref{eq:32_mode_eqn_ThreeHalf} can be brought into an approximately-diagonal form by performing a time-dependent transformation\footnote{The transformation between basis is nothing more than a Bogoliubov  transformation.  } 
\begin{align}\label{eq:change_basis_32}
	\begin{pmatrix} \chi_{+,\ThreeHalf,k} \\ \chi_{-,\ThreeHalf,k} \end{pmatrix} = \calU_{\ThreeHalf} \begin{pmatrix} \chi_{A,\ThreeHalf,k} \\ \chi_{B,\ThreeHalf,k} \end{pmatrix} 
	\qquad \text{where} \qquad 
	\calU_{\ThreeHalf} = \begin{pmatrix} \cos\varphi_{\ThreeHalf}  & \sin\varphi_{\ThreeHalf}  \\ -\sin\varphi_{\ThreeHalf} & \cos\varphi_{\ThreeHalf}  \end{pmatrix} 
	\per
\end{align} 
The matrix $\mathcal{A}_{\ThreeHalf}$ is diagonalized, $\mathcal{U}_{\ThreeHalf}\, {\cal A}_{\ThreeHalf}\,{\mathcal{U}}^T_{\ThreeHalf} =\mathrm{diag}(\omega_{\ThreeHalf,k}, -\omega_{\ThreeHalf,k})$ if the rotation angle $\varphi_{\ThreeHalf}$ satisfies the equalities $\cos\varphi_{\ThreeHalf} = \sqrt{\omega_{\ThreeHalf,k}+am}/\sqrt{2\omega_{\ThreeHalf,k}}$ and $\sin\varphi_{\ThreeHalf} = \sqrt{\omega_{\ThreeHalf,k}-am} / \sqrt{2\omega_{\ThreeHalf,k}}$.  
In the quasi-diagonal basis, the mode equation \pref{eq:32_mode_eqn_ThreeHalf} becomes 
\begin{align}\label{eq:Diracnewbasis32}
	\begin{pmatrix} \partial_\eta\chi_{+,\ThreeHalf,k}\\ \partial_\eta\chi_{-,\ThreeHalf,k} \end{pmatrix} 
	& = \begin{pmatrix} -i\omega_{\ThreeHalf,k} & \partial_\eta\varphi_{\ThreeHalf}  \\ -\partial_\eta\varphi_{\ThreeHalf} & i\omega_{\ThreeHalf,k} \end{pmatrix} \begin{pmatrix} \chi_{+,\ThreeHalf,k} \\ \chi_{-,\ThreeHalf,k} \end{pmatrix}    \com
\end{align}
where $\omega_{\ThreeHalf,k}$ is given by \eref{eq:evaluesThreeHalf} and where 
\begin{align}
\partial_\eta\varphi_{\ThreeHalf} = -\frac{1}{2}\frac{ka^2Hm}{\omega_{\ThreeHalf,k}^2} 
\per
\end{align}
Notice how the mode functions are diagonal for $\partial_\eta\varphi_{\ThreeHalf} = 0$, and the solutions are $\chi_{\pm,\ThreeHalf,k} = e^{\mp i \omega_{\ThreeHalf,k}\eta}$.  
The off-diagonal terms vanish for $m = 0$, which reflects the fact that $m$ is the order parameter for the breaking of Weyl invariance in the helicity-3/2 modes.  
Additionally $\partial_\eta\varphi_{\ThreeHalf} \to 0$ as $a\to0$ at the early times and as $H \to 0$ at late times.  In other words $\chi_{+,\ThreeHalf,k}$ would describe positive-frequency solutions and $\chi_{-,\ThreeHalf,k}$ would describe negative-frequency solutions \textit{if the evolution is adiabatic} ($\partial_\eta\varphi_{\ThreeHalf}\to0$).   The function $\partial_\eta\varphi_{\ThreeHalf}$ drives mixing between positive- and negative-frequency solutions.  Thus an initial condition consisting of only positive-frequency modes can evolve into an admixture of negative frequency modes, which is a signal of particle production.

\subsection{Helicity-1/2 modes}

Now we turn to the helicity-1/2 modes that obey \eref{eq:32_mode_eqn_OneHalf}.  
It is convenient to rewrite $\mathcal{A}_{\Half}$ as
\begin{align}\label{eq:AHALFH}
\mathcal{A}_{\Half} = \begin{pmatrix} am & \cs\,k\,e^{i\zeta} \\ \cs\,k\,e^{-i\zeta} & -am \end{pmatrix} \com
\end{align}
where the time-dependent phase $\zeta(\eta)$ obeys $\cos\zeta=C_A/\cs$ and $\sin\zeta=C_B/\cs$.  Since $\mathcal{A}_{\Half}$ is Hermitian, it can be diagonalized by a unitary transformation 
\begin{align}\label{eq:change_basis_12}
	\begin{pmatrix} \chi_{+,\Half,k} \\ \chi_{-,\Half,k} \end{pmatrix} = \calU_{\Half} \begin{pmatrix} \chi_{A,\Half,k} \\ \chi_{B,\Half,k} \end{pmatrix} 
	\qquad \text{where} \qquad 
	\mathcal{U}_{\Half} = \begin{pmatrix}  \cos\varphi_{\Half}\ e^{-i\zeta/2} & \sin\varphi_{\Half}\ e^{i\zeta/2} \\ -\sin\varphi_{\Half}\ e^{-i\zeta/2} & \cos\varphi_{\Half}\ e^{i\zeta/2} \end{pmatrix} 
	\com
\end{align}
and where the time-dependent angle $\varphi_{\Half}(\eta)$ should be chosen to satisfy $\cos\varphi_{\Half} = \sqrt{\omega_{\Half,k}+am}/\sqrt{2\omega_{\Half,k}}$ and $\sin\varphi_{\Half} = \sqrt{\omega_{\Half,k}-am} / \sqrt{2\omega_{\Half,k}}$, which gives $\mathcal{U}_{\Half} \mathcal{A}_{\Half} \mathcal{U}_{\Half}^{\,\dagger} = \mathrm{diag}(\omega_{\Half,k},-\omega_{\Half,k})$.  
This transformation puts the mode equation \pref{eq:32_mode_eqn_OneHalf} into an approximately-diagonal form
\begin{align}\label{eq:Diracnewbasis12}
	\begin{pmatrix} \partial_\eta\chi_{+,\Half,k} \\ \partial_\eta\chi_{-,\Half,k} \end{pmatrix} 
	& = \begin{pmatrix} -i\omega_{\Half,k} & \partial_\eta\varphi_{\Half} \\ -\partial_\eta\varphi_{\Half} & i\omega_{\Half,k} \end{pmatrix}
	\begin{pmatrix} \chi_{+,\Half,k} \\ \chi_{-,\Half,k} \end{pmatrix} 
	+ i\dfrac{\partial_\eta\zeta}{2\omega_{\Half,k}} \begin{pmatrix}-a\,m& \cs\,k \\ -\cs\,k & a\,m \end{pmatrix} \begin{pmatrix} \chi_{+,\Half,k} \\ \chi_{-,\Half,k} \end{pmatrix} \com
\end{align}
where the derivatives are given by 
\begin{align}
	\partial_\eta\varphi_{\Half} & = -\frac{1}{2} \frac{kam}{\cs\omega_{\Half,k}^2} \Bigl[ a H \cs^2 - \bigl( C_A\partial_\eta C_A + C_B\partial_\eta C_B \bigr) \Bigr] \nn
	\partial_\eta\zeta & = \cs^{-2} \bigl( C_A\,\partial_\eta C_B - C_B\, \partial_\eta C_A \bigr) 
\per
\end{align}
Recall that $C_A$ and $C_B$ were defined in \eref{eq:CA_CB_def}.  
At late times, the spacetime becomes asymptotically Minkowski meaning $a \to 1$, $H \to 0$, and $R \to 0$, and if $m \to \const$ as well, then $C_A \to 1$ and $C_B \to 0$ and both of the derivatives above vanish.  
At early times, in the quasi-de Sitter era, both of the derivatives are also small, since $H$, $R$, and $m$ are slowly varying.  
Similar to the helicity-3/2 case, in the asymptotic early and late times $\chi_{\pm,\Half,k}$ describe positive and negative frequency modes, and for the study of gravitational particle production, we will be interested in the value of $\chi_{-,\Half,k}$ at late times.

\subsection{Quantization}

In the quantized theory, the mode functions discussed previously, along with a set of ladder operators, are used to construct the field operator.  
The field operator and its conjugate momentum are required to obey an anticommutation relation, which imposes a normalization condition on the mode functions.  
In terms of the non-diagonal mode equations \pref{eq:32_mode_eqn_chiA_chiB}, the normalization conditions are 
\begin{subequations}\label{eq:normalization}
\begin{align}
	|\chi_{A,\ThreeHalf,k}|^2 + |\chi_{B,\ThreeHalf,k}|^2 = 1 
	\qquad \text{and} \qquad 
	|\chi_{A,\Half,k}|^2 + |\chi_{B,\Half,k}|^2 = 1 
	\com
\end{align}
and in terms of the quasi-diagonal mode equations from \erefs{eq:Diracnewbasis32}{eq:Diracnewbasis12} these conditions are 
\begin{align}
	|\chi_{+,\ThreeHalf,k}|^2 + |\chi_{-,\ThreeHalf,k}|^2 = 1 
	\qquad \text{and} \qquad 
	|\chi_{+,\Half,k}|^2 + |\chi_{-,\Half,k}|^2 = 1 
	\per
\end{align}
\end{subequations}
The relative plus sign between each of the pairs of terms is a consequence of Fermi-Dirac statistics, and it implies $|\chi|^2 \leq 1$ for each mode function.

\section{Catastrophic gravitational particle production} \label{sec:GPP}

We now turn our attention to the gravitational production of spin-3/2 particles at the end of inflation.  
In order to calculate the spectrum and total number of gravitationally-produced particles, we solve the mode equations subject to the Bunch-Davies initial condition, and we read off the late-time behavior of the mode functions.  
To begin this section, we first describe how the initial conditions were chosen and how the spectrum was extracted from the late-time solutions, and then we present our numerical results.

\subsection{Initial conditions \& particle number spectrum}

For the helicity-3/2 modes, the mode functions are required to obey the Bunch-Davies initial condition
\begin{align}\label{eq:BD_32_pm}
	\lim_{\eta \to -\infty} \chi_{+,\ThreeHalf,k}(\eta) = 1
	\qquad \text{and} \qquad 
	\lim_{\eta \to -\infty} \chi_{-,\ThreeHalf,k}(\eta) = 0
	\com
\end{align}
such that only positive-frequency modes are present at early times.  In terms of the mode functions in the non-diagonal basis, we have 
\begin{align}\label{eq:BD_32_AB}
	\lim_{\eta \to -\infty} \chi_{A,\ThreeHalf,k}(\eta) = 1 / \sqrt{2}
	\qquad \text{and} \qquad 
	\lim_{\eta \to -\infty} \chi_{B,\ThreeHalf,k}(\eta) = 1 / \sqrt{2}
	\com
\end{align}
which is obtained by inverting the transformation in \eref{eq:change_basis_32}, using the initial condition in \eref{eq:BD_32_pm}, and noting that $a \to 0$ and $\omega_{\ThreeHalf,k}\to k$ as $\eta \to -\infty$.  Evolution under the mode equation will populate the negative-frequency modes.  We are interested in the solutions at late times, meaning that any given mode $k$ has become nonrelativistic $k \ll am$ and passed inside the horizon $k \gg aH$.  
Having solved the mode equations, the spectrum of helicity-3/2 particles that results from gravitational particle production is then calculated from the late-time amplitude of the negative-frequency modes as~\cite{Chung:2011ck}
\begin{align}\label{eq:nk_32}
	n_{\ThreeHalf,k} = \frac{k^3}{2\pi^2} \, |\beta_{\ThreeHalf,k}|^2 
	\qquad \text{for} \qquad 
	|\beta_{\ThreeHalf,k}|^2 \equiv \lim_{\eta \to \infty} |\chi_{-,\ThreeHalf,k}(\eta)|^2 
	\per
\end{align}
Here $n_{\ThreeHalf,k}$ is the comoving number density of helicity-3/2 particles per logarithmically-spaced wavenumber interval.  
As a matter of numerical stability, we find it easier to solve the non-diagonal mode equations \pref{eq:32_mode_eqn_ThreeHalf}.  
Then the late-time solution is $\chi_{-,\ThreeHalf,k} \approx \chi_{B,\ThreeHalf,k}$, since the dispersion relation \pref{eq:evaluesThreeHalf} is approximately $\omega_{\ThreeHalf,k} \approx am$, and $\varphi_{\ThreeHalf} \approx 0$ in the transformation from \eref{eq:change_basis_32}.  

We also implement the Bunch-Davies initial condition for the helicity-1/2 modes, which is now written as 
\begin{align}\label{eq:BD_12_pm}
	\lim_{\eta \to -\infty} \chi_{+,\Half,k}(\eta) = 1
	\qquad \text{and} \qquad 
	\lim_{\eta \to -\infty} \chi_{-,\Half,k}(\eta) = 0
	\per
\end{align}
After the mode equations are solved, the spectrum of helicity-1/2 particles is extracted by calculating 
\begin{align}\label{eq:nk_12}
	n_{\Half,k} = \frac{k^3}{2\pi^2} \, |\beta_{\Half,k}|^2 
	\qquad \text{for} \qquad 
	|\beta_{\Half,k}|^2 \equiv \lim_{\eta \to \infty} |\chi_{-,\Half,k}(\eta)|^2 
	\per
\end{align}
To express the initial condition in terms of the non-diagonal basis mode functions, we invert the transformation in \eref{eq:change_basis_12} and use \eref{eq:BD_12_pm}, which gives 
\begin{align}\label{eq:Halfinitial}
	\lim_{\eta\to-\infty} \, \chi_{A,\Half,k} & = \lim_{\eta\to-\infty} \, \frac{1}{2}\left( \sqrt{1+\frac{C_A}{\cs}} + i\sqrt{1-\frac{C_A}{\cs}} \right) \nn
	\lim_{\eta\to-\infty} \, \chi_{B,\Half,k} & = \lim_{\eta\to-\infty} \, \frac{1}{2}\left( \sqrt{1+\frac{C_A}{\cs}} - i\sqrt{1-\frac{C_A}{\cs}} \right) 
	\per
\end{align}
Although $\varphi_{\Half} \to 0$ at early times, the phase $\zeta$ depends nontrivially on $C_A$ and $\cs$, which leads to the expression above.  
Once again, for numerical stability we solve the non-diagonal mode equations \pref{eq:32_mode_eqn_OneHalf} and extract the late-time solution as $\chi_{-,\Half,k} \approx \chi_{B,\Half,k}$.  

To close this section we remark upon the impact of Fermi-Dirac statistics and Pauli blocking.  Recall that the Fermi-Dirac statistics of the Rarita-Schwinger field leads to normalization conditions on the mode functions \pref{eq:normalization}, implying $|\chi_{-,\Half,k}|^2 \leq 1 $ and $|\chi_{-,\Half,k}|^2 \leq 1$.  In other words, the occupation number cannot exceed $1$ for each Fourier mode and each helicity, which is the phenomenon of Pauli blocking.  
As a result the spectra from \erefs{eq:nk_32}{eq:nk_12} are capped by $n_{\ThreeHalf,k} = n_{\Half,k} \leq k^3 / 2\pi^2$.  

In the next subsection we will observe that a ``catastrophic'' particle production causes these spectra to saturate at their upper limit for arbitrarily large $k$ up to the UV cutoff.  

\subsection{Catastrophic particle production--numerical results}

Using the mode equations and initial conditions described above, we solve for the evolution of the mode functions and extract the spectrum $n_k$ of gravitationally produced particles.  The total number of particles in a region of space with physical radius $R$ is given by,
\begin{equation}
N_{\rm tot}(R,t) = \frac{4}{3a^3(t)} \pi R^3 \int_{1/R} \frac{{\rm d}k}{k}  \, n_k .
\label{eq:Ntot}
\end{equation}
This is divergent if $n_k$ grows with increasing $k$, or equivalently, if $|\beta_k|^2$ falls more slowly than $k^{-3}$. The particle number at late times (when the particles are non-relativistic) in a bounded region of space is a physical observable, of the kind advocated for by \cite{Senatore:2012nq}.\footnote{From \cite{Senatore:2012nq}, p.5., ``This procedure corresponds to counting particles present in the region after reheating and using the
enclosed number of particles as a measure of the physical size of the region... In the late universe
we could use the number of dark matter particles or equivalently the mass in dark matter particles
enclosed in the volume. Not only is this measure of size a well defined physical choice, but the
amplitude of perturbations at a given mass scale is directly related to the abundance of objects of
that mass that will form in the late Universe." } Thus a UV divergence of $N$ is physical, and its implications must be understood. We refer to this divergent particle number as ``catastrophic'' particle production.

We compute the particle production
for both the helicity-1/2 and the helicity-3/2 modes, separately.  
For simplicity, both aesthetic and computational, our numerical study focuses on the canonical massive Rarita-Schwinger model, namely a spin-$3/2$ field with constant mass.  
For the background spacetime, we assume a quadratic inflaton potential at the end of inflation.  

\begin{figure}[t!]
\begin{center}
\includegraphics[width=0.495\textwidth]{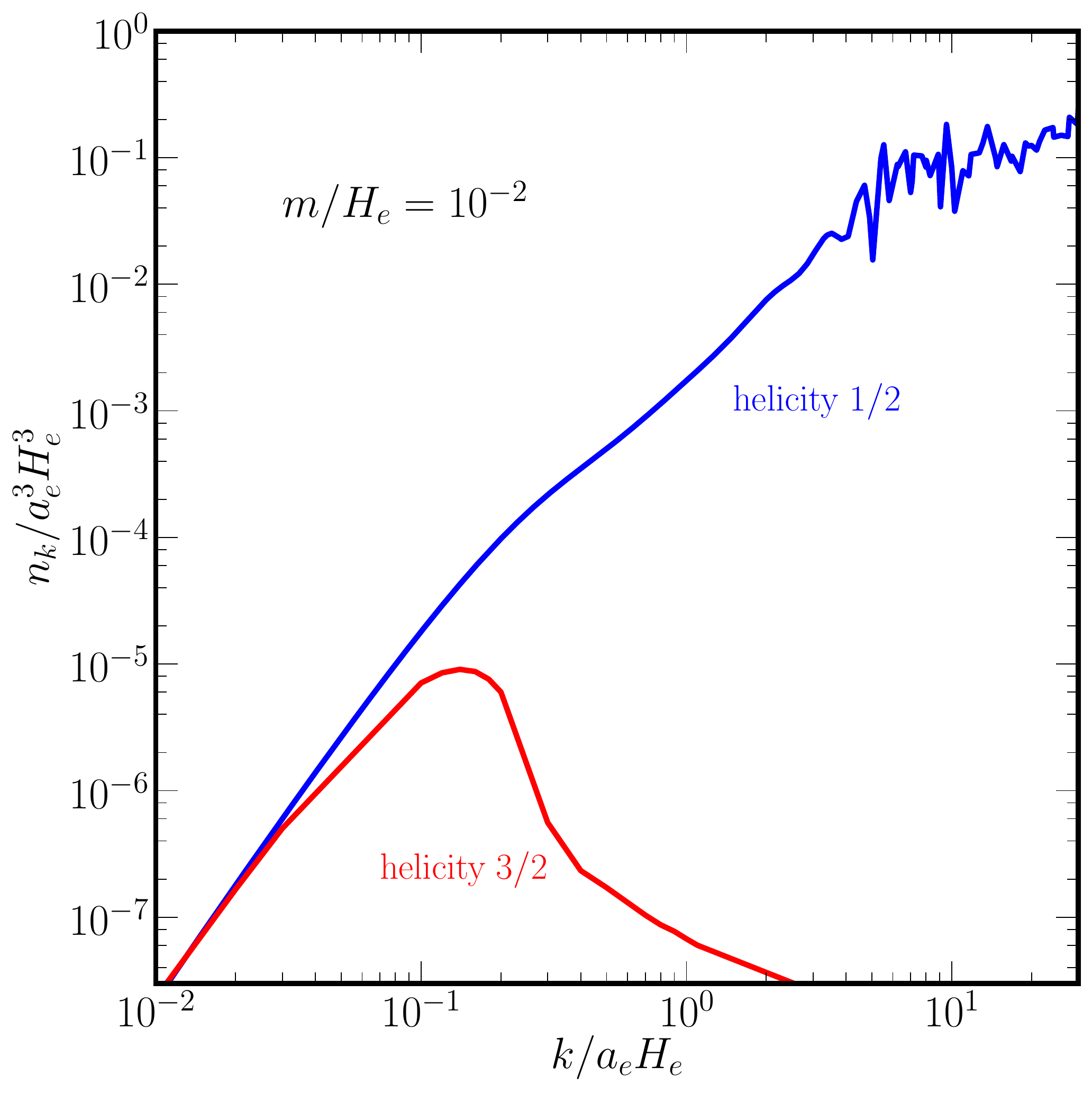}
\includegraphics[width=0.495\textwidth]{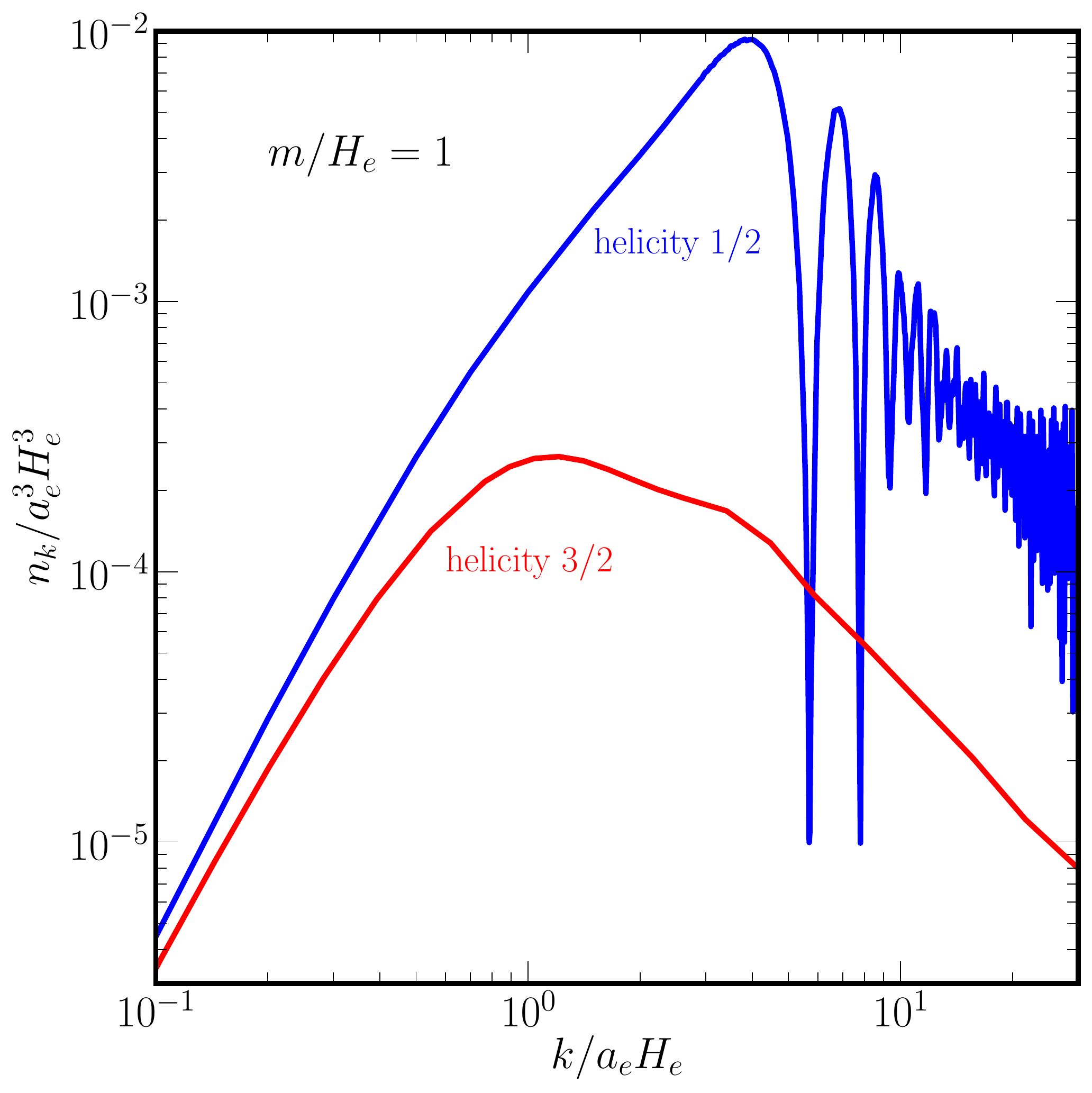}
\caption{Values of $n_k/a_e^3H_e^3 = k^3|\beta_k|^2/2\pi^2$ as a function of $k$ for two values of $m/H_e$ for helicity-3/2 and helicity-1/2. (The helicity-3/2 results are identical to the results for a Dirac fermion.) If $m/H_e=0.01$ then $\cs^2$ vanishes during the evolution (see \fref{fig:r23_gang}) and catastrophic particle production results.  If $m/H_e=1.0$ the $\cs^2$ does not vanish in the evolution (but is less than unity).  There is no catastrophic GPP, but particle production is enhanced over helicity-3/2 which has $\cs^2=1$. The oscillations for $k\gtrsim 5$ can be compared to the oscillations seen in Figure 3 of Giudice, Riotto, and Tkachev \cite{Giudice:1999am}. \label{fig:spectrum_vanilla} }  
\end{center}
\end{figure}

Numerical examples are shown in \fref{fig:spectrum_vanilla}.  
The spectrum of helicity-3/2 modes are represented by the red curves in the two panels.  
At low-$k$ the spectra rise like $n_k \propto k^3$, corresponding to to $|\beta_k|^2 \approx 1$, which is consistent with these modes being maximally occupied and subject to Pauli blocking.  
At high-$k$ the spectra drop off, which is consistent with the evolution of these modes remaining approximately adiabatic.  
Since the mode equation for the helicity-3/2 modes of the Rarita-Schwinger field \pref{eq:32_field_eqn_3d} is identical to the mode equation for a spin-1/2 Dirac field, we can compare our numerical results against previous studies in the literature~\cite{Chung:2011ck}, and by doing so we find good agreement for the shape and amplitude of the spectrum.  

The spectrum of helicity-1/2 modes are given by the blue curves in each panel.  
In the right panel where the Rarita-Schwinger field's mass is $m = H_e$, we observe that the spectrum peaks at $k \sim \mathrm{few} \times a_e H_e$, and drops off for larger $k$. 
However, in the left panel where the Rarita-Schwinger field is lighter, here only $m = 0.01 H_e$, we observe that the spectrum continues to rise well past $k = 2 a_e H_e$.  
Numerical limitations prevent us from evaluating the spectrum for larger $k$, but we expect that the spectrum will continue to rise as $n_k \propto k^3$ indefinitely.  
We justify this claim in the following subsections, where we argue that the time-dependent sound speed of the helicity-1/2 modes plays a crucial role in amplifying the high-$k$ spectrum, which saturates the Pauli blocking limit.  
By contrast, the helicity-3/2 modes have a static sound speed and a time-dependent effective mass, which is responsible for their particle production.  

Let us reiterate the main message of \fref{fig:spectrum_vanilla}.  
The spectrum of helicity-1/2 modes in a spin-3/2 Rarita-Schwinger field with constant mass $m = 0.01 H_e$ appears to grow without bound as $n_k \propto k^3$.  We refer to this UV-dominated spectrum as ``catastrophic'' gravitational particle production.  
The total number of particles produced is calculated by integrating this spectrum over all $k$, Eq.~\eqref{eq:Ntot}, and this quantity acquires a power-law sensitivity to the UV cutoff.  
Similarly the net energy of these particles is UV dominated.  
One might argue that the calculation is not self-consistent in this regime, since we are solving for the evolution of a spectator field in the external background induced by the inflaton, and neglecting the backreaction of the produced particles on the classical background, which can be expected to be significant once their energy densities become comparable.
Further, any theory containing gravity is an effective field theory with a cutoff at (or below) the Planck scale, so the divergent particle number should also be cutoff at some scale. We will instead argue that the calculation displaying catastrophic GPP is inconsistent because effective field theory itself has broken down, precisely due to the on-shell particles with momenta equal to the cutoff. This breakdown of the EFT that manifests itself only when the gravitino is quantized is reminiscent of swampland conjectures that deal with EFT's that become inconsistent when gravity is quantized. We discuss this point further in \sref{sec:conc}.  

On the other hand, for the helicity-3/2 modes or for the helicity-1/2 modes with $m/H_e \gtrsim 1$ we do not observe catastrophic GPP.  
Here we claim that the numerical calculation is robust.  
If the Rarita-Schwinger field is stable, then its spin-3/2 particle provides a dark matter candidate, and the GPP calculation here furnishes a prediction for the dark matter relic abundance.  
We will explore this dark matter model further in future work.  

\subsection{Conditions for a vanishing sound speed}

An important feature of the mode equations described above is that the sound speed $\cs^2 = C_A^2 + C_B^2$ will oscillate at the end of inflation, and that these oscillations can allow $\cs$ to vanish multiple times.  
We will see shortly that this behavior is a crucial determinant in whether catastrophic GPP takes place.  
We begin here by deriving an analytic expression for the squared sound speed and determining the conditions under which it vanishes at the end of inflation.  

We assume that at the end of inflation the energy density is dominated by an inflaton field $\phi$ with potential $V(\phi)$.\footnote{The potential at the end of inflation may well differ from the potential that determines curvature fluctuations on scales probed by observations.}  
Using the Friedmann equations, $R = -\Mpl^{-2}(\rho-3p)$ and $H^2 = \Mpl^{-2} \rho/3$, \eref{eq:ca2pcb2} becomes
\begin{align}\label{eq:r2again}
	\cs^2 = \frac{(p-3m^2\Mpl^2)^2}{(\rho+3m^2\Mpl^2)^2} + \frac{\Mpl^4(2 a^{-1}\partial_\eta\, m)^2}{(\rho+3m^2\Mpl^2)^2} \per
\end{align}
Recall that $\cs^2$ enters the dispersion relation for the helicity-1/2 mode \pref{eq:evaluesHalf}, which is written as $\omega_{\Half,k}^2 = \cs^2k^2+a^2m^2$.  
At late times $\rho \to 0$, $p \to 0$, and if $\partial_\eta m \to 0$ as well, then $\cs^2 \to 1$, giving the usual dispersion relation.  
At early times, during the quasi-de Sitter phase of inflation, $p \approx - \rho$ and if also $\partial_\eta m \approx 0$ then $\cs^2 \approx 1$.  
However, near the end of inflation, the sound speed is expected to deviate from $1$.  

To better understand how the sound speed varies with time, we need to specify a model for the inflaton so that $\rho(\eta)$ and $p(\eta)$ can be calculated, and we must also specify how the spin-3/2 field's mass varies with time, so that $\partial_\eta m$ can be calculated.  For simplicity we consider a quadratic potential, and the inflaton's energy density and pressure are given by 
\begin{align}
	\rho = \half\dot{\phi}^2 + \half m^2_\phi\phi^2
	\qquad \text{and} \qquad 
	\rho = \half\dot{\phi}^2 - \half m^2_\phi\phi^2 
	\per
\end{align}
Here and below ``dot'' denotes $d/dt$ where $t$ is cosmic time.  
The inflaton mass is related to $H_e$ by $m_\phi = 2H_e\Mpl/\phi_e$.  Again, we emphasize that this is the potential that describes the oscillation of the inflaton about its minimum after inflation, and it need not describe the inflaton potential when the scale factor was about 50 $e$-folds from the end of inflation when scales important for observable curvature fluctuations crossed the Hubble radius.  For the spin-3/2 field's mass, we consider the simplest scenario first and assume that it is a constant so that $\partial_\eta m = 0$.  
Then it is possible for the squared sound speed \pref{eq:r2again} to vanish when $p = 3 m^2 \Mpl^2$.  

\begin{figure}[!t]
\begin{center}
\includegraphics[width=0.65\textwidth]{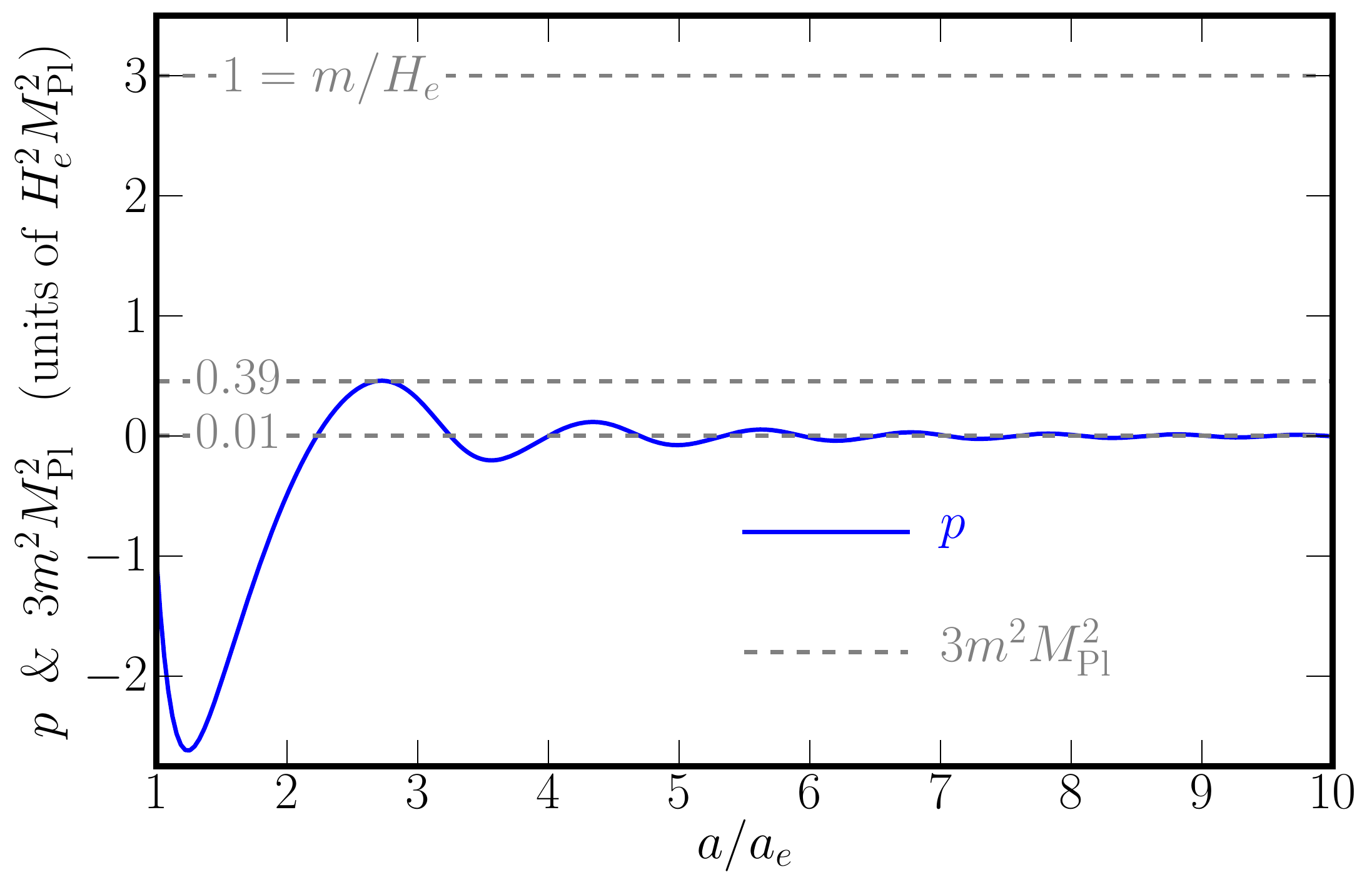}\\
\caption{The evolution of the pressure at the end of inflation in the chaotic inflation model.  For a given choice of the spin-3/2 field's mass $m$, the sound speed $\cs$ vanishes when $p = 3 m^2 \Mpl^2$ corresponding to intersections of the blue curve with the corresponding gray-dashed curve.  
\label{fig:p}}  
\end{center}
\end{figure}

\begin{figure}[ph!]
\begin{center}
\includegraphics[width=0.81\textwidth]{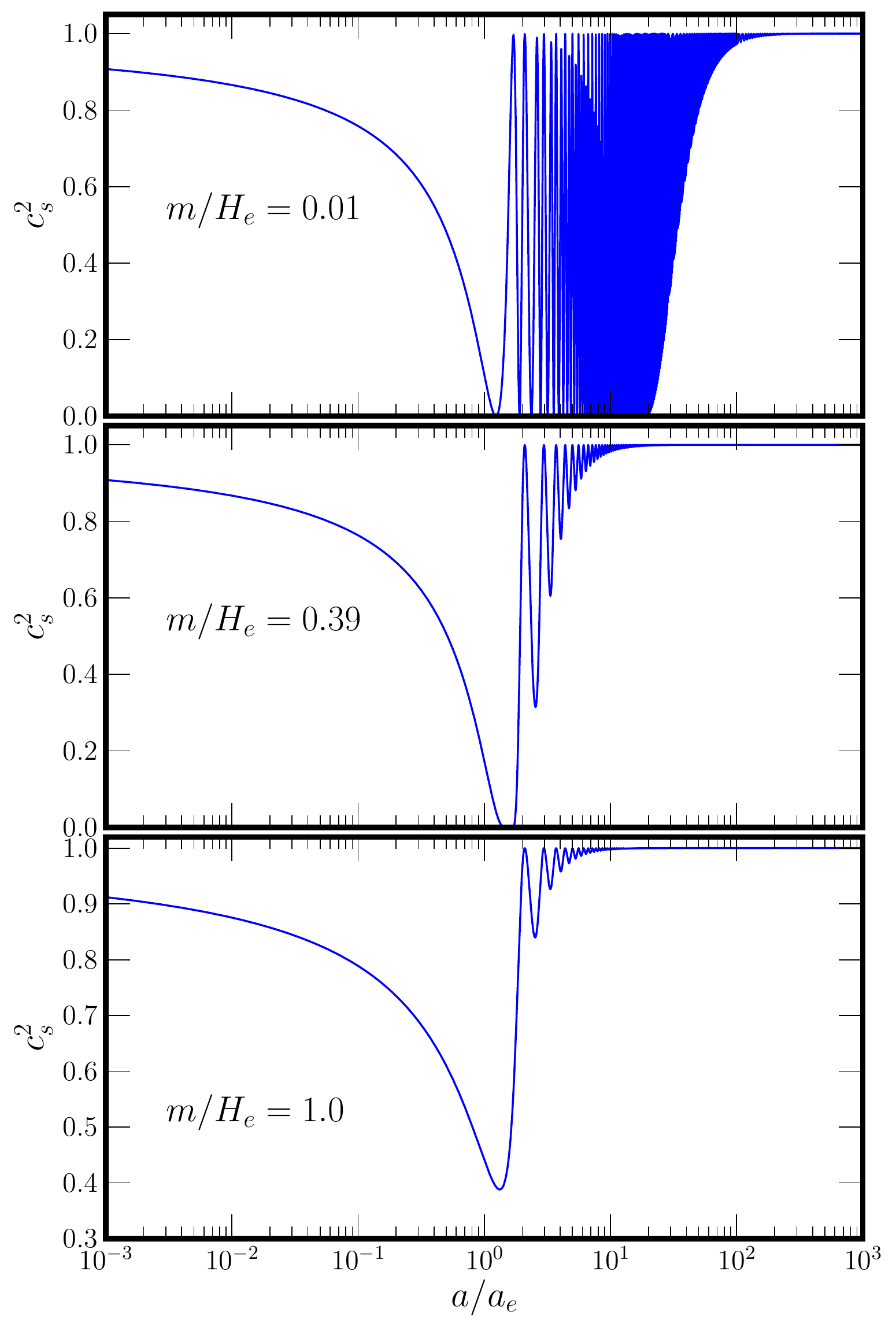}\\
\caption{The evolution of $\cs^2$ as a function of $a/a_e$ for three values of $m/H_e$.  As illustrated in the upper panel, for $m/H_e\ll1$ after inflation, $\cs^2=0$ many times in the oscillation of the inflaton field.  If $m/H_e \simeq 0.39$, while $\cs^2$ still oscillates it vanishes only once (see the middle panel).  As seen in the lower panel, if $m/H_e\gtrsim0.39$  the oscillations become less important and the sound speed never vanishes.  (Note the different lower limit in the bottom panel.)
\label{fig:r23_gang}}  
\end{center}
\end{figure}

It is easy to see that the sound speed cannot vanish \textit{during} inflation: since $\rho + 3p < 0$ and $\rho>0$, it follows that $p < 0$ and $\cs^2 > 0$.  We study the evolution of $\cs(\eta)$ after inflation by numerically solving the inflaton's field equation to calculate the pressure $p$.  In \fref{fig:p} we show the pressure $p$ as as a function of $a/a_e$, and we compare with $3 m^2 \Mpl^2$ for several different values of the spin-3/2 field's mass $m$.  If the spin-3/2 field's mass is static, $\partial_\eta m = 0$, then the sound speed $\cs$ vanishes when the pressure $p$ equals $3 m^2 \Mpl^2$ [see \eref{eq:r2again}].  The blue curve shows the pressure $p$ (in units of $H_e^2 \Mpl^2$) and the gray-dashed lines show values of $3 m^2 \Mpl^2$ (same units) for different choices of the spin-3/2 field's mass $m/H_e$.  For a given mass choice, the sound speed vanishes $\cs = 0$ whenever the blue and gray-dashed curves intersect.  In this example, for $m/H_e\gtrsim 0.39$ the sound speed never vanishes, for $m/H_e \simeq 0.39$ it vanishes just once, and for $m/H_e \ll 0.39$ it vanishes many times as the pressure oscillates at the end of inflation.

The evolution of the sound speed is illustrated directly in \fref{fig:r23_gang}, which presents $\cs^2$ for the same three values of $m/H_e$ that appear in \fref{fig:p}.  Every time $\cs^2$ touches zero the sound speed vanishes.  Note that the upper limit to the sound speed is unity, so there is no superluminal propagation.  As the pressure damps away with time, it eventually decreases below the threshold $|p| < 3 m^2 \Mpl^2$ and subsequently the sound speed has no more zero-crossings and it remains positive.  

\subsection{Non-adiabatic particle production from vanishing sound speed}

Here we seek to clarify how the vanishing sound speed leads to efficient production of the helicity-1/2 modes.  
In general, particle production results from non-adiabatic mode evolution.  
If the mode equation takes the form of $\partial_\eta^2 \chi_k + \omega_k^2(\eta) \chi_k = 0$ with time-dependent dispersion relation $\omega_k(\eta)$, then 
\begin{equation}\label{eq:Ak_def}
	A_k(\eta) \equiv \frac{\partial_\eta\omega_k}{\omega_k ^2} 
\end{equation}
provides a dimensionless measure of the nonadiabaticity.  When the dispersion relation changes rapidly and $|A_k| \gg 1$, efficient particle production results.  
(See, e.g., \cite{Kofman:1997yn,Amin:2014eta} in the context of preheating.)  

The mode equation for the helicity-1/2 modes \pref{eq:2nd_order_mode_eqns} has the dispersion relation $\omega_{k,\Half}^2 = \cs^2 k^2 + a^2 m^2$ from \eref{eq:evaluesHalf}. For simplicity, we consider the case of a constant mass, as studied in the numerics. The corresponding adiabaticity parameter is 
\begin{equation}\label{eq:shutoff}
	A_k = \frac{a^3 H m^2 + \cs (\partial_\eta \cs) k^2}{(a^2 m^2 + \cs^2 k^2)^{3/2}} 
\end{equation}
for $\partial_\eta m = 0$ and $\partial_\eta a = a^2 H$.  As illustrated in Fig.~\ref{fig:r23_gang}, if $m \lesssim 0.39 H_e$, the sound speed $\cs$ will vanish, and may do so many times. At the moments when the sound speed vanishes, the adiabaticity parameter is given by,
\begin{equation}
	A_k|_{\cs=0} = \frac{H}{m} \com
	\label{eq:Akvs0}
\end{equation}
which is manifestly larger than unity since $m/H<1$ is a requirement for $\cs$ to ever vanish.  

Strikingly, Eq.~\eqref{eq:Akvs0} is  independent of $k$, which implies that particles of arbitrarily high momentum can be produced. This is precisely the ``catastrophrophic particle production.'' This is very different from the standard situation of particle production, e.g., of spin-0 fields with $\cs = 1$ and oscillating nonzero mass $m^2(\eta) > 0$, where large values of $k$ act to {\it shut-off} the violation of adiabaticity by making the denominator in \eref{eq:shutoff} large.  

In fact, the violation of adiabaticity is even more severe than suggested by Eq.~\eqref{eq:Akvs0}. The maximum value of $A_k$, and hence maximum violation of adiabaticity, occurs shortly before and shortly after $\cs=0$, when $\cs$ is small but finite. This occurs when the second term in the numerator of Eq.~\eqref{eq:shutoff} is dominant, while the first term is dominant in the denominator of Eq.~\eqref{eq:shutoff}. In this case, one finds $A_k$ is given by,
\begin{equation}\label{eq:Ak_approx}
	A_k \approx \frac{\cs (\partial_\eta \cs) k^2}{a^3 m^3} \per
\end{equation}
This approximation is valid while $k$ lies in the range $\cs^2 \ll a^2 m^2/k^2 \ll \cs \, |\partial_\eta \cs|/aH$, which occurs, for arbitrarily large $k$, for a transient period before $\cs$ vanishes, and again for a period after. Combining this constraint with Eq.~\eqref{eq:Ak_approx}, we find,
\begin{equation}\label{eq:Ak_range}
	\frac{H}{m} \ll  |A_k|  \ll \frac{|\partial_t \ln \cs|}{H} \, \frac{H}{m}
	\com
\end{equation}
which brackets the nonadiabaticity away from $\cs=0$ for a mode of arbitrary $k$, including arbitrarily large $k$, for example, $k$ equal to the UV cutoff of the theory.   This occurs for a very short period of time, reminiscent of preheating, wherein adiabaticity is violated as the inflaton crosses zero~\cite{Kofman:1997yn}. In particular, non-adiabatic particle production from a rapidly-varying mass parameter has been studied in Refs.~\cite{Amin:2015ftc,Garcia:2019icv,Garcia:2020mwi}, where it is observed that an arbitrarily rapid change in the mass parameter leads to particle production in arbitrarily high-$k$ modes.

Before closing this section, we make a few additional comments. First, we note that the role of the vanishing sound speed in adiabaticity violation, Eq.~\eqref{eq:Akvs0}, sheds light on the impact of backreaction on gravitino production.  Since the production of gravitinos of all $k$ occurs simultaneously (for an example time evolution, we refer the reader to \cite{Kolb:2021nob}), there is no intermediary time period wherein the backreaction of low-$k$ modes on the background cosmology can prevent the production of the high-$k$ modes. This indicates that the breakdown of effective field theory, to be discussed in Sec.~\ref{sec:GSC} and Sec.~\ref{sec:conc}, is robust to the inclusion of backreaction.

We also note that in the limit $\Mpl^2 \rightarrow \infty$ with $H$ held fixed, which in the context of supergravity corresponds to the decoupling limit (see, e.g., \cite{FARAKOS2013322}), the sound speed is unity. There is still a (non-catastrophic) gravitational production of the helicity-1/2 mode, in agreement with the gravitino-goldstino equivalence theorem~\cite{Fayet:1986zc,Casalbuoni:1988kv,Casalbuoni:1988qd}, as studied in detail in \cite{Kallosh:2000ve}.

Finally, we note that the small gravitino sound speed is not a signal that the theory is becoming strongly coupled.  This behavior should be contrasted with the effective field theory of inflation~\cite{Cheung:2007st} (see also \cite{Delacretaz:2016nhw}) wherein the reduced sound speed ($c_s ^2 <1$) of the Goldstone boson of spontaneously broken time-translation invariance is generated by interactions,  and the value of sound speed is related by symmetry requirements to the interaction strength, as they both descend from the term $(1+ g^{00})^2$.
In that case, the limit $\cs^2 \to 0$ causes the theory to become strongly coupled and lowers the cutoff as $\Lambda \sim \cs^{5/4}$.  Similarly, in $P(X)$  theories \cite{ArmendarizPicon:2000dh,ArmendarizPicon:2000ah}, the reduced sound speed is induced by irrelevant operators, via $P_{,X}$ and $P_{,XX}$. In contrast with both of these, the reduced sound speed of the gravitino follows from the canonical kinetic term for $\psi^\mu$, and the constraint equations $D_{\mu}{\cal R}^{\mu}=$ and $\gamma_\mu {\cal R}^\mu=0$, as derived in Sec.~\ref{sec:Spin_32}. Thus, in contrast with the EFT of inflation, for the gravitino there is no connection between vanishing sound speed and strong coupling.

\section{The situation in supergravity} \label{sec:SUGRA}

In the spirit of effective field theory, one might hope that the catastrophic particle production exhibited by the Rarita-Schwinger model with $m \lesssim H_e$ would be cured by UV completion. Fortunately, the UV completion is in hand: the UV completion of the massive Rarita-Schwinger field is supergravity, and the UV completion of supergravity is string theory.

Indeed, the construction of a consistent quantum field theory for spin-3/2 particles in a general spacetime was once thought to be problematic \cite{Velo:1969bt,Velo:1970ur} but the issues are resolved if the spin 3/2 field is coupled to a supercurrent as in supergravity (SUGRA).  As we have alluded to many times, the physical spin-3/2 field in SUGRA is known as the gravitino.  A salient feature of SUGRA models is that the gravitino mass is related to the energy density and pressure and can be time dependent. The expression for the gravitino sound speed, \eref{eq:r2again}, applies independent of SUGRA considerations, and indeed matches that given throughout the SUGRA literature \cite{Giudice:1999yt,Giudice:1999am,Kallosh:1999jj}. On the other hand, whether $\cs$ vanishes or not (hence, catastrophic GPP or not) is dependent on the supergravity model.

Some discussion of SUGRA model building can be found in \aref{app:SUGRA} for those unfamiliar with SUGRA.  In SUGRA models there are several masses to keep track of; so when discussing SUGRA models we will denote the mass of the spin-3/2 gravitino by $m_{3/2}$.  Also, in SUGRA $m_{3/2}$ need not be real, so we will often encounter $|m_{3/2}|$.  Here, we first consider a model with a single chiral superfield, then consider more realistic models with more than one superfield.

\subsection{Models with a Single Chiral Superfield}

As a starting point, let us consider a single chiral superfield oscillating in its potential. This is the model studied by Giudice, Riotto, and Tkachev \cite{Giudice:1999yt,Giudice:1999am}  and Kallosh, Linde, with Van Proeyen \cite{Kallosh:1999jj} when discussing gravitino GPP (and discussed again  by Kallosh, Linde, Van Proeyen, and Kofman \cite{Kallosh:2000ve}).  Denoting as $\Phi$ the complex scalar component of a chiral superfield by ${\bf \Phi}$,  and $\bar{\Phi}$ the scalar field's  complex conjugate, we consider as a model the K\"{a}hler potential $K$ and superpotential $W$ given by, 
\begin{equation}\label{eq:model1}
    K(\Phi, \bar{\Phi}) = \Phi \bar{\Phi} \;\; , \;\; W(\Phi) = \half m_\phi \Phi^2 \per
\end{equation}
We may express the complex scalar $\Phi$ in terms of real fields as  $\Phi=\tfrac{1}{\sqrt{2}}\, \phi\, e^{i\sigma/\Mpl}$.  With this choice of the K\"{a}hler potential and the superpotential, and after setting $\sigma = 0$, the gravitino mass is given by [see \eref{eq:gravitinomass}], 
\begin{align}\label{eq:m321}
	m_{3/2} = e^{K(\Phi,\bar{\Phi})/2\Mpl^2} \frac{W(\Phi)}{\Mpl^2} = e^{\phi^2/4\Mpl^2} \, \frac{m_\phi}{4\Mpl^2} \, \phi^2 \com
\end{align}
where the middle expression is general for single-superfield models and the final expression is particular to the above choices of $K$ and $W$.  For example, $\phi$ may be the inflaton field, and its evolution during inflation leads to a time-dependence for the gravitino mass $m_{3/2} = m_{3/2}(t)$.  The scalar-field Lagrangian is ,
\begin{equation}
    {\mathcal L} =  \frac{1}{2}(\partial \phi)^2 + \frac{1}{2 \Mpl^2}\phi^2 (\partial \sigma)^2 - V(\phi) \com
\end{equation}
with potential given by \eref{eq:pot} 
\begin{align}\label{eq:v321}
 V(\phi) & = e^{K/\Mpl^2} \left( \left|\partial_\Phi W + \frac{W}{\Mpl^2} \partial_\Phi K \right|^2-3\frac{|W|^2}{\Mpl^2} \right)
 =  e^{\phi^2/2\Mpl^2} \frac{1}{2} \ m_\phi^2 \phi^2 \left(1 + \frac{\phi^2}{8\Mpl^2} + \frac{\phi^4}{16\Mpl^4} \right) \com
\end{align} 
where again the middle expression is general for single-superfield models and the final expression obtains for $K$ and $W$ of \eref{eq:model1}.  In the limit $\phi \ll \Mpl$ the potential is simply $V = \tfrac{1}{2}m_\phi ^2 \phi^2$.   

As a momentary digression, we note that this model is not complete. In particular, the potential is independent of the angular field $\sigma$, and thus $\sigma$ is massless. In general, one must introduce additional superfields to stabilize $\sigma$, and additional terms in the K\"{a}hler potential to stabilize the fields added to stabilize $\sigma$ (see, e.g., \cite{Kallosh:2010ug,Kallosh:2010xz}). Furthermore, in order to describe our universe the model must at least be able to accommodate the Standard Model of particle physics, which includes the Higgs field, and any supersymmetric embedding of the Standard Model generally introduces many scalar fields.
 
The shortcomings aside, the model has a seemingly miraculous feature: the sound speed $\cs^2=1$ at all times! We can see this by noting that the energy density $\rho$ and pressure $p$ contributed by $\phi$ are $\rho = \tfrac{1}{2} \dot{\phi}^2 + V(\phi)$ and $p=\tfrac{1}{2}\dot{\phi}^2-V(\phi)$, and that $4 \Mpl^4\left(a^{-1}\partial_\eta m_{3/2}\right)^2 = 4 \Mpl^2\dot{m}_{3/2}^2$ is given by
\begin{align}\label{eq:mdot}
4\Mpl^4\dot{m}_{3/2}^2 = e^{\phi^2/2\Mpl^2} \, m_\phi^2 \, \phi^2  \left( 1+ \frac{\phi^2}{2\Mpl^2} + \frac{\phi^4}{16\Mpl^2} \right) \dot{\phi}^2 =  2V\dot{\phi}^2 + 6 m_{3/2}^2 \Mpl^2 \dot{\phi}^2 \per
\end{align}
Using \pref{eq:mdot} and the expressions for $\rho$ and $p$ in \eref{eq:r2again} results in $\cs^2=1$!  To recap: $\cs=1$ is a result of the cancellation of the time dependence of $\rho$ and $p$ with the time dependence of $\dot{m}_{3/2}^2$.  

One might ask if the conspiracy that leads to $\cs^2=1$ is general in SUGRA, whether it is a feature of all models with a single chiral superfield, or is it particular to this model of a single superfield with the K\"{a}hler potential and superpotential of \eref{eq:model1}. We now demonstrate that $\cs^2=1$ for all models with a single chiral superfield (as shown long ago in \cite{Kallosh:1999jj,Kallosh:2000ve}), and in general is not true in models with multiple superfields. 

\subsection{Models with Multiple Chiral Superfields} \label{sec:nsuperfields}

We now consider the sound speed \eref{eq:r2again} in a model of $N$ number of chiral superfields, ${\bf \Phi}^I$, with field index $I=1...N$. For a thorough discussion of the gravitino equations of motion in models of arbitrary superfields, we refer the reader to \cite{Kallosh:2000ve}. It was strongly emphasized in \cite{Kallosh:2000ve} that to compute the particle production in models with more than one chiral superfield, one must carefully track all of the spin-$1/2$ fields in the theory (in contrast to what was done in \cite{BasteroGil:2000je}). Here we content ourselves to consider only the sound speed, with a vanishing sound speed considered to be a diagnostic for catastrophic particle production.

The first step is to find a general expression for $\dot{m}_{3/2}$.  The time derivative can be written as a directional covariant derivative in field space. We write the time-derivative as a field-space derivative as,
\begin{align}
\frac{\rm d}{{\rm d} t} = \sum_{I} \left[ \dot{\Phi}^I \nabla_I + \dot{\bar{\Phi}}^I \nabla_{\bar{I}}  \right] ,
\end{align}
where $\nabla$ represents a directional covariant derivative in field space on a complex manifold with metric $G_{I\Bar{J}}$ defined in \eref{eq:kahlermetric}.   When operating on a scalar function $f(\Phi,\Bar{\Phi})$, the directional derivative is $\nabla_I f(\Phi,\Bar{\Phi}) = \partial f(\Phi,\Bar{\Phi})/\partial\Phi^I = f,_I$ and $\nabla_{\Bar{I}} f(\Phi,\Bar{\Phi}) = \partial f(\Phi,\Bar{\Phi})/\partial\Bar{\Phi}^I = f,_{\Bar{I}}$.  The bar over an index denotes an operation with respect to $\Bar{\Phi}$.

For the gravitino mass given by $\Mpl^2\,m_{3/2} = e^{K(\Phi,\Bar{\Phi})/2\Mpl^2}\ W(\Phi)$ [see \eref{eq:gravitinomass}], we express $\Mpl^2\,\dot{m}_{3/2}$ as
\begin{align}\label{eq:mdoto}
\Mpl^2\,\dot{m}_{3/2} & = e^{K(\Phi,\Bar{\Phi})/2\Mpl^2} \sum_{I=1}^n  \left( \dot{\Phi}^I \partial_I W(\Phi) \, + \frac{1}{2\Mpl^2} \dot{\Phi}^I \, K,_I \, W(\Phi) + \frac{1}{2\Mpl^2} \dot{\Bar{\Phi}}^I \, K,_{\bar{I}} \, W(\Phi) \right)  \per
\end{align}
For the models we consider, $K$ is symmetric under the interchange $\Phi^I\longleftrightarrow \bar{\Phi}^I$. We further assume that the imaginary parts of $\Phi^I$ are constant and we set them equal to zero, i.e., for each complex scalar $\bar{\Phi}^I$, we consider only Re$\Phi$ or $|\Phi|$ to be dynamical, as is the case in common supergravity cosmological models.  These properties yield $K_{,I} = K_{,\Bar{I}}$ and $\dot{\Bar{\Phi}}^I = \dot{\Phi}^I$, and results in
\begin{align}
\Mpl^4 \left|\,\dot{m}_{3/2}\right|^2 & = e^{K(\Phi,\Bar{\Phi})/\Mpl^2} \left| \sum_I \dot{\Phi}^I  \left[ \partial_{I} + \Mpl^{-2}\, K_{,I} \right]W(\Phi)\right|^2 = e^{K(\Phi,\Bar{\Phi})/\Mpl^2} \left| \sum_I   \dot{\Phi}^I  D_I W(\Phi) \right|^2 \com
\end{align}
where $D_I \equiv \partial_I + \Mpl^{-2}\,K_{,I}$.

The next step is to express the energy density and pressure as $\rho = \sum_I|\dot\Phi^I|^2 + V$ and $p = \sum_I|\dot\Phi^I|^2 - V$ where $V = e^{K/\Mpl^2} \sum_I |D_I W|^2 - 3\Mpl^2 |m_{3/2}|^2$.  Using these expressions, along with the expression for $\Mpl^2\,\dot{m}_{3/2}$ in \eref{eq:mdoto}, leads to the result
\begin{align}\label{eq:r2m1}
\cs^2 = 1 - 4\frac{e^{K/\Mpl^2}}{(\rho+3\Mpl^2 m_{3/2})^2} \left[ \left(\sum_I  |\dot{\Phi}^I|^2 \right)  \left( \sum_{J} |D_JW|^2 \right) - \left|\sum_I {\dot{\Phi}}^I  D_{I} W \right|^2 \right] \com
\end{align}
which is not equal to unity except for special cases.  In particular, the sum-of-squares and  the square-of-a-sum only cancel in the trivial case where the ``sum'' is over 1 element, but not in general when there is more than one field.  \textit{This means that in general the $\cs^2=1$ result is peculiar to the case of only one chiral superfield} \cite{Kallosh:1999jj,Kallosh:2000ve}. The above expression matches the corresponding expression given in \cite{BasteroGil:2000je}.

We can simplify this further by defining the field space vectors $\vec{\dot{\Phi}}$ and $\vec{F}$, which have components $\dot{\Phi}^I$ and $e^{K/2} D_{I}W$ respectively. In this notation, one finds, 
\begin{align}
\rho + 3 M_{pl}^2 m_{3/2}^2  =   \vec{\dot{\Phi}} \cdot \vec{\dot{\Phi}} +  \vec{F} \cdot \vec{F} \com
\end{align}  
where the dot product is with respect to the metric $G_{I \bar{J}}$, and similarly
\begin{align}
\sum_I {\dot{\Phi}}^I  D_{I} W  =e ^{-K/2\Mpl^2} \vec{\dot{\Phi}} \cdot \vec{F} \com \quad
\sum_{J} |D_JW|^2 = e^{-K/\Mpl^2}\vec{F} \cdot \vec{F} \com \quad
\sum_I  |\dot{\Phi}^I|^2  = \vec{\dot{\Phi}}  \cdot \vec{\dot{\Phi}} \per
\end{align}
From this we can express $\cs$ as 
\begin{align}
\cs^2 = 1 - 4  \frac{ (\vec{\dot{\Phi}}  \cdot \vec{\dot{\Phi}} ) (\vec{F} \cdot \vec{F} ) - (\vec{\dot{\Phi}} \cdot \vec{F})^2 }{(\vec{\dot{\Phi}} \cdot \vec{\dot{\Phi}} +  \vec{F} \cdot \vec{F})^2}.
\end{align}
This has a simple geometric interpretation as an angle in field space. To appreciate this, consider the simple case of a flat field-space metric $G_{I \bar{J}}=\delta_{I {\bar{J}}}$. In this case, the dot product has the usual property $\vec{\dot{\Phi}} \cdot \vec{F}= |\vec{\dot{\Phi}}| |\vec{F}| \,\cos( \theta)$, where $\theta$ is the angle between the two vectors.  The above can be written as,
\begin{align}
\label{eq:soundspeed}
\cs^2 = 1 -  \frac{ 4 |\vec{\dot{\Phi}}|^2 |\vec{F}|^2  }{(|\vec{\dot{\Phi}}|^2 +|\vec{F}|^2)^2} \left( 1 - \cos^2 (\theta) \right) \per
\end{align}
Thus the gravitino is ``slow'', i.e., $\cs^2 <1$, in any case when $\cos(\theta) \neq 1$. As an interesting side note, the above is proof of subluminal propagation of gravitinos in supergravity: since $\cos^2(\theta) -1 \leq 0$, and $f(x,y)=4 xy/(x + y)^2 \leq 1$, we see that $\cs^2 \leq 1$ in general supergravity models: gravitino propagation is sub-luminal, as expected. 

One may also appreciate from \eref{eq:soundspeed} that a vanishing sound speed $\cs^2=0$ corresponds to a relatively simple limit. Consider a case wherein the cosmological evolution, namely the field space trajectory as measured by $\vec{\dot{\Phi}}$, is orthogonal to the breaking of supersymmetry, namely the vector ${\vec{F}}$. In this $\cos(\theta)=0$. The sound speed vanishes when $|\dot{\Phi}|=|\vec{F}|$. In summary, 
\begin{equation}
\label{eq:vs2condition}
\vec{\dot{\Phi}} \cdot \vec{F} =0 \;\; {\rm and} \;\; |\vec{\dot{\Phi}}|= | \vec{F} |  \,\, {\rm implies} \,\, \cs^2 = 0 \per
\end{equation}
Importantly, in contrast with what one might expect given that the single superfield model has $\cs^2=1$ at all times, the above suggests that supergravity does not a priori protect against catastrophic particle production. To understand this in more detail, we now consider particular supergravity cosmology constructions.

\subsection{Nilpotent Superfield Models}

An extensively studied class of supergravity models is that of nilpotent superfields \cite{Ferrara:2014kva}. These models contain a superfield ${\bf S}(x,\theta)$ that obeys a superspace constraint equation,
\begin{equation}
{\bf S}^2(x,\theta) = 0 \per
\end{equation} 
These models are reviewed in Appendix \ref{app:SUGRA}. They provide the effective field theory of the KKLT construction for de Sitter in string theory \cite{Kachru:2003aw}, where $S$ encodes the spontaneous breaking of supersymmetry by an anti-D3 brane \cite{Kallosh:2014wsa, Bergshoeff:2015jxa, Vercnocke:2016fbt,GarciadelMoral:2017vnz,Cribiori:2019hod}.  The constraint on ${\bf S}$ in turn imposes that the scalar component $S$ is a fermion bilinear, and the bosonic sector of theory corresponds to setting $S$ equal to $0$.  The cosmology of these models was developed in, e.g., \cite{Ferrara:2014kva,McDonough:2016der,Kallosh:2017wnt}.

In simple models of this type one always has the property that the field evolution is orthogonal to the SUSY breaking, since the field predominantly responsible for SUSY breaking has no dynamical scalar component. The sound speed is given by \eref{eq:r2m1}, with the simplification that scalar component $S$ of the nilpotent superfield, ${\bf S}$, is vanishing.  A simple model realization is the following \cite{McDonough:2016der,Kallosh:2017wnt},
\begin{equation}
\label{eq:nilpotentkahler}
W = M S+W_0 \Mpl \com  K = \Mpl^2 \frac{S \bar{S}}{f(\Phi,\bar{\Phi})} - 3 \alpha \Mpl^2 \log \frac{ \Phi + \bar{\Phi}}{ \sqrt{ 4 \Phi \bar{\Phi}}}
\end{equation}
where $M$, $f$, and $W_0$ have mass dimension $2$, and $\alpha$ is dimensionless. The last term of the K\"{a}hler potential describes half-plane coordinates on hyperbolic field-space manifold with Ricci scalar $R=-2/3\alpha$; a setup that naturally leads to $\alpha$--attractor inflation \cite{Ferrara:2013rsa,Kallosh:2013yoa,Kallosh:2014rga}. The $S$-dependence of $W$ breaks supersymmetry, while the $\Phi$ dependence of the K\"{a}hler potential generates a scalar potential for Re$\Phi$ and Im$\Phi$.

We consider a phase of inflation that proceeds along the $|\Phi|\equiv \phi/\sqrt{2}$ direction; the phase of $\Phi$ has a super-Hubble mass independent of the values of $W_0$ and $M$ or the precise form of $f(\Phi, \bar{\Phi})$ \cite{McDonough:2016der,Kallosh:2017wnt}, and thus can be self-consistently set to $0$. In this background, the K\"{a}hler potential vanishes, $K=0$. The $F$-term vector ${\vec{F}}$ is given by ${\vec{F}} \equiv (F^{\Phi},F^S) = (0 , M \Mpl)$, the gravitino mass is given by $m_{3/2}= W_0/\Mpl$, and the potential is given by $V(\Phi,\bar{\Phi}) = - 3 W_0 ^2  +M^2 f(\Phi,\bar{\Phi})/\Mpl^2$.

As a simple example, we consider $f(\Phi,\bar{\Phi})= \Mpl^2 + \lambda (v^2- 2|\Phi|^2 )^2 M^{-2}$,  such that the potential becomes
\begin{equation}
V = \Lambda + \lambda (v^2-\phi^2 )^2 \com
\end{equation}
with cosmological constant $\Lambda$ given by $\Lambda \equiv (M^2 - 3 W_0 ^2)$.  This model exhibits $\alpha$-attractor inflation \cite{Ferrara:2013rsa,Kallosh:2013yoa,Kallosh:2014rga}, that is well described in terms of the canonically normalized field $ \varphi\equiv - \sqrt{6 \alpha} \Mpl \log (\phi/\Mpl)$.
 Indeed, the inflaton potential may be written as,
\begin{equation}
V_{\rm infl}(\varphi) = V_0 \left( 1 - \frac{\Mpl ^2}{v^2} \,e^{- \sqrt{\frac{2}{3 \alpha}} \varphi/\Mpl} \right)^2,
\end{equation}
where $V_0 = \lambda v^4$. As shown in \cite{Kallosh:2013hoa}, the inflationary predictions for $n_s$ and $r$ are independent of $v$, and thus $v$ may be taken as a free parameter that specifies the position of the minimum of the potential. The inflatino mass is given by \cite{Hasegawa:2017nks},
\begin{align}
m_{\widetilde{\phi}}=& m_{3/2} \left( K_{\Phi\Phi} - (K_{\Phi\bar{\Phi}})^{-1}K_{\Phi\Phi\bar{\Phi}}K_{\Phi} + \frac{1}{3} (K_{\Phi})^2 \right) .
\end{align}
At the end of inflation, when $\phi=v$, this is,
\begin{equation}
|m_{\widetilde{\phi}}| =  \frac{3 m_{3/2} \alpha \Mpl^2}{2 v^2} .
\end{equation}
For $v/\sqrt{\alpha} \ll M_{pl}$, the inflatino is heavy, and can be consistently neglected in studying the gravitino. 

Interestingly, the gravitino mass and SUSY breaking have effectively been sequestered into the cosmological constant, and thus are (naively) decoupled from the Hubble parameter during inflation.    However, at the end of inflation, we find ourselves in an identical situation to the constant-mass massive Rarita-Schwinger. The oscillations of $|\dot{\Phi}|$, or equivalently, $\dot{\varphi}$, have amplitude of approximately $\Mpl H_e$, while the SUSY breaking has amplitude $|\vec{F}| =M$, and the field space evolution and SUSY breaking are orthogonal at all times, $\vec{\dot{\Phi}} \cdot \vec{F}=0$.  If $M  \lesssim H_e \Mpl$, then the condition for vanishing $c_s^2$, \eref{eq:vs2condition}, is satisfied once every oscillation, and catastrophic particle production occurs. Meanwhile, the observed near-vanishing of the cosmological constant, $\Lambda\simeq0$, dictates that $M \simeq \sqrt{3} m_{3/2} \Mpl$.  From this one finds that for $m_{3/2} \lesssim H_e$ there is catastrophic particle production in the nilpotent superfield model \eref{eq:nilpotentkahler}.

Somewhat surprisingly, the bound $m_{3/2} \gtrsim H$ to avoid the catastrophe is identical to that found in \sref{sec:GPP} without any of the machinery of supergravity. This demonstrates that supergravity in itself,  despite introducing new degrees of freedom, does not always resolve the catastrophic particle production.

\subsection{Orthogonal Constrained Superfield Models} \label{sec:constrained}

A final class of models we consider is that of orthogonal constrained superfields \cite{Kallosh:2014hxa,Carrasco:2015iij,Kallosh:2019apq, Ferrara:2015tyn}. These models supplement the nilpotent superfield model with an additional superfield ${\bf \Phi}$ satisfying the orthogonality constraint,
\begin{equation}
{\bf S} \cdot ({\bf \Phi} - {\bf \bar{\Phi}})=0.
\end{equation}
For a model comprised solely of ${\bf \Phi}$ and ${\bf S}$, in the unitary gauge the dynamical fields are only a single real scalar, and the gravitino, making this an ideal playground for cosmological model building \cite{Carrasco:2015iij,Kallosh:2019apq, Ferrara:2015tyn}. Catastrophic gravitino production in this class of models has been noted previously in Ref.\ \cite{Hasegawa:2017hgd}. The sound speed in this type of models has also been studied in e.g., Ref.\ \cite{Kahn:2015mla}.

For simplicity we restrict ourselves to two fields. Consider the model \cite{Kallosh:2014hxa,Carrasco:2015iij,Kallosh:2019apq, Ferrara:2015tyn},
\begin{align}
W = f(\Phi)S + g(\Phi)\Mpl  \;\; , \;\; K = \frac{1}{2}(\Phi - \bar{\Phi})^2 + S \bar{S} ,
\end{align}
with nilpotency constraint ${\bf S}^2 = 0$ and orthogonal constraint ${\bf S} \cdot ({\bf \Phi} - {\bf \bar{\Phi}})=0$. The superpotential $W = f(\Phi) S + g(\Phi)\Mpl$ is the most general possible superpotential for these two fields given the superfield constraints. The second constraint removes the $D_{\Phi}W$ contribution to $V$, and correspondingly, removes the $D_{\Phi}W$ contribution to the first term in the square brackets \eref{eq:r2m1}. Furthermore, the orthogonality constraint, in the unitary gauge, removes Im$\Phi$ from the spectrum \cite{Carrasco:2015iij}, which implies $K=0$. More explicitly: 
\begin{align}
\label{eq:A2ortho}
(\cs^2) _{\rm orthogonal}=  1 - \frac{4}{(\rho + 3 m_{3/2})^2} \left[ \left(\sum_I |\dot{\Phi}^I|^2 \right)  \left( \sum_{J\neq \Phi} |D_JW|^2 \right) - \left|\sum_I {\dot{\Phi}}^I  D_{I} W \right|^2 \right] \per
\end{align}
Again we consider $\Phi=\bar{\Phi}=\phi/\sqrt{2}$. The scalar potential is
\begin{align}
V = |D_S W|^2 - 3 \Mpl^{-2}|W|^2 =  \left( |f|^2 - 3 |g|^2 \right)  \per
\end{align}
Note that $D_{\Phi}W$ does not contribute to $V$ due to the additional constraint equation of the orthogonal models. The gravitino mass is $m_{3/2}(\Phi) = g(\Phi)/\Mpl$, which has time derivative 
\begin{align}
\dot{m}_{3/2} = \Mpl^{-1}\ \dot{\Phi} \frac{\partial g(\Phi)}{ \partial {\Phi}} \per  
\end{align}
We find the sound speed is
\begin{align}
\label{eq:r2ortho}
\cs^2 (\Phi, \bar{\Phi})= 1 - \frac{4 \dot{\Phi}^2}{(\dot{\Phi}^2 + f^2)^2 }\left[f^2 -   2 \Mpl^2  \left(\frac{\partial g(\Phi)}{ \partial {\Phi}}\right)^2 \right]  \per
\end{align}
This applies to general $f(\Phi)$ and $g(\Phi)$. 

As a simple example, consider a constant mass gravitino. That is, consider $g(\Phi)= m_{3/2} \Mpl$.  We can express the potential $V$ so as to absorb the gravitino mass into the cosmological constant, as $V = \Lambda + \hat{f}^2(\phi)$, where $\Lambda \equiv f^2(0) - 3 m_{3/2}^2$ and $\hat{f}^2(\phi) = f^2(\phi) -f^2(0)$. The sound speed can be computed from the above or directly from Eq.~\eqref{eq:r2again}. It is given by
\begin{align}
\label{eq:vs2orthosimple}
\cs^2 = \left( \frac{p_\phi }{\rho_\phi } \right)^2 \com
\end{align}
with $p_\phi= \dot{\Phi}^2 -  |\hat{f}(\Phi)|^2$ and $\rho_\phi=\dot{\Phi}^2 + |\hat{f}(\Phi)|^2$. The above is simply the equation of state of the field Re$\Phi\equiv \phi/\sqrt{2}$. If $\phi$ is oscillating, the above crosses zero any time the pressure crosses zero, leading to catastrophic particle production. Moreover, this applies for any choice of the gravitino mass, provided we set $\Lambda=0$ up to small corrections. Notably, this model with a constant gravitino mass can {\it not} be saved by the restriction to $m_{3/2}>H$, since the gravitino mass is absorbed into $\Lambda$ and we take $\Lambda \simeq 0$ up to very small corrections. 

Thus we see that orthogonal constrained superfield models with a constant-mass gravitino have the catastrophic particle production detailed in \sref{sec:GPP} for any mass of the gravitino. However, this is not the fate of all orthogonal constrained superfield models.

To understand the diverse possibilities, we consider a toy model that interpolates between the differing cases. We consider a constrained orthogonal superfield, and take a superpotential similar to those proposed in Ref.\ \cite{Carrasco:2015iij}. We consider the model, 
\begin{align}
W = f(\Phi)S + g(\Phi)\Mpl \;\; , \;\; K = \frac{1}{2}(\Phi - \bar{\Phi})^2 + S \bar{S} ,
\end{align}
satisfying the superspace constraints, $ {\bf S}^2 = 0$ and ${\bf S} \cdot ({\bf \Phi} - {\bf \bar{\Phi}})=0$, and with 
\begin{align}
\label{eq:orthomodel}
f(\Phi) = \sqrt{m^2 \Phi^2 +a}  \quad \text{and} \quad
g(\Phi) = \sqrt{ \frac{2}{\Mpl^2} \alpha^2 m^2 \Phi^4 + b}
\end{align}
The motivation for this seemingly ad hoc model is to parametrize in a simple way the cosmological constant, the Hubble constant, the time-dependence of the gravitino mass, and the size of the gravitino mass in vacuum. These are parametrized by $a$, $m$, $\alpha$, and $b$ respectively.

We take $\Phi = \bar{\Phi}=\phi/\sqrt{2}$. The scalar potential is given by
\begin{align}
V = (a-3b)+ \frac{1}{2}m^2 \phi^2 - \frac{3 }{8\Mpl^2} \alpha^2 m^2 \phi^4 \per
\end{align}
The first term is the cosmological constant. Demanding that it be near-zero sets $a\simeq 3b$, which we take to be case in what follows.  This leaves as model parameters $\{ \alpha,b,m\}$. 

The gravitino mass is
\begin{align}
\Mpl \ m_{3/2}=\sqrt{ \frac{1}{8\Mpl^2} \alpha^2 m^2 \phi^4 + b} \per
\end{align}
In the vacuum $\phi=0$, we have $m_{3/2}=b^{1/2}/\Mpl$. The time-dependence of $m_{3/2}$ is controlled by the size of $\alpha$. The gravitino sound speed in general is given by, \eref{eq:r2ortho}
\begin{align}
\cs^2  = 1-\frac{8 \dot{\phi}^2 \left(3b-\dfrac{2 \alpha ^4 m^4 \phi^6}{8 b+\alpha ^2 m^2 \phi^4}+\dfrac{m^2 \phi ^2}{2}\right)}{\left[2 \left(3b+\dfrac{m^2 \phi ^2}{2}\right)+\dot{\phi}^2\right]^2} \per
\end{align}

The minimum of the potential is locally a $m^2 \phi^2$ type \footnote{One may introduce additional corrections to $f(\Phi)$ to enforce the potential be positive definite at large $\phi$, but these terms are not relevant provided $\phi$ is restricted to small oscillations.}. We consider $m \gg H$ such that the oscillations are much faster than Hubble, to a good approximation we can neglect the expansion and consider the background solution $\phi(t) = \phi_0 \cos (m t)$. 

\begin{figure}[t!]
\begin{center}
\includegraphics[width=0.6\textwidth]{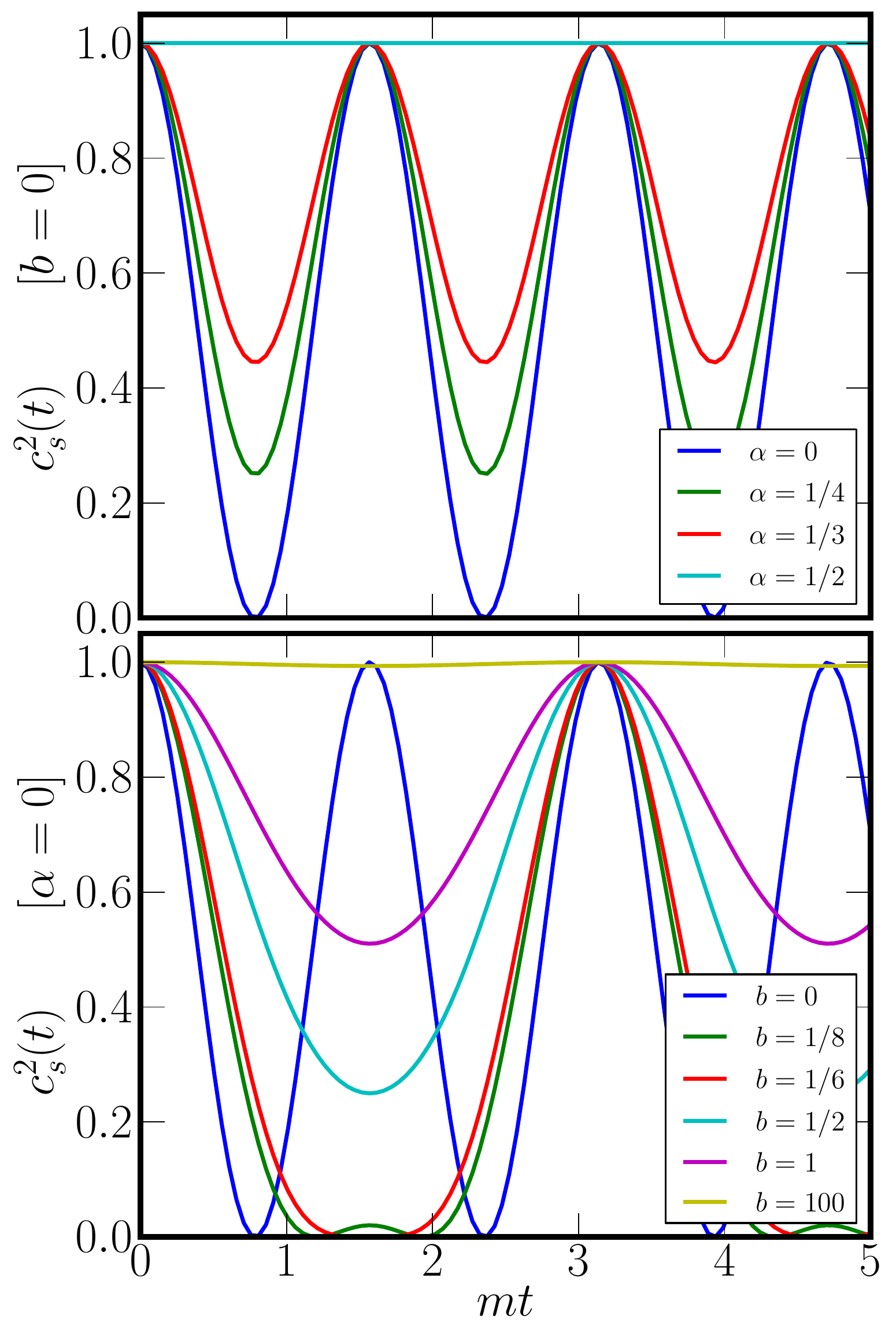}
\end{center}
\caption{ Evolution of the gravitino sound speed in a toy background with $\phi = \phi_0 \cos(m t)$. Here we fix $m=\phi_0=1$. [Top Panel] For $b=0$ but $\alpha \neq 0$, from \eref{eq:vsbequalzero} we find the zeros are lifted to $\cs^2 = \alpha^2$. [Bottom Panel:] For $\alpha=0$, from \eref{eq:vsalphaequalzero} the sound speed has zeros whenever $m_{3/2}<H$ . All other regions of parameter space interpolate between these examples.}
\label{fig:a_b}
\end{figure}

First consider a constant gravitino mass, $\alpha=0$. We find the sound speed is
\begin{align}\label{eq:vsalphaequalzero}
\cs^2(t) |_{\alpha=0} =  \frac{\left[6 b+m^2 \phi_0^2 \cos (2 m t)\right]^2}{\left(6 b+m^2 \phi_0^2\right)^2} \per
\end{align}
This has zeros at $t_*$, defined by
\begin{align}
\cos(2 m t_*) = - 6 \frac{b}{m^2 \phi_0 ^2} \per
\end{align}
Since $|\cos(x)| \leq 1 $, we deduce that $\cs^2(t)$ has zeros if $b< m^2 \phi^2/6$. Incidentally, the Friedmann equation reads $H^2 \simeq (3 M_{Pl}^2)^{-1}(m^2 \phi^2/2) = m^2 \phi^2/6\Mpl^2$, and $m_{3/2}=g=\sqrt{b}$, and hence we find zeros of $\cs^2$ whenever $m_{3/2}<H$. This can be seen in the top panel of \fref{fig:a_b}.

Now let us consider a time-varying gravitino mass. For simplicity, take a vanishing vacuum gravitino mass, $b=0$, and maintain $a=3b$. We find
\begin{align}\label{eq:vsbequalzero}
\cs^2(t) |_{b=0} = \frac{1}{2} \left[ (1+ 4\alpha^2)+ (1- 4 \alpha^2) \cos(4 mt) \right]\per 
\end{align}
Note that  we demand $\alpha \leq 1/2$ to ensure casuality. When $\alpha=0$,  this exhibits zeros at $ t = n \pi/4m$. However, for $\alpha \neq 0$, we see
\begin{align}
{\rm min}\,\cs^2(t) |_{b=0} = 4 \alpha^2 \per
\end{align}
Thus the vanishing of the sound speed is cured by turning on a field-dependence of the gravitino mass. This can be seen in the bottom panel of \fref{fig:a_b}.

Finally, we note that all of these examples have $D_{S}W \neq 0 $ and thus are within the regime of validity of the dS supergravity theory \cite{Bergshoeff:2015tra,Hasegawa:2015bza,Kallosh:2015sea, Kallosh:2015tea, Schillo:2015ssx,Freedman:2017obq}, and thus ostensibly in the regime of effective field theory. As a concrete example, we fix $\alpha=b=0$ and $a=\Lambda$. Then we find for the sound speed
\begin{align}
\cs^2 = \frac{\left(2 \Lambda +m^2 \phi_0^2 \cos (2 m t)\right)^2}{\left(2 \Lambda+m^2 \phi_0^2\right)^2} = \frac{p^2}{\rho^2} \per
\end{align}
We see $\cs$ has zeros if $m^2 \phi_0 ^2/(2\Lambda)> 1$. Meanwhile, for SUSY breaking along the $S$-direction
\begin{align}
|D_{S}W|^2 = f^2= \Lambda + \frac{1}{2} m^2 \phi^2 = \Lambda + \frac{1}{2} m^2 \phi_0 ^2 \cos(mt)^2   \com
\end{align}
is given at the zeros of $\cs$ by
\begin{align}
|D_S W|^2 |_{\cs=0} = \frac{\Lambda}{2} + \frac{1}{4}m^2 \phi_0 ^2 \com
\end{align}
which is manifestly positive, even in the limit $a\rightarrow 0$. Thus $D_{S}W$ is well behaved.

\subsection{Corrections to the K\"{a}hler Potential}

In the previous section we observed that a non-zero time dependence of the gravitino mass can lift the zeros of the sound speed $c_s^2$. In the simple model specified by Eq.~\eqref{eq:orthomodel}, the parameter $\alpha$ controls the time-dependence, and as can be appreciated from Fig.~\ref{fig:a_b}, raises the minimum value of $\cs^2$ from $0$ to $\alpha^2$.

This phenomenon can arise in a simple fashion due to small corrections to the K\"{a}hler potential. We note that while the superpotential is protected from perturbative corrections by non-renormalization theorems \cite{Grisaru:1979wc,Seiberg:1993vc}, the K\"{a}hler potential is not.  In the context of string theory, the K\"{a}hler potential receives corrections from both the $\alpha'$ and string loop expansions. The leading corrections in string compactifications have been computed in e.g. \cite{Becker:2002nn,Berg:2005ja}.

Corrections to the K\"{a}hler potential play a similar role to the parameter $\alpha$ of the previous section (see Fig.~\ref{fig:a_b}). As an illustrative example, consider the following K\"{a}hler potential,
\begin{equation}
K(\Phi,\bar{\Phi},S,\bar{S}) = K_0(\Phi,\bar{\Phi},S,\bar{S})+ \delta K(\Phi,\bar{\Phi})
\end{equation}
with
\begin{equation}
K_0 = \frac{1}{2}(\Phi - \bar{\Phi})^2 + S \bar{S} \;\; , \;\; \delta K = \frac{1}{2}c^2 (\Phi + \bar{\Phi})^2
\end{equation}
Here $K_0$ has a shift symmetry for Re$\Phi$, that is broken by $\delta K$, with the breaking parametrized by the parameter $c \ll 1$. In this case, the kinetic terms are no longer canonical, but can be made canonical by a simple rescaling. The potential gets a small contribution that is subdominant as long as $c^2  \ll 1$. The gravitino mass is now given by,
\begin{equation}
m_{3/2} = e^{c^2 \phi^2/2\Mpl^2}  \frac{W(\Phi)}{\Mpl^2} ,
\end{equation}
which has inherited a field-dependence from the breaking of the shift-symmetry of Re$\Phi$. 

Consider an orthogonal constrained superfield model with $g(\Phi)=g_0 \rm \Mpl^3 $, with $g_0$ a dimensionless constant.  The time-dependence of the gravitino mass is given by,
\begin{equation}
\dot{m}_{3/2} = \frac{c^2 \phi \dot{\phi}}{\Mpl} g_0
\end{equation}
The sound speed for $c=0$ is given by Eq.~\eqref{eq:vs2orthosimple}. When $c\neq0$, we instead find,
\begin{equation}
\cs^2 = \frac{p^2 _\phi }{\rho^2 _\phi} +  \frac{c^4 \Mpl^2 \phi^2 \dot{\phi}^2}{\rho_\phi ^2}  e^{c^2 \phi^2/\Mpl^2}.
\end{equation}
For a potential with a minimum that is locally quadratic, such that one can approximate $\phi \simeq \phi_0 \cos(m t)$, one finds the would-be zeros  of $\cs^2$, which occur when $p_{\phi}=0$ (at $m t=\pi/4$), are lifted to,
\begin{equation}
\cs^2 |_{p_\phi=0} = \frac{c^4}{m^2} e^{c^2 \phi^2/2 \Mpl^2}.
\end{equation}
Thus we see that corrections to the K\"{a}hler potential, much like the parameter $\alpha$ in Fig.~\ref{fig:a_b}, lifts the zeros of the sound speed, making it positive definite. Incidentally, corrections to the K\"{a}hler potential which mix $S$ and $\Phi$ play a crucial role in the inflation scenarios of \cite{McDonough:2016der,Kallosh:2017wnt}. As mentioned above, the leading corrections to the K\"{a}hler potential in compactifications of string theory are known and have been computed, see e.g.  \cite{Becker:2002nn,Berg:2005ja}.

\section{The Gravitino Swampland Conjecture (GSC)} \label{sec:GSC}

From the exercises of the previous section we deduce that supergravity does not a priori cure the vanishing of the sound speed seen in the Rarita-Schwinger model.  We take $\cs^2 = 0$ as a diagnostic for the catastrophe. Hence we deduce that SUGRA does not necessarily cure the catastrophic particle production: There are a plethora of supergravity models which do exhibit the catastrophe, and a plethora of models which do not. This breakdown of theory when the gravitino is quantized (which we will expand on shortly) suggests a possible link to the swampland \cite{Vafa:2005ui} (for a review, see \cite{Palti:2019pca,Brennan:2017rbf}).

The swampland is the set of four-dimensional effective field theories, which are self-consistent as an effective field theory coupled to classical gravity, but become inconsistent when gravity is quantized. From this inconsistency one infers that the effective theory does not have a UV completion in quantum gravity. The leading candidate of quantum gravity is superstring theory, and in this context, quantizing gravity means quantizing not only the graviton but also the gravitino(s). In this work we have seen examples of field theories that falter when the gravitino is quantized. Motivated by this, we propose the Gravitino Swampland Conjecture:

\begin{adjustwidth}{30pt}{30pt}
{\it The sound speed\footnote{In a general theory of many interacting fields, the scalar sound speed $c_s$ may be understood as the determinant of the matrix of sound speeds of all fields kinetically coupled to the gravitino and with mass below the UV cutoff. }  of gravitino(s) must be positive-definite, $ \cs^2 > 0$, at all points in moduli space and for all initial conditions, in all 4d effective field theories that are low-energy limits of quantum gravity. }
\end{adjustwidth}

\noindent The GSC conjecture forbids the constant-mass massive Rarita-Schwinger field with $m_{3/2} \lesssim H_e$ where $H_e$ is the Hubble expansion rate at the end of inflation. This also forbids the supergravity models shown to exhibit the vanishing sound speed, such as simple nilpotent superfield models with $m_{3/2}\lesssim H_e$ and orthogonal constrained superfield models with constant $m_{3/2}$.  

This conjecture is substantiated by the prominent proposals for moduli stabilization in string theory, namely the KKLT \cite{Kachru:2003aw} and Large Volume \cite{Balasubramanian:2005zx} scenarios. In both cases, $m_{3/2} > H$ is required for the radial field to be stabilized during inflation \cite{Kallosh:2004yh,Conlon:2008cj,Linde:2020mdk}. In the opposite limit, the inflationary vacuum energy destabilizes the compactification, and the radial field will experience a runaway.  In string theory setups which do allow $m_{3/2}<H$, such as the KL model \cite{Kallosh:2004yh}, consistency with the conjecture requires only the inclusion of small corrections to the K\"{a}hler potential, which are ubiquitious in string theory.

A detailed discussion of the Gravitino Swampland Conjecture, the cosmological implications, and evidence from string theory, can be found in \cite{Kolb:2021nob}.

\section{Discussion and conclusions\label{sec:conc}}

In this work we have considered the dynamics of massive spin-3/2 fields in curved spacetime. Considering simple cosmological backgrounds, we have computed the gravitational particle production, following the procedure developed in the context of gravitational production of dark matter \cite{Chung:1998zb,Chung:1998ua,Chung:1998rq,Kolb:1998ki, Chung:2001cb,Chung:2004nh,Chung:2011ck,Ema:2019yrd, Herring:2020cah, Herring:2020cah,Kolb:2020fwh,Kallosh:1999jj,Giudice:1999yt,Giudice:1999am}. We have found the surprising and striking result that the Rarita-Schwinger model, namely a constant mass spin-3/2 particle, has divergent (``catastrophic'') particle production if the spin-3/2 particle (the ``gravitino'') is cosmologically light, namely lighter than the Hubble parameter of the cosmological spacetime, $m_{3/2} \lesssim H$. The physical effect driving the production is a vanishing sound speed $\cs^2 =0$ for the helicity-1/2 gravitino, which can be used as a diagnostic for the catastrophic particle production. The particle number at late times (when the particles are non-relativistic) in a bounded region of space is a physical observable: all observers should agree on how many billiard balls there are on the pool table. Thus we are forced to take this catastrophe seriously.

The production of particles of arbitrarily-high momentum implies a breakdown of effective field theory, since an infinite tower of irrelevant operators, for example a set of operators labelled by $n$,
\begin{equation}
{\cal O}_n = c_n m_{3/2 }\frac{\partial^{n}}{\Lambda^{n}} \left( \bar{\psi}^\mu  \psi_\mu \right)  \sim c_n m_{3/2}  \int d^3k   \left( \frac{k}{\Lambda}\right)^n \bar{\psi} ^\mu _{\bf k} \psi_{\mu{\bf k}} ,
\end{equation}
make a non-negligible contribution to the equations of motion if particles with momenta near cutoff, $k \sim \Lambda$, are produced. The relevance of an infinite tower of operators is indicative of a total breakdown of effective field theory. This feature, in addition to the physical mechanism underlying it, distinguishes the catastrophic production from the conventional gravitino problem \cite{KHLOPOV1984265,ELLIS1984181}. It bears strong resemblance to the Planck-suppressed operators in large field inflation, which lead to the $\eta$-problem \cite{Copeland:1994vg}, and which is the domain of the Swampland Distance Conjecture \cite{Hebecker:2018vxz,Garbrecht:2006az}. The SDC, in effect, states that no single EFT can describe regions of moduli space separated by Planckian distances, and thus that any EFT which purports to do so (i.e., a theory without any new degrees of freedom entering the theory as a scalar field traverses a Planckian distance in field space) is in the ``swampland'' \cite{Vafa:2005ui}. 

Faced with this, one may impose an ad hoc UV cutoff on the Rarita-Schwinger model, to parametrize our ignorance of UV physics, and postulate that the particle production computation can only be computed up to the cutoff, and moreover that the calculation can be trusted all the way up to the cutoff, despite the argument above. However, for the Rarita-Schwinger model we are fortunate to have the UV completion already in hand: supergravity (SUGRA), and string theory. In this work we have studied the gravitino sound speed in single field SUGRA models, multifield SUGRA models, nilpotent constrained SUGRA models, and orthogonal constrained SUGRA models. We have shown SUGRA does not provide a single answer: it is model dependent whether $\cs^2 =0$, and hence whether there is catastrophic particle production. In simple cases, with the exception of orthogonal models, the restriction to gravitino masses greater than Hubble, $m_{3/2 } > H$, is a sufficient condition to avoid catastrophic production. In the orthogonal models, one may lift the zeros of $\cs^2$ by additional terms in the superpotential or K\"{a}hler potential which induce a time-dependence of the gravitino mass. 

From this we conclude that the new degrees of freedom of SUGRA do not a priori save the Rarita-Schwinger model with $m_{3/2} \lesssim H$. In supersymmetric theories of quantum gravity, such as superstring theory, quantizing the gravitino is part and parcel with quantizing the graviton. This motivates an extension of the string swampland program  \cite{Vafa:2005ui} to effective field theories that become inconsistent when the gravitino is quantized, and along these lines, we have proposed the Gravitino Swampland Conjecture. Since the conjecture specifically pertains to the gravitino sound speed, this conjecture applies to theories with arbitrary field content and scalar potential. This thus applies to a plethora of example models beyond those studied here.

This work illustrates the power of gravitational particle production to uncover new physics. It complements nicely the power of GPP to generate the observed dark matter abundance, and thereby provides a substantial puzzle piece in understanding physics beyond the standard model. In future work we will develop the GPP of spin-3/2 fields as a dark matter model, focusing on the region of parameter space where the production is large but finite. It will also be interesting to develop other observational signatures, such as the CMB non-Gaussianity. The non-Gaussianity of higher-spin fermions, spin-$s+1/2$ with $s>2$, has been studied in detail in \cite{Alexander:2019vtb}, and it will be interesting to understand in detail how the spin-$3/2$ case differs. We leave this, and other possibilities, to future work.

\section*{Acknowledgements}

The authors thank Mustafa Amin, Sylvester James Gates, Jr., Gian Giudice, Wayne Hu, Andrei Linde, Zhen Liu, Antonio Riotto, Marco Scalisi, Leonardo Senatore, Igor Tkachev, and Lian-Tao Wang for helpful comments.  The work of E.W.K.\ and E.M.\ was supported in part by the US Department of Energy contract DE-FG02-13ER41958.

\begin{appendices}

\section{Frame Fields \label{app:framefields}} 

We will be concerned with quantum field theory in the expanding universe.  To promote a familiar flat-space quantum field theory to one in curved space we start with a relativistic field theory in Minkowski space.  For fields that transform under Lorentz transformations as scalars, vectors, or tensors, the procedure is to replace in the action or field equations the Minkowski metric $\eta_{\mu\nu}$ by $g_{\mu\nu}$, replace all tensors by objects that behave as tensors under general coordinate transformations, and replace all derivatives $\partial_\mu$ with covariant derivatives $\nabla_\mu$; i.e., for the derivative of a scalar field the covariant derivative is just $\partial_\mu$, and for a vector field $V_\mu$ the covariant derivative is $\nabla_\mu V_\nu = \partial_\mu V_\nu - \Gamma^\alpha_{\mu\nu} V_\alpha$ and $\nabla_\mu V^\nu = \partial_\mu V^\nu + \Gamma^\nu_{\mu\alpha} V^\alpha$, and similarly for tensors. 
 
This prescription fails for fields with half-integer spin, such as the Dirac electron field, which transform as spinors under (infinitesimal) Lorentz transformations.  Another approach is required to promote a Minkowski field theory with spinor fields to curved space.  The frame field\footnote{The frame field was originally introduced by Cartan who called them \textit{rep\'{e}res mobiles} (moving frames); in German they are referred to as \textit{vierbeins} (four-leg), and in the English literature they are usually called \textit{tetrads} (Greek for ``set of four'').  We will follow the notation of Freedman and Van Proeyen \cite{Freedman:2012zz}, and in the spirit of Cartan refer to them as ``frame fields.''} formalism is an elegant and general formalism of general relativity which will allow a field theory with spinors to be extended to curved spacetime.

To start, erect at every spacetime point $X$ a set of local inertial coordinates $y^\alpha_X$.  At that spacetime point in terms of the local inertial coordinates the metric is simply the Minkowski metric $\eta_{\alpha\beta}$.\footnote{The early-letter Greek alphabet will refer to the local inertial frame with coordinates $y^\alpha_X$, $\alpha=0,\cdots,3$, and early-letter Latin alphabet for the spatial part $y^a_X$, $a=1,2,3$.  Mid-alphabet Greek letters will refer to coordinates in the noninertial system $x^\mu$, $\mu = 0,\cdots,3$, and mid-alphabet Latin letters for the spatial part $x^i$, $i=1,2,3$.} This implies that the metric in a general noninertial coordinate system can be expressed it terms of the Minkowski metric $\eta_{\alpha\beta}$ as
\begin{align}\label{eq:vierbeindef}
g_{\mu\nu}(x) & = \tensor{e}{^\alpha_\mu}(x)\,\tensor{e}{^\beta_\nu}(x)\,\eta_{\alpha\beta}\com \ \ \mathrm{with} \nn
\tensor{e}{^\alpha_\mu}(x) & \equiv \left(\frac{\partial y_X^\alpha}{\partial x^\mu}\right)_{x=X} \com
\end{align}
where $\tensor{e}{^\alpha_\mu}(x)$ is the frame field. Early-alphabet indices are raised/lowered by $\eta_{\alpha\beta}$ while mid-alphabet indices are raised/lowered by $g_{\mu\nu}$.   The frame field transforms as a covariant vector, so it should be regarded as four orthonormal covariant vectors (one timelike and three spacelike) rather than as a single tensor. By using frame fields, general objects (like spinors) can be converted into proper local, Lorentz-transforming tensors with the additional spacetime dependence absorbed by the frame fields.

To promote a Minkowski-space field theory to curved spacetime, contract vectors and tensors into frame fields (e.g., $V_\alpha \rightarrow \tensor{e}{_\alpha^\mu}\,V_\mu$) and replace derivatives $\partial_\alpha$ by $\nabla_\alpha$, where 
\begin{align}
\nabla_\alpha = \tensor{e}{_\alpha^\mu}\,\partial_\mu + \tensor{e}{_\alpha^\mu}\,\Gamma_\mu \com
\end{align}
with the spin connection $\Gamma_\mu(x)$ defined as
\begin{align}\label{eq:spinconnectionapp}
\Gamma_\mu(x) = \half \Sigma^{\alpha\beta}\,\tensor{e}{_\alpha^\nu}\,g_{\sigma\nu}\, \left(\partial_\mu\, \tensor{e}{_\beta^\sigma} + \Gamma^\sigma_{\mu\rho}\tensor{e}{_\beta^\rho} \right) \per
\end{align}
In the above expression for $\Gamma_\mu$, the quantity $\Sigma^{\alpha\beta}$ is the generator of the Lorentz group associated with the representation under which the field transforms.

The stress-energy tensor can also be found by varying the geometry via the frame field
\begin{align}
\tensor{T}{_\alpha^\mu} = \frac{1}{|e|} \frac{\delta S_\mathrm{M}}{\delta \tensor{e}{^\alpha_\mu}} \com
\end{align}
where $|e|$ is the determinant of the frame field.  Then, $T^{\mu\nu} = -\eta^{\alpha\beta}\, \tensor{e}{_\alpha^\mu}\, \tensor{T}{_\beta^\nu}$.

The frame-field formalism can be used for any field (scalar, vector, etc.) but we will only employ it for half-integer spins.   The interested reader should consult a fuller discussion of frame-field formalism (e.g., \cite{BirrellDavies:1982, Weinberg:1972kfs, Wald:106274}).

\section{Projection Operators \label{app:Projection}}

The four spinor fields $\psi_{0,\kvec}\ldots\psi_{3,\kvec}$ separately obey the Dirac equation with canonically normalized kinetic terms.  However, they are not independent as there are two constraint equations.  Here we find two orthogonal combinations of $\psi_{0,\kvec}\ldots\psi_{3,\kvec}$ that are helicity eigenstates, have correctly normalized kinetic terms, and satisfy the Dirac equation. Since $\psi_{\mu,\kvec}$ is a vector-spinor, helicity projection operators are constructed from projection operators for vectors and spinors.  To simplify the analysis we will employ the freedom to choose the momentum in the $z$ direction.  

First, the projection operators for the helicity $\pm1/2$ states of a spinor (with momentum in the $z$ direction) are
\begin{align}\label{eq:spinorprojector}
S_\pm = \frac{1}{2}\left(1\pm i \gamma^1\gamma^2\right) \per
\end{align} 
We note $S_+$ and $S_-$ are real and satisfy $S_+ S_+=S_+$, $S_+ S_-=0$ and $S_-S_-=S_-$.  The vector projectors for helicity $s$ are constructed from polarization vectors $(\epsilon_s)_\mu$, with $s=\pm1,0$.  For $z$-directed momentum, the polarization vectors are
\begin{align}
(\epsilon_\pm)_\mu & = \dfrac{1}{\sqrt{2}}\ \left(0,\pm 1,i,0\right) \, \qquad \mathrm{and} \qquad (\epsilon_0)_\mu = \dfrac{1}{m}\left(k,0,0,-\sqrt{k^2+m^2}\right) \per
\end{align}
Note the normalization of the vectors: $(\epsilon_s^*)^\mu (\epsilon_r)_\mu = -\delta_{sr}$.  The spin-1, helicity-$s$ projectors $\tensor{(M_s)}{^\mu_\nu}$ are formed from the polarization vectors:
\begin{align}\label{eq:vectorprojector}
\tensor{(M_s)}{^\mu_\nu} \equiv -(\epsilon_s^*)^\mu \, (\epsilon_s)_\nu \qquad (s=\pm1,0) \per
 \end{align}
This is a well-behaved helicity projector:  $\tensor{(M_s)}{^\mu_\rho} \, \tensor{(M_r)}{^\rho_\nu} = \delta_{rs}\, \tensor{(M_s)}{^\mu_\nu}$.
Putting together the spinor projectors \pref{eq:spinorprojector} and vector projectors \pref{eq:vectorprojector} we find the projectors for helicities $+3/2,-3/2,+1/2,-1/2$:
\begin{align}
\tensor{(P_{\pm3/2})}{^\mu_\nu} & = S_\pm \, \tensor{(M_\pm)}{^\mu_\nu} \nn
\tensor{(P_{\pm1/2})}{^\mu_\nu} & = S_\pm \, \tensor{(M_0)}{^\mu_\nu}  + S_\mp \, \tensor{(M_\pm)}{^\mu_\nu} \per
\end{align}
One can easily demonstrate $\tensor{(P_s)}{^\mu_\nu} \, \tensor{(P_r)}{^\mu_\nu} = \delta_{rs}\,\tensor{(P_s)}{^\mu_\nu}$.  

Now we define $\psi_{{\ThreeHalf,\kvec}}$ as
\begin{align}\label{eq:projthreehalf}
\psi_{{\ThreeHalf,\kvec}} & = -  {(\epsilon_+^*)}^\nu\, S_+\, \tensor{(M_+)}{^\mu_\nu}\, \psi_{\mu,\kvec} + {(\epsilon_-^*)}^\nu\, S_-\, \tensor{(M_-)}{^\mu_\nu}\, \psi_{\mu,\kvec} \nn
\psi_{{\ThreeHalf,\kvec}} & = \dfrac{1}{\sqrt{2}}\left( \psi_{1,\kvec} + \gamma^1\gamma^2\psi_{2,\kvec} \right) \com
\end{align}
and we define $\psi_{\Half,\kvec}$ as
\begin{align}\label{eq:projonehalf}
\psi_{{\Half,\kvec}} & = - \sqrt{3}\, ({\epsilon_+ ^*})^\nu\, S_-\, \tensor{(M_+)}{^\mu_\nu}\, \psi_{\mu,\kvec} + \sqrt{3}\, ({\epsilon_-^*})^\nu\, S_+\, \tensor{(M_-)}{^\mu_\nu}\, \psi_{\mu,\kvec} \nn
\psi_{{\Half,\kvec}} & = \dfrac{\sqrt{6}}{2}\left( \psi_{1,\kvec}-\gamma^1\gamma^2\psi_{2,\kvec} \right) \per
\end{align}
So defined, $\psi_{{\ThreeHalf,\kvec}}$ and $\psi_{{\Half,\kvec}}$ are constructed from helicity projections, have canonical kinetic terms, are orthogonal, and obey the Dirac equation.

\section{Supergravity \label{app:SUGRA}}

We consider ${\cal N}=1$ supergravity in $d=4$ dimensions with $N$ chiral superfields. We denote chiral superfields by boldface, e.g., $\bm{\Phi}$.   The field content of $\bm{\Phi}$ consists of ${\bf \Phi}(x,\theta) = \Phi(x) + \sqrt{2}\,\theta \chi^{\Phi}(x) + \theta \theta F^{\Phi}(x)$ where $x$ represents the spacetime coordinates which form the bosonic coordinates on superspace, and $\theta$ is the Grassman-valued fermionic superspace coordinates. Here $\Phi(x)$ is a complex scalar, $\chi$ is a chiral fermion, and $F$ is a non-dynamical complex scalar; i.e., an auxiliary field. When dealing with a set of fields, we use a capital Roman alphabet superscript, as in ${\bf \Phi}^I(x,\theta)$ with scalar component $\Phi^{I}(x)$. We denote the corresponding anti-chiral superfields as ${\Bar{\bf \Phi}}^{\Bar{I}}$. 

After integrating out auxiliary fields, the Lagrangian for the scalar components $\Phi(x)$ is given by $ {\mathcal{L}} = \mathcal{L}_{\rm kinetic} - V(\Phi,\Bar{\Phi})$.  The kinetic term is given by $\mathcal{L}_{\rm kinetic} =  G_{I \Bar{J}} g^{\mu \nu } \partial_{\mu} \Phi^I \partial_{\nu} \Bar{\Phi}^{\bar{J}}$.  This defines a non-linear sigma model with a target space metric $G^{I \bar{J}}$. The target space metric is specified by derivatives of a real scalar, 
\begin{align}\label{eq:kahlermetric}
G_{I \bar{J}} \equiv \frac{\partial}{\partial \Phi^{I}} \frac{\partial}{\partial {\bar{\Phi}}^{\bar{J}}} K (\Phi, \Bar{\Phi}) \per
\end{align}
The above property ensures that the target space manifold is a K\"{a}hler manifold. For a review of the differential geometry aspects we refer the reader to Green, Schwartz, and Witten \cite{Green:1987sp}. We refer to the scalar $K(\Phi, \bar{\Phi})$  as the K\"{a}hler potential. 

For ease of notation, we henceforth denote partial derivative by a comma, 
\begin{align}
X_{,I} \equiv \frac{\partial}{\partial \Phi^{I}} \;\; , \;\; X_{,\bar{I}} \equiv \frac{\partial}{\partial \bar{\Phi}^{\bar{I}}} \per
\end{align}
In this notation the metric on field space is given by $G_{I \bar{J}} = K_{, I \bar{J}}$.
Canonical kinetic terms correspond to a flat field space manifold, e.g., in Cartesian coordinates, $G_{I {\bar{J}}} = \delta_{I \bar{J}}$.

There are conventional coordinate systems (i.e., bases for the fields $\Phi^I$) that lead to canonical kinetic terms for the real scalars. The first is the shift-symmetric K\"{a}hler potential
\begin{align}
K = \frac{1}{2}\left( \Phi + \Bar{\Phi} \right)^2 \;\; {\rm or} \;\; K = \frac{1}{2}\left( \Phi -  \bar{\Phi} \right)^2 \com
\end{align}
where the $+$ choice endows $K$ with a shift symmetry in Im$\Phi$, and the $-$ choice with a shift symmetry in Re$\Phi$.  Expanding the complex scalar in terms of real fields $\phi$ and $a$ as
\begin{align}
\Phi = \frac{\phi + i a}{\sqrt{2}}
\end{align}
one finds canonical kinetic terms for $\phi$ and $a$. Note the factor of $1/\sqrt{2}$ is necessary for the correct normalization of the kinetic terms.  An alternative coordinate system leading to canonical kinetic terms is,
\begin{align}
K = \Phi \bar{\Phi} \per
\end{align}
Decomposing the complex scalar as
\begin{align}
\Phi = \frac{1}{\sqrt{2}}\phi e^{i a} \com
\end{align}
again one finds canonical kinetic terms, properly normalized by the factor of $1/\sqrt{2}$.

In string theory the former choice appears in the Large Volume Scenario (LVS) \cite{Balasubramanian:2005zx} as the approximate K\"{a}hler potential for the K\"{a}hler moduli parametrizing the small cycles, and the latter choice is the approximate K\"{a}hler potential for $D$-brane moduli \cite{Baumann:2007ah,DeWolfe:2002nn}. In string scenarios the K\"{a}hler potential is given by the log of the inverse volume, $K = 2 \log \, {\cal V}^{-1}$, where ${\cal V}$ is the volume of the compactification, and depends on all K\"{a}hler moduli. In KKLT \cite{Kachru:2003aw}, which has a single K\"{a}hler modulus, the K\"{a}hler potential is given by $K = - 3 \, \log \left( T + \bar{T}\right)$, where Re\,$T$ is the volume modulus of the compactification. This choice of K\"{a}hler potential generates a negative Ricci scalar on the field space, and non-canonical kinetic terms. Generalizing the prefactor from $3$ to $3 \alpha$, with $\alpha$ a positive real parameter, leads to the $\alpha$-attractor models of inflation \cite{Ferrara:2013rsa,Kallosh:2013yoa,Kallosh:2014rga}.

We now turn to the potential energy of the bosonic components of the chiral superfields ${\bf \Phi}^I$.  After integrating out the auxiliary fields $F^I$, one finds
\begin{align}
V(\Phi,\bar{\Phi}) = e^{K(\Phi,\Bar{\Phi})/\Mpl^2} \left( G^{I \Bar{J}} D_{I} W D_{\Bar J} \Bar{W}(\Bar{\Phi}) - 3 \Mpl^{-2} W \Bar{W}\right) \com
\end{align}
where $W(\Phi)$ is a holomorphic function referred to as the superpotential, and where $D_{I}$ denotes a K\"{a}hler covariant derivative: $D_{I}\equiv \partial_{I} + \Mpl^{-2}K_{,I}$ and $D_{\bar{I}} \equiv \partial_{\bar{I}} + \Mpl^{-2}K_{,\bar{I}}$,
where we again note that the comma subscript denotes a partial derivative. This can be compactly written as,
\begin{align}\label{eq:pot}
V = e^{K/\Mpl^2}\left( |DW|^2 - 3 \Mpl^{-2}|W|^2\right) \com
\end{align}
where the norm of $D_{I}W$ has been taken with respect to the metric $G_{I \bar{J}}$. This may alternately be written as $V = F_I F^I  - 3\Mpl^{2} |m_{3/2}|^2$,
where $F_I = D_{I}W$ is the VEV of the auxiliary fields, and $m_{3/2}$ is the gravitino mass:
\begin{align}\label{eq:gravitinomass}
m_{3/2}(\Phi) \equiv e^{K/2 M_{\rm Pl}} \Mpl^{-2}W(\Phi) \per
\end{align}
This corresponds to the massive Rarita-Schwinger field with a real mass via the identification of the mass with $|m_{3/2}|$.  The $F_{I}$ comprise the components of a field space vector $\vec{F}$, and the direction of $\vec{F}$ is referred to as the ``direction of supersymmetry breaking.''

We note that the recent SUGRA literature has focused on models of constrained superfields. The most well studied example is that of a nilpotent superfield ${\bf S}(x,\theta)$, satisfying the constraint equation ${\bf S}^2(x,\theta) = 0$. These models are the effective field theory describing KKLT in the case that all bosonic degrees of freedom have been frozen out, e.g., by placing the anti-D3 brane directly on an O3 orientifold plane. The cosmology of these models was developed in e.g., \cite{Ferrara:2014kva,McDonough:2016der,Kallosh:2017wnt}. Solving the above order-by-order in $\theta$ imposes a relation between the scalar component $S$ and fermionic component $\chi^S$: $S = \chi^S \chi^S/2 F^S$.  This implies that the bosonic sector of the theory corresponds to imposing $ S = 0 $. Note that this solution only exists for $F \neq 0$. Note also that  $\langle F \rangle=D_{S}W$. Thus, the nilpotent superfield framework only applies when $D_{S}W\neq0$. We will discuss this further below.

The recent gravitino literature \cite{Hasegawa:2017hgd} has focused on supergravity models wherein an additional constraint is on imposed at the level of the full supergravity action, i.e., before integrating out the auxiliary fields. These so-called ``orthogonal constrained superfield'' models were were developed in \cite{Carrasco:2015iij,Kallosh:2019apq,Ferrara:2015tyn}, and in \cite{Vercnocke:2016fbt} were shown to be the effective field theory of an anti-D3 brane in a KKLT compactification where the scalar fields, parametrizing the position of the brane, are explicitly kept in the spectrum. These models add an additional constraint, in terms of a chiral superfield ${\bf \Phi}$: ${\bf S}(x,\theta) \cdot \left( {\bf \Phi} - \bar{\bf \Phi}\right) = 0$.  This constraint removes the $D_{\Phi}W$ contribution to the scalar potential \cite{Carrasco:2015iij,Kallosh:2019apq,Ferrara:2015tyn}. For example, in a model with superpotential $W = f(\Phi) S + g(\Phi)$, satisfying constraints ${\bf S}^2(x,\theta) =0$ and ${\bf S}(x,\theta) \cdot \left( {\bf \Phi} - \bar{\bf \Phi}\right) = 0$, the scalar potential is given by
\begin{align}
V(\Phi,\bar{\Phi}) = e^{K/\Mpl^2}\left( |D_S W(\Phi)|^2 - 3 \Mpl^{-2}|g(\Phi)|^2\right) = e^{K/\Mpl^2} (|f|^2 - 3 |g|^2 ) .
\end{align}
where we note the would be $D_{\Phi}W = g'$ is absent from the scalar potential.

Before we proceed further, we note a limitation of the constrained superfield models.  The discussion of constrained superfield models above, and in the cosmology literature, has been generally limited to the bosonic truncation of de Sitter supergravity \cite{Bergshoeff:2015tra,Hasegawa:2015bza,Kallosh:2015sea, Kallosh:2015tea, Schillo:2015ssx,Freedman:2017obq}. As emphasized in  \cite{Kallosh:2018wme}, the construction and self-consistency of the supergravity theory demands that $D_{S}W \neq 0$ \cite{Bergshoeff:2015tra,Hasegawa:2015bza, Kallosh:2015sea,Kallosh:2015tea, Schillo:2015ssx}.  For example, for $W(T,S)= W(T) + S f(T)$ one has $D_S\, W(T,S) \big |_{S=0}= f(T) \neq 0$. This requirement can be appreciated from two perspectives. The first is that the full Lagrangian contains fermionic interactions that scale with $1/f$,  i.e., there are fermionic interactions of the form
\begin{align}
\mathcal{L}_{\rm dS-SUGRA} \supset \mathcal{L}_{int} \sim \frac{1}{f(T)} \bar{\chi} \chi \bar{\chi} \chi \com
\end{align}
where $\chi$ can be the fermionic component of ${\bf S}$ or other fields. For the explicit interactions see \cite{Bergshoeff:2015tra,Hasegawa:2015bza,Kallosh:2015sea,Kallosh:2015tea, Schillo:2015ssx,Freedman:2017obq}. This would suggest $f\rightarrow 0$ corresponds to a strong-coupling limit of the theory. However, this too is misleading as to the severity of the problem. Recall that that the nilpotency condition ${\bf S}^2=0$ translates into 3 equations: 
\begin{align}
S^2(x)=0, \qquad S(x) \chi^S=0,  \qquad 2 S (x) F - \chi^S \chi^S=0 \ .
\end{align}
If $F=D_SW\neq 0$ one may find the non-trivial solution $S=\chi^S \chi^S/(2 F)$.  Inserting this back in the action leads to dS supergravity. On the other hand, if $F=D_SW=0$ there is only the trivial solution: $S(x)=0,\ \chi^S(x) = 0,\ \text{and}\ F=0$. Substituting this back in the full supergravity action one finds `textbook' supergravity with no trace of the ${\bf S}$ superfield, i.e., one does not arrive at dS SUGRA. At present there is no supergravity theory which can continuously interpolate between textbook and dS supergravity.

\end{appendices}

\bibliographystyle{JHEP}
\bibliography{spin1.5}

\providecommand{\href}[2]{#2}\begingroup\raggedright\begin{thebibliography}{100}

\bibitem{Schrodinger:1939:PVE}
E.~Schr{\"o}dinger, \emph{{The proper vibrations of the expanding Universe}},
  \href{https://doi.org/https://doi.org/10.1016/S0031-8914(39)90091-1}{\emph{PHYSICA}
  {\bfseries 6} (1939) 899--912}.

\bibitem{Chung:1998zb}
D.~J.~H. Chung, E.~W. Kolb and A.~Riotto, \emph{{Superheavy dark matter}},
  \href{https://doi.org/10.1103/PhysRevD.59.023501}{\emph{Phys. Rev.}
  {\bfseries D59} (1998) 023501},
  [\href{https://arxiv.org/abs/hep-ph/9802238}{{\ttfamily hep-ph/9802238}}].

\bibitem{Chung:1998ua}
D.~J.~H. Chung, E.~W. Kolb and A.~Riotto, \emph{{Nonthermal supermassive dark
  matter}}, \href{https://doi.org/10.1103/PhysRevLett.81.4048}{\emph{Phys. Rev.
  Lett.} {\bfseries 81} (1998) 4048--4051},
  [\href{https://arxiv.org/abs/hep-ph/9805473}{{\ttfamily hep-ph/9805473}}].

\bibitem{Chung:1998rq}
D.~J.~H. Chung, E.~W. Kolb and A.~Riotto, \emph{{Production of massive
  particles during reheating}},
  \href{https://doi.org/10.1103/PhysRevD.60.063504}{\emph{Phys. Rev.}
  {\bfseries D60} (1999) 063504},
  [\href{https://arxiv.org/abs/hep-ph/9809453}{{\ttfamily hep-ph/9809453}}].

\bibitem{Kolb:1998ki}
E.~W. Kolb, D.~J.~H. Chung and A.~Riotto, \emph{{WIMPzillas!}},
  \href{https://doi.org/10.1063/1.59655}{\emph{AIP Conf. Proc.} {\bfseries 484}
  (1999) 91--105}, [\href{https://arxiv.org/abs/hep-ph/9810361}{{\ttfamily
  hep-ph/9810361}}].

\bibitem{Kuzmin:1998uv}
V.~Kuzmin and I.~Tkachev, \emph{{Ultrahigh-energy cosmic rays, superheavy long
  living particles, and matter creation after inflation}},
  \href{https://doi.org/10.1134/1.567858}{\emph{JETP Lett.} {\bfseries 68}
  (1998) 271--275}, [\href{https://arxiv.org/abs/hep-ph/9802304}{{\ttfamily
  hep-ph/9802304}}].

\bibitem{Chung:2001cb}
D.~J.~H. Chung, P.~Crotty, E.~W. Kolb and A.~Riotto, \emph{{On the
  Gravitational Production of Superheavy Dark Matter}},
  \href{https://doi.org/10.1103/PhysRevD.64.043503}{\emph{Phys. Rev.}
  {\bfseries D64} (2001) 043503},
  [\href{https://arxiv.org/abs/hep-ph/0104100}{{\ttfamily hep-ph/0104100}}].

\bibitem{Chung:2004nh}
D.~J.~H. Chung, E.~W. Kolb, A.~Riotto and L.~Senatore, \emph{{Isocurvature
  constraints on gravitationally produced superheavy dark matter}},
  \href{https://doi.org/10.1103/PhysRevD.72.023511}{\emph{Phys. Rev.}
  {\bfseries D72} (2005) 023511},
  [\href{https://arxiv.org/abs/astro-ph/0411468}{{\ttfamily
  astro-ph/0411468}}].

\bibitem{Chung:2011ck}
D.~J.~H. Chung, L.~L. Everett, H.~Yoo and P.~Zhou, \emph{{Gravitational Fermion
  Production in Inflationary Cosmology}},
  \href{https://doi.org/10.1016/j.physletb.2012.04.066}{\emph{Phys. Lett.}
  {\bfseries B712} (2012) 147--154},
  [\href{https://arxiv.org/abs/1109.2524}{{\ttfamily 1109.2524}}].

\bibitem{Ema:2019yrd}
Y.~Ema, K.~Nakayama and Y.~Tang, \emph{{Production of Purely Gravitational Dark
  Matter: The Case of Fermion and Vector Boson}},
  \href{https://doi.org/10.1007/JHEP07(2019)060}{\emph{JHEP} {\bfseries 07}
  (2019) 060}, [\href{https://arxiv.org/abs/1903.10973}{{\ttfamily
  1903.10973}}].

\bibitem{Herring:2020cah}
N.~Herring and D.~Boyanovsky, \emph{{Gravitational production of nearly thermal
  fermionic dark matter}},
  \href{https://doi.org/10.1103/PhysRevD.101.123522}{\emph{Phys. Rev. D}
  {\bfseries 101} (2020) 123522},
  [\href{https://arxiv.org/abs/2005.00391}{{\ttfamily 2005.00391}}].

\bibitem{Kolb:2020fwh}
E.~W. Kolb and A.~J. Long, \emph{{Completely Dark Photons from Gravitational
  Particle Production During Inflation}},
  \href{https://arxiv.org/abs/2009.03828}{{\ttfamily 2009.03828}}.

\bibitem{Kallosh:1999jj}
R.~Kallosh, L.~Kofman, A.~D. Linde and A.~Van~Proeyen, \emph{{Gravitino
  production after inflation}},
  \href{https://doi.org/10.1103/PhysRevD.61.103503}{\emph{Phys. Rev.}
  {\bfseries D61} (2000) 103503},
  [\href{https://arxiv.org/abs/hep-th/9907124}{{\ttfamily hep-th/9907124}}].

\bibitem{Kallosh:2000ve}
R.~Kallosh, L.~Kofman, A.~D. Linde and A.~Van~Proeyen, \emph{{Superconformal
  symmetry, supergravity and cosmology}},
  \href{https://doi.org/10.1088/0264-9381/17/20/308}{\emph{Class. Quant. Grav.}
  {\bfseries 17} (2000) 4269--4338},
  [\href{https://arxiv.org/abs/hep-th/0006179}{{\ttfamily hep-th/0006179}}].

\bibitem{Giudice:1999yt}
G.~F. Giudice, I.~Tkachev and A.~Riotto, \emph{{Nonthermal production of
  dangerous relics in the early universe}},
  \href{https://doi.org/10.1088/1126-6708/1999/08/009}{\emph{JHEP} {\bfseries
  08} (1999) 009}, [\href{https://arxiv.org/abs/hep-ph/9907510}{{\ttfamily
  hep-ph/9907510}}].

\bibitem{Giudice:1999am}
G.~F. Giudice, A.~Riotto and I.~Tkachev, \emph{{Thermal and nonthermal
  production of gravitinos in the early universe}},
  \href{https://doi.org/10.1088/1126-6708/1999/11/036}{\emph{JHEP} {\bfseries
  11} (1999) 036}, [\href{https://arxiv.org/abs/hep-ph/9911302}{{\ttfamily
  hep-ph/9911302}}].

\bibitem{BasteroGil:2000je}
M.~Bastero-Gil and A.~Mazumdar, \emph{{Gravitino production in hybrid
  inflationary models}},
  \href{https://doi.org/10.1103/PhysRevD.62.083510}{\emph{Phys. Rev. D}
  {\bfseries 62} (2000) 083510},
  [\href{https://arxiv.org/abs/hep-ph/0002004}{{\ttfamily hep-ph/0002004}}].

\bibitem{Albornoz:2017yup}
N.~L. Gonz\'alez~Albornoz, A.~Schmidt-May and M.~von Strauss, \emph{{Dark
  matter scenarios with multiple spin-2 fields}},
  \href{https://doi.org/10.1088/1475-7516/2018/01/014}{\emph{JCAP} {\bfseries
  01} (2018) 014}, [\href{https://arxiv.org/abs/1709.05128}{{\ttfamily
  1709.05128}}].

\bibitem{Alexander:2020gmv}
S.~Alexander, L.~Jenks and E.~McDonough, \emph{{Higher Spin Dark Matter}},
  \href{https://arxiv.org/abs/2010.15125}{{\ttfamily 2010.15125}}.

\bibitem{Chen:2009we}
X.~Chen and Y.~Wang, \emph{{Large non-Gaussianities with Intermediate Shapes
  from Quasi-Single Field Inflation}},
  \href{https://doi.org/10.1103/PhysRevD.81.063511}{\emph{Phys. Rev. D}
  {\bfseries 81} (2010) 063511},
  [\href{https://arxiv.org/abs/0909.0496}{{\ttfamily 0909.0496}}].

\bibitem{Arkani-Hamed:2015bza}
N.~Arkani-Hamed and J.~Maldacena, \emph{{Cosmological Collider Physics}},
  \href{https://arxiv.org/abs/1503.08043}{{\ttfamily 1503.08043}}.

\bibitem{Chen:2015lza}
X.~Chen, M.~H. Namjoo and Y.~Wang, \emph{{Quantum Primordial Standard Clocks}},
  \href{https://doi.org/10.1088/1475-7516/2016/02/013}{\emph{JCAP} {\bfseries
  02} (2016) 013}, [\href{https://arxiv.org/abs/1509.03930}{{\ttfamily
  1509.03930}}].

\bibitem{Arkani-Hamed:2018kmz}
N.~Arkani-Hamed, D.~Baumann, H.~Lee and G.~L. Pimentel, \emph{{The Cosmological
  Bootstrap: Inflationary Correlators from Symmetries and Singularities}},
  \href{https://doi.org/10.1007/JHEP04(2020)105}{\emph{JHEP} {\bfseries 04}
  (2020) 105}, [\href{https://arxiv.org/abs/1811.00024}{{\ttfamily
  1811.00024}}].

\bibitem{Baumann:2019oyu}
D.~Baumann, C.~Duaso~Pueyo, A.~Joyce, H.~Lee and G.~L. Pimentel, \emph{{The
  Cosmological Bootstrap: Weight-Shifting Operators and Scalar Seeds}},
  \href{https://arxiv.org/abs/1910.14051}{{\ttfamily 1910.14051}}.

\bibitem{Baumann:2020dch}
D.~Baumann, C.~Duaso~Pueyo, A.~Joyce, H.~Lee and G.~L. Pimentel, \emph{{The
  Cosmological Bootstrap: Spinning Correlators from Symmetries and
  Factorization}},  \href{https://arxiv.org/abs/2005.04234}{{\ttfamily
  2005.04234}}.

\bibitem{Lee:2016vti}
H.~Lee, D.~Baumann and G.~L. Pimentel, \emph{{Non-Gaussianity as a Particle
  Detector}}, \href{https://doi.org/10.1007/JHEP12(2016)040}{\emph{JHEP}
  {\bfseries 12} (2016) 040},
  [\href{https://arxiv.org/abs/1607.03735}{{\ttfamily 1607.03735}}].

\bibitem{Alexander:2019vtb}
S.~Alexander, S.~J. Gates, L.~Jenks, K.~Koutrolikos and E.~McDonough,
  \emph{{Higher Spin Supersymmetry at the Cosmological Collider: Sculpting SUSY
  Rilles in the CMB}},
  \href{https://doi.org/10.1007/JHEP10(2019)156}{\emph{JHEP} {\bfseries 10}
  (2019) 156}, [\href{https://arxiv.org/abs/1907.05829}{{\ttfamily
  1907.05829}}].

\bibitem{Wang:2020uic}
Y.~Wang and Y.~Zhu, \emph{{Cosmological Collider Signatures of Massive Vectors
  from Non-Gaussian Gravitational Waves}},
  \href{https://doi.org/10.1088/1475-7516/2020/04/049}{\emph{JCAP} {\bfseries
  04} (2020) 049}, [\href{https://arxiv.org/abs/2001.03879}{{\ttfamily
  2001.03879}}].

\bibitem{McDonough:2020tqq}
E.~McDonough, \emph{{The Cosmological Heavy Ion Collider: Fast Thermalization
  after Cosmic Inflation}},
  \href{https://doi.org/10.1016/j.physletb.2020.135755}{\emph{Phys. Lett. B}
  {\bfseries 809} (2020) 135755},
  [\href{https://arxiv.org/abs/2001.03633}{{\ttfamily 2001.03633}}].

\bibitem{Harigaya:2013vwa}
K.~Harigaya and K.~Mukaida, \emph{{Thermalization after/during Reheating}},
  \href{https://doi.org/10.1007/JHEP05(2014)006}{\emph{JHEP} {\bfseries 05}
  (2014) 006}, [\href{https://arxiv.org/abs/1312.3097}{{\ttfamily 1312.3097}}].

\bibitem{Harigaya:2014waa}
K.~Harigaya, M.~Kawasaki, K.~Mukaida and M.~Yamada, \emph{{Dark Matter
  Production in Late Time Reheating}},
  \href{https://doi.org/10.1103/PhysRevD.89.083532}{\emph{Phys. Rev. D}
  {\bfseries 89} (2014) 083532},
  [\href{https://arxiv.org/abs/1402.2846}{{\ttfamily 1402.2846}}].

\bibitem{Mukaida:2015ria}
K.~Mukaida and M.~Yamada, \emph{{Thermalization Process after Inflation and
  Effective Potential of Scalar Field}},
  \href{https://doi.org/10.1088/1475-7516/2016/02/003}{\emph{JCAP} {\bfseries
  02} (2016) 003}, [\href{https://arxiv.org/abs/1506.07661}{{\ttfamily
  1506.07661}}].

\bibitem{Ema:2016oxl}
Y.~Ema, K.~Mukaida, K.~Nakayama and T.~Terada, \emph{{Nonthermal Gravitino
  Production after Large Field Inflation}},
  \href{https://doi.org/10.1007/JHEP11(2016)184}{\emph{JHEP} {\bfseries 11}
  (2016) 184}, [\href{https://arxiv.org/abs/1609.04716}{{\ttfamily
  1609.04716}}].

\bibitem{Harigaya:2019tzu}
K.~Harigaya, K.~Mukaida and M.~Yamada, \emph{{Dark Matter Production during the
  Thermalization Era}},
  \href{https://doi.org/10.1007/JHEP07(2019)059}{\emph{JHEP} {\bfseries 07}
  (2019) 059}, [\href{https://arxiv.org/abs/1901.11027}{{\ttfamily
  1901.11027}}].

\bibitem{Brandenberger:2019njw}
R.~Brandenberger, R.~Namba and R.~O. Ramos, \emph{{Kinetic Equilibration after
  Preheating}},  \href{https://arxiv.org/abs/1908.09866}{{\ttfamily
  1908.09866}}.

\bibitem{Litsa:2020rsm}
A.~Litsa, K.~Freese, E.~I. Sfakianakis, P.~Stengel and L.~Visinelli,
  \emph{{Large Density Perturbations from Reheating to Standard Model particles
  due to the Dynamics of the Higgs Boson during Inflation}},
  \href{https://arxiv.org/abs/2009.14218}{{\ttfamily 2009.14218}}.

\bibitem{Addazi:2016bus}
A.~Addazi and M.~Y. Khlopov, \emph{{Dark matter and inflation in $R+\zeta
  R^{2}$ supergravity}},
  \href{https://doi.org/10.1016/j.physletb.2016.12.044}{\emph{Phys. Lett. B}
  {\bfseries 766} (2017) 17--22},
  [\href{https://arxiv.org/abs/1612.06417}{{\ttfamily 1612.06417}}].

\bibitem{Addazi:2017kbx}
A.~Addazi and M.~Y. Khlopov, \emph{{Dark Matter from Starobinsky
  Supergravity}}, \href{https://doi.org/10.1142/S0217732317400028}{\emph{Mod.
  Phys. Lett. A} {\bfseries 32} (2017) Mod.Phys.Lett.},
  [\href{https://arxiv.org/abs/1702.05381}{{\ttfamily 1702.05381}}].

\bibitem{Hasegawa:2017hgd}
F.~Hasegawa, K.~Mukaida, K.~Nakayama, T.~Terada and Y.~Yamada, \emph{{Gravitino
  Problem in Minimal Supergravity Inflation}},
  \href{https://doi.org/10.1016/j.physletb.2017.02.030}{\emph{Phys. Lett. B}
  {\bfseries 767} (2017) 392--397},
  [\href{https://arxiv.org/abs/1701.03106}{{\ttfamily 1701.03106}}].

\bibitem{Hasegawa:2017nks}
F.~Hasegawa, K.~Nakayama, T.~Terada and Y.~Yamada, \emph{{Gravitino problem in
  inflation driven by inflaton-polonyi K\"ahler coupling}},
  \href{https://doi.org/10.1016/j.physletb.2017.12.038}{\emph{Phys. Lett. B}
  {\bfseries 777} (2018) 270--274},
  [\href{https://arxiv.org/abs/1709.01246}{{\ttfamily 1709.01246}}].

\bibitem{Giddings:2001yu}
S.~B. Giddings, S.~Kachru and J.~Polchinski, \emph{{Hierarchies from fluxes in
  string compactifications}},
  \href{https://doi.org/10.1103/PhysRevD.66.106006}{\emph{Phys. Rev. D}
  {\bfseries 66} (2002) 106006},
  [\href{https://arxiv.org/abs/hep-th/0105097}{{\ttfamily hep-th/0105097}}].

\bibitem{Kachru:2003aw}
S.~Kachru, R.~Kallosh, A.~D. Linde and S.~P. Trivedi, \emph{{De Sitter vacua in
  string theory}},
  \href{https://doi.org/10.1103/PhysRevD.68.046005}{\emph{Phys. Rev. D}
  {\bfseries 68} (2003) 046005},
  [\href{https://arxiv.org/abs/hep-th/0301240}{{\ttfamily hep-th/0301240}}].

\bibitem{Balasubramanian:2005zx}
V.~Balasubramanian, P.~Berglund, J.~P. Conlon and F.~Quevedo,
  \emph{{Systematics of moduli stabilisation in Calabi-Yau flux
  compactifications}},
  \href{https://doi.org/10.1088/1126-6708/2005/03/007}{\emph{JHEP} {\bfseries
  03} (2005) 007}, [\href{https://arxiv.org/abs/hep-th/0502058}{{\ttfamily
  hep-th/0502058}}].

\bibitem{Bousso:2000xa}
R.~Bousso and J.~Polchinski, \emph{{Quantization of four form fluxes and
  dynamical neutralization of the cosmological constant}},
  \href{https://doi.org/10.1088/1126-6708/2000/06/006}{\emph{JHEP} {\bfseries
  06} (2000) 006}, [\href{https://arxiv.org/abs/hep-th/0004134}{{\ttfamily
  hep-th/0004134}}].

\bibitem{Susskind:2003kw}
L.~Susskind, \emph{{The Anthropic landscape of string theory}},
  \href{https://arxiv.org/abs/hep-th/0302219}{{\ttfamily hep-th/0302219}}.

\bibitem{Kallosh:2014wsa}
R.~Kallosh and T.~Wrase, \emph{{Emergence of Spontaneously Broken Supersymmetry
  on an Anti-D3-Brane in KKLT dS Vacua}},
  \href{https://doi.org/10.1007/JHEP12(2014)117}{\emph{JHEP} {\bfseries 12}
  (2014) 117}, [\href{https://arxiv.org/abs/1411.1121}{{\ttfamily 1411.1121}}].

\bibitem{Bergshoeff:2015jxa}
E.~A. Bergshoeff, K.~Dasgupta, R.~Kallosh, A.~Van~Proeyen and T.~Wrase,
  \emph{{$ \overline{\mathrm{D}3} $ and dS}},
  \href{https://doi.org/10.1007/JHEP05(2015)058}{\emph{JHEP} {\bfseries 05}
  (2015) 058}, [\href{https://arxiv.org/abs/1502.07627}{{\ttfamily
  1502.07627}}].

\bibitem{Vercnocke:2016fbt}
B.~Vercnocke and T.~Wrase, \emph{{Constrained superfields from an anti-D3-brane
  in KKLT}}, \href{https://doi.org/10.1007/JHEP08(2016)132}{\emph{JHEP}
  {\bfseries 08} (2016) 132},
  [\href{https://arxiv.org/abs/1605.03961}{{\ttfamily 1605.03961}}].

\bibitem{GarciadelMoral:2017vnz}
M.~P. Garcia~del Moral, S.~Parameswaran, N.~Quiroz and I.~Zavala,
  \emph{{Anti-D3 branes and moduli in non-linear supergravity}},
  \href{https://doi.org/10.1007/JHEP10(2017)185}{\emph{JHEP} {\bfseries 10}
  (2017) 185}, [\href{https://arxiv.org/abs/1707.07059}{{\ttfamily
  1707.07059}}].

\bibitem{Cribiori:2019hod}
N.~Cribiori, C.~Roupec, T.~Wrase and Y.~Yamada, \emph{{Supersymmetric
  anti-D3-brane action in the Kachru-Kallosh-Linde-Trivedi setup}},
  \href{https://doi.org/10.1103/PhysRevD.100.066001}{\emph{Phys. Rev. D}
  {\bfseries 100} (2019) 066001},
  [\href{https://arxiv.org/abs/1906.07727}{{\ttfamily 1906.07727}}].

\bibitem{KHLOPOV1984265}
M.~Khlopov and A.~Linde, \emph{Is it easy to save the gravitino?},
  \href{https://doi.org/https://doi.org/10.1016/0370-2693(84)91656-3}{\emph{Physics
  Letters B} {\bfseries 138} (1984) 265--268}.

\bibitem{ELLIS1984181}
J.~Ellis, J.~E. Kim and D.~Nanopoulos, \emph{Cosmological gravitino
  regeneration and decay},
  \href{https://doi.org/https://doi.org/10.1016/0370-2693(84)90334-4}{\emph{Physics
  Letters B} {\bfseries 145} (1984) 181--186}.

\bibitem{Benakli:2014bpa}
K.~Benakli, L.~Darm\'e and Y.~Oz, \emph{{The Slow Gravitino}},
  \href{https://doi.org/10.1007/JHEP10(2014)121}{\emph{JHEP} {\bfseries 10}
  (2014) 121}, [\href{https://arxiv.org/abs/1407.8321}{{\ttfamily 1407.8321}}].

\bibitem{Benakli:2015mbb}
K.~Benakli and L.~Darm\'e, \emph{{Off-trail SUSY}}, {\emph{PoS} {\bfseries
  PLANCK2015} (2015) 019}, [\href{https://arxiv.org/abs/1511.02044}{{\ttfamily
  1511.02044}}].

\bibitem{Kahn:2015mla}
Y.~Kahn, D.~A. Roberts and J.~Thaler, \emph{{The goldstone and goldstino of
  supersymmetric inflation}},
  \href{https://doi.org/10.1007/JHEP10(2015)001}{\emph{JHEP} {\bfseries 10}
  (2015) 001}, [\href{https://arxiv.org/abs/1504.05958}{{\ttfamily
  1504.05958}}].

\bibitem{Eberl:2020fml}
H.~Eberl, I.~D. Gialamas and V.~C. Spanos, \emph{{Gravitino thermal production
  revisited}},  \href{https://arxiv.org/abs/2010.14621}{{\ttfamily
  2010.14621}}.

\bibitem{Nilles:2001ry}
H.~P. Nilles, M.~Peloso and L.~Sorbo, \emph{{Nonthermal production of
  gravitinos and inflatinos}},
  \href{https://doi.org/10.1103/PhysRevLett.87.051302}{\emph{Phys. Rev. Lett.}
  {\bfseries 87} (2001) 051302},
  [\href{https://arxiv.org/abs/hep-ph/0102264}{{\ttfamily hep-ph/0102264}}].

\bibitem{Nilles:2001fg}
H.~P. Nilles, M.~Peloso and L.~Sorbo, \emph{{Coupled fields in external
  background with application to nonthermal production of gravitinos}},
  \href{https://doi.org/10.1088/1126-6708/2001/04/004}{\emph{JHEP} {\bfseries
  04} (2001) 004}, [\href{https://arxiv.org/abs/hep-th/0103202}{{\ttfamily
  hep-th/0103202}}].

\bibitem{Dalianis:2017okk}
I.~Dalianis and F.~Farakos, \emph{{Constrained superfields from inflation to
  reheating}},
  \href{https://doi.org/10.1016/j.physletb.2017.09.020}{\emph{Phys. Lett. B}
  {\bfseries 773} (2017) 610--615},
  [\href{https://arxiv.org/abs/1705.06717}{{\ttfamily 1705.06717}}].

\bibitem{Vafa:2005ui}
C.~Vafa, \emph{{The String landscape and the swampland}},
  \href{https://arxiv.org/abs/hep-th/0509212}{{\ttfamily hep-th/0509212}}.

\bibitem{Palti:2019pca}
E.~Palti, \emph{{The Swampland: Introduction and Review}},
  \href{https://doi.org/10.1002/prop.201900037}{\emph{Fortsch. Phys.}
  {\bfseries 67} (2019) 1900037},
  [\href{https://arxiv.org/abs/1903.06239}{{\ttfamily 1903.06239}}].

\bibitem{Brennan:2017rbf}
T.~D. Brennan, F.~Carta and C.~Vafa, \emph{{The String Landscape, the
  Swampland, and the Missing Corner}},
  \href{https://doi.org/10.22323/1.305.0015}{\emph{PoS} {\bfseries TASI2017}
  (2017) 015}, [\href{https://arxiv.org/abs/1711.00864}{{\ttfamily
  1711.00864}}].

\bibitem{Kolb:2021nob}
E.~W. Kolb, A.~J. Long and E.~Mcdonough, \emph{{The Gravitino Swampland
  Conjecture}},  \href{https://arxiv.org/abs/2103.10437}{{\ttfamily
  2103.10437}}.

\bibitem{LL}
L.~D. Landau and E.~M. Lifshitz, \emph{The Classical Theory of Fields}.
\newblock Pergamon Press, Oxford, 4~ed., 1980.

\bibitem{MTW}
C.~W. {Misner}, K.~S. {Thorne} and J.~A. {Wheeler}, \emph{{Gravitation}}.
\newblock W.H.~Freeman and Co., San Francisco, 1973.

\bibitem{Rarita:1941mf}
W.~Rarita and J.~Schwinger, \emph{{On a theory of particles with half integral
  spin}}, \href{https://doi.org/10.1103/PhysRev.60.61}{\emph{Phys. Rev.}
  {\bfseries 60} (1941) 61}.

\bibitem{Freedman:2012zz}
D.~Z. Freedman and A.~Van~Proeyen, \emph{{Supergravity}}.
\newblock Cambridge Univ. Press, Cambridge, UK, 5, 2012.

\bibitem{Senatore:2012nq}
L.~Senatore and M.~Zaldarriaga, \emph{{On Loops in Inflation II: IR Effects in
  Single Clock Inflation}},
  \href{https://doi.org/10.1007/JHEP01(2013)109}{\emph{JHEP} {\bfseries 01}
  (2013) 109}, [\href{https://arxiv.org/abs/1203.6354}{{\ttfamily 1203.6354}}].

\bibitem{Kofman:1997yn}
L.~Kofman, A.~D. Linde and A.~A. Starobinsky, \emph{{Towards the theory of
  reheating after inflation}},
  \href{https://doi.org/10.1103/PhysRevD.56.3258}{\emph{Phys. Rev. D}
  {\bfseries 56} (1997) 3258--3295},
  [\href{https://arxiv.org/abs/hep-ph/9704452}{{\ttfamily hep-ph/9704452}}].

\bibitem{Amin:2014eta}
M.~A. Amin, M.~P. Hertzberg, D.~I. Kaiser and J.~Karouby,
  \emph{{Nonperturbative Dynamics Of Reheating After Inflation: A Review}},
  \href{https://doi.org/10.1142/S0218271815300037}{\emph{Int. J. Mod. Phys. D}
  {\bfseries 24} (2014) 1530003},
  [\href{https://arxiv.org/abs/1410.3808}{{\ttfamily 1410.3808}}].

\bibitem{Amin:2015ftc}
M.~A. Amin and D.~Baumann, \emph{{From Wires to Cosmology}},
  \href{https://doi.org/10.1088/1475-7516/2016/02/045}{\emph{JCAP} {\bfseries
  02} (2016) 045}, [\href{https://arxiv.org/abs/1512.02637}{{\ttfamily
  1512.02637}}].

\bibitem{Garcia:2019icv}
M.~A.~G. Garcia, M.~A. Amin, S.~G. Carlsten and D.~Green, \emph{{Stochastic
  Particle Production in a de Sitter Background}},
  \href{https://doi.org/10.1088/1475-7516/2019/05/012}{\emph{JCAP} {\bfseries
  05} (2019) 012}, [\href{https://arxiv.org/abs/1902.09598}{{\ttfamily
  1902.09598}}].

\bibitem{Garcia:2020mwi}
M.~A.~G. Garcia, M.~A. Amin and D.~Green, \emph{{Curvature Perturbations From
  Stochastic Particle Production During Inflation}},
  \href{https://doi.org/10.1088/1475-7516/2020/06/039}{\emph{JCAP} {\bfseries
  06} (2020) 039}, [\href{https://arxiv.org/abs/2001.09158}{{\ttfamily
  2001.09158}}].

\bibitem{FARAKOS2013322}
F.~Farakos and A.~Kehagias, \emph{Decoupling limits of sgoldstino modes in
  global and local supersymmetry},
  \href{https://doi.org/https://doi.org/10.1016/j.physletb.2013.06.001}{\emph{Physics
  Letters B} {\bfseries 724} (2013) 322--327}.

\bibitem{Fayet:1986zc}
P.~Fayet, \emph{{Lower Limit on the Mass of a Light Gravitino from e+ e-
  Annihilation Experiments}},
  \href{https://doi.org/10.1016/0370-2693(86)90626-X}{\emph{Phys. Lett. B}
  {\bfseries 175} (1986) 471--477}.

\bibitem{Casalbuoni:1988kv}
R.~Casalbuoni, S.~De~Curtis, D.~Dominici, F.~Feruglio and R.~Gatto, \emph{{A
  GRAVITINO - GOLDSTINO HIGH-ENERGY EQUIVALENCE THEOREM}},
  \href{https://doi.org/10.1016/0370-2693(88)91439-6}{\emph{Phys. Lett. B}
  {\bfseries 215} (1988) 313--316}.

\bibitem{Casalbuoni:1988qd}
R.~Casalbuoni, S.~De~Curtis, D.~Dominici, F.~Feruglio and R.~Gatto,
  \emph{{High-Energy Equivalence Theorem in Spontaneously Broken
  Supergravity}}, \href{https://doi.org/10.1103/PhysRevD.39.2281}{\emph{Phys.
  Rev. D} {\bfseries 39} (1989) 2281}.

\bibitem{Cheung:2007st}
C.~Cheung, P.~Creminelli, A.~L. Fitzpatrick, J.~Kaplan and L.~Senatore,
  \emph{{The Effective Field Theory of Inflation}},
  \href{https://doi.org/10.1088/1126-6708/2008/03/014}{\emph{JHEP} {\bfseries
  03} (2008) 014}, [\href{https://arxiv.org/abs/0709.0293}{{\ttfamily
  0709.0293}}].

\bibitem{Delacretaz:2016nhw}
L.~V. Delacretaz, V.~Gorbenko and L.~Senatore, \emph{{The Supersymmetric
  Effective Field Theory of Inflation}},
  \href{https://doi.org/10.1007/JHEP03(2017)063}{\emph{JHEP} {\bfseries 03}
  (2017) 063}, [\href{https://arxiv.org/abs/1610.04227}{{\ttfamily
  1610.04227}}].

\bibitem{ArmendarizPicon:2000dh}
C.~Armendariz-Picon, V.~F. Mukhanov and P.~J. Steinhardt, \emph{{A Dynamical
  solution to the problem of a small cosmological constant and late time cosmic
  acceleration}},
  \href{https://doi.org/10.1103/PhysRevLett.85.4438}{\emph{Phys. Rev. Lett.}
  {\bfseries 85} (2000) 4438--4441},
  [\href{https://arxiv.org/abs/astro-ph/0004134}{{\ttfamily
  astro-ph/0004134}}].

\bibitem{ArmendarizPicon:2000ah}
C.~Armendariz-Picon, V.~F. Mukhanov and P.~J. Steinhardt, \emph{{Essentials of
  k essence}}, \href{https://doi.org/10.1103/PhysRevD.63.103510}{\emph{Phys.
  Rev. D} {\bfseries 63} (2001) 103510},
  [\href{https://arxiv.org/abs/astro-ph/0006373}{{\ttfamily
  astro-ph/0006373}}].

\bibitem{Velo:1969bt}
G.~Velo and D.~Zwanziger, \emph{{Propagation and quantization of
  Rarita-Schwinger waves in an external electromagnetic potential}},
  \href{https://doi.org/10.1103/PhysRev.186.1337}{\emph{Phys. Rev.} {\bfseries
  186} (1969) 1337--1341}.

\bibitem{Velo:1970ur}
G.~Velo and D.~Zwanziger, \emph{{Noncausality and other defects of interaction
  lagrangians for particles with spin one and higher}},
  \href{https://doi.org/10.1103/PhysRev.188.2218}{\emph{Phys. Rev.} {\bfseries
  188} (1969) 2218--2222}.

\bibitem{Kallosh:2010ug}
R.~Kallosh and A.~Linde, \emph{{New models of chaotic inflation in
  supergravity}},
  \href{https://doi.org/10.1088/1475-7516/2010/11/011}{\emph{JCAP} {\bfseries
  11} (2010) 011}, [\href{https://arxiv.org/abs/1008.3375}{{\ttfamily
  1008.3375}}].

\bibitem{Kallosh:2010xz}
R.~Kallosh, A.~Linde and T.~Rube, \emph{{General inflaton potentials in
  supergravity}}, \href{https://doi.org/10.1103/PhysRevD.83.043507}{\emph{Phys.
  Rev. D} {\bfseries 83} (2011) 043507},
  [\href{https://arxiv.org/abs/1011.5945}{{\ttfamily 1011.5945}}].

\bibitem{Ferrara:2014kva}
S.~Ferrara, R.~Kallosh and A.~Linde, \emph{{Cosmology with Nilpotent
  Superfields}}, \href{https://doi.org/10.1007/JHEP10(2014)143}{\emph{JHEP}
  {\bfseries 10} (2014) 143},
  [\href{https://arxiv.org/abs/1408.4096}{{\ttfamily 1408.4096}}].

\bibitem{McDonough:2016der}
E.~McDonough and M.~Scalisi, \emph{{Inflation from Nilpotent K\"ahler
  Corrections}},
  \href{https://doi.org/10.1088/1475-7516/2016/11/028}{\emph{JCAP} {\bfseries
  1611} (2016) 028}, [\href{https://arxiv.org/abs/1609.00364}{{\ttfamily
  1609.00364}}].

\bibitem{Kallosh:2017wnt}
R.~Kallosh, A.~Linde, D.~Roest and Y.~Yamada, \emph{{$ \overline{D3} $ induced
  geometric inflation}},
  \href{https://doi.org/10.1007/JHEP07(2017)057}{\emph{JHEP} {\bfseries 07}
  (2017) 057}, [\href{https://arxiv.org/abs/1705.09247}{{\ttfamily
  1705.09247}}].

\bibitem{Ferrara:2013rsa}
S.~Ferrara, R.~Kallosh, A.~Linde and M.~Porrati, \emph{{Minimal Supergravity
  Models of Inflation}},
  \href{https://doi.org/10.1103/PhysRevD.88.085038}{\emph{Phys. Rev.}
  {\bfseries D88} (2013) 085038},
  [\href{https://arxiv.org/abs/1307.7696}{{\ttfamily 1307.7696}}].

\bibitem{Kallosh:2013yoa}
R.~Kallosh, A.~Linde and D.~Roest, \emph{{Superconformal Inflationary
  $\alpha$-Attractors}},
  \href{https://doi.org/10.1007/JHEP11(2013)198}{\emph{JHEP} {\bfseries 11}
  (2013) 198}, [\href{https://arxiv.org/abs/1311.0472}{{\ttfamily 1311.0472}}].

\bibitem{Kallosh:2014rga}
R.~Kallosh, A.~Linde and D.~Roest, \emph{{Large field inflation and double
  $\alpha$-attractors}},
  \href{https://doi.org/10.1007/JHEP08(2014)052}{\emph{JHEP} {\bfseries 08}
  (2014) 052}, [\href{https://arxiv.org/abs/1405.3646}{{\ttfamily 1405.3646}}].

\bibitem{Kallosh:2013hoa}
R.~Kallosh and A.~Linde, \emph{{Universality Class in Conformal Inflation}},
  \href{https://doi.org/10.1088/1475-7516/2013/07/002}{\emph{JCAP} {\bfseries
  07} (2013) 002}, [\href{https://arxiv.org/abs/1306.5220}{{\ttfamily
  1306.5220}}].

\bibitem{Kallosh:2014hxa}
R.~Kallosh, A.~Linde and M.~Scalisi, \emph{{Inflation, de Sitter Landscape and
  Super-Higgs effect}},
  \href{https://doi.org/10.1007/JHEP03(2015)111}{\emph{JHEP} {\bfseries 03}
  (2015) 111}, [\href{https://arxiv.org/abs/1411.5671}{{\ttfamily 1411.5671}}].

\bibitem{Carrasco:2015iij}
J.~J.~M. Carrasco, R.~Kallosh and A.~Linde, \emph{{Minimal supergravity
  inflation}}, \href{https://doi.org/10.1103/PhysRevD.93.061301}{\emph{Phys.
  Rev. D} {\bfseries 93} (2016) 061301},
  [\href{https://arxiv.org/abs/1512.00546}{{\ttfamily 1512.00546}}].

\bibitem{Kallosh:2019apq}
R.~Kallosh and Y.~Yamada, \emph{{Simple sinflaton-less $\alpha$-attractors}},
  \href{https://doi.org/10.1007/JHEP03(2019)139}{\emph{JHEP} {\bfseries 03}
  (2019) 139}, [\href{https://arxiv.org/abs/1901.09046}{{\ttfamily
  1901.09046}}].

\bibitem{Ferrara:2015tyn}
S.~Ferrara, R.~Kallosh and J.~Thaler, \emph{{Cosmology with orthogonal
  nilpotent superfields}},
  \href{https://doi.org/10.1103/PhysRevD.93.043516}{\emph{Phys. Rev.}
  {\bfseries D93} (2016) 043516},
  [\href{https://arxiv.org/abs/1512.00545}{{\ttfamily 1512.00545}}].

\bibitem{Bergshoeff:2015tra}
E.~A. Bergshoeff, D.~Z. Freedman, R.~Kallosh and A.~Van~Proeyen, \emph{{Pure de
  Sitter Supergravity}}, \href{https://doi.org/10.1103/PhysRevD.93.069901,
  10.1103/PhysRevD.92.085040}{\emph{Phys. Rev.} {\bfseries D92} (2015) 085040},
  [\href{https://arxiv.org/abs/1507.08264}{{\ttfamily 1507.08264}}].

\bibitem{Hasegawa:2015bza}
F.~Hasegawa and Y.~Yamada, \emph{{Component action of nilpotent multiplet
  coupled to matter in 4 dimensional $ \mathcal{N}=1 $ supergravity}},
  \href{https://doi.org/10.1007/JHEP10(2015)106}{\emph{JHEP} {\bfseries 10}
  (2015) 106}, [\href{https://arxiv.org/abs/1507.08619}{{\ttfamily
  1507.08619}}].

\bibitem{Kallosh:2015sea}
R.~Kallosh, \emph{{Matter-coupled de Sitter Supergravity}},
  \href{https://doi.org/10.1134/S0040577916050068}{\emph{Theor. Math. Phys.}
  {\bfseries 187} (2016) 695--705},
  [\href{https://arxiv.org/abs/1509.02136}{{\ttfamily 1509.02136}}].

\bibitem{Kallosh:2015tea}
R.~Kallosh and T.~Wrase, \emph{{De Sitter Supergravity Model Building}},
  \href{https://doi.org/10.1103/PhysRevD.92.105010}{\emph{Phys. Rev.}
  {\bfseries D92} (2015) 105010},
  [\href{https://arxiv.org/abs/1509.02137}{{\ttfamily 1509.02137}}].

\bibitem{Schillo:2015ssx}
M.~Schillo, E.~van~der Woerd and T.~Wrase, \emph{{The general de Sitter
  supergravity component action}}, \href{https://doi.org/10.1002/prop201500074,
  10.1002/prop.201500074}{\emph{Fortsch. Phys.} {\bfseries 64} (2016)
  292--302}, [\href{https://arxiv.org/abs/1511.01542}{{\ttfamily 1511.01542}}].

\bibitem{Freedman:2017obq}
D.~Z. Freedman, D.~Roest and A.~Van~Proeyen, \emph{{Off-shell Poincare
  Supergravity}}, \href{https://doi.org/10.1007/JHEP02(2017)102}{\emph{JHEP}
  {\bfseries 02} (2017) 102},
  [\href{https://arxiv.org/abs/1701.05216}{{\ttfamily 1701.05216}}].

\bibitem{Grisaru:1979wc}
M.~T. Grisaru, W.~Siegel and M.~Rocek, \emph{{Improved Methods for
  Supergraphs}},
  \href{https://doi.org/10.1016/0550-3213(79)90344-4}{\emph{Nucl. Phys. B}
  {\bfseries 159} (1979) 429}.

\bibitem{Seiberg:1993vc}
N.~Seiberg, \emph{{Naturalness versus supersymmetric nonrenormalization
  theorems}}, \href{https://doi.org/10.1016/0370-2693(93)91541-T}{\emph{Phys.
  Lett. B} {\bfseries 318} (1993) 469--475},
  [\href{https://arxiv.org/abs/hep-ph/9309335}{{\ttfamily hep-ph/9309335}}].

\bibitem{Becker:2002nn}
K.~Becker, M.~Becker, M.~Haack and J.~Louis, \emph{{Supersymmetry breaking and
  alpha-prime corrections to flux induced potentials}},
  \href{https://doi.org/10.1088/1126-6708/2002/06/060}{\emph{JHEP} {\bfseries
  06} (2002) 060}, [\href{https://arxiv.org/abs/hep-th/0204254}{{\ttfamily
  hep-th/0204254}}].

\bibitem{Berg:2005ja}
M.~Berg, M.~Haack and B.~Kors, \emph{{String loop corrections to Kahler
  potentials in orientifolds}},
  \href{https://doi.org/10.1088/1126-6708/2005/11/030}{\emph{JHEP} {\bfseries
  11} (2005) 030}, [\href{https://arxiv.org/abs/hep-th/0508043}{{\ttfamily
  hep-th/0508043}}].

\bibitem{Kallosh:2004yh}
R.~Kallosh and A.~D. Linde, \emph{{Landscape, the scale of SUSY breaking, and
  inflation}}, \href{https://doi.org/10.1088/1126-6708/2004/12/004}{\emph{JHEP}
  {\bfseries 12} (2004) 004},
  [\href{https://arxiv.org/abs/hep-th/0411011}{{\ttfamily hep-th/0411011}}].

\bibitem{Conlon:2008cj}
J.~P. Conlon, R.~Kallosh, A.~D. Linde and F.~Quevedo, \emph{{Volume Modulus
  Inflation and the Gravitino Mass Problem}},
  \href{https://doi.org/10.1088/1475-7516/2008/09/011}{\emph{JCAP} {\bfseries
  0809} (2008) 011}, [\href{https://arxiv.org/abs/0806.0809}{{\ttfamily
  0806.0809}}].

\bibitem{Linde:2020mdk}
A.~Linde, \emph{{KKLT without AdS}},
  \href{https://doi.org/10.1007/JHEP05(2020)076}{\emph{JHEP} {\bfseries 05}
  (2020) 076}, [\href{https://arxiv.org/abs/2002.01500}{{\ttfamily
  2002.01500}}].

\bibitem{Copeland:1994vg}
E.~J. Copeland, A.~R. Liddle, D.~H. Lyth, E.~D. Stewart and D.~Wands,
  \emph{{False vacuum inflation with Einstein gravity}},
  \href{https://doi.org/10.1103/PhysRevD.49.6410}{\emph{Phys. Rev. D}
  {\bfseries 49} (1994) 6410--6433},
  [\href{https://arxiv.org/abs/astro-ph/9401011}{{\ttfamily
  astro-ph/9401011}}].

\bibitem{Hebecker:2018vxz}
A.~Hebecker and T.~Wrase, \emph{{The Asymptotic dS Swampland Conjecture - a
  Simplified Derivation and a Potential Loophole}},
  \href{https://doi.org/10.1002/prop.201800097}{\emph{Fortsch. Phys.}
  {\bfseries 67} (2019) 1800097},
  [\href{https://arxiv.org/abs/1810.08182}{{\ttfamily 1810.08182}}].

\bibitem{Garbrecht:2006az}
B.~Garbrecht, C.~Pallis and A.~Pilaftsis, \emph{{Anatomy of F(D)-Term Hybrid
  Inflation}}, \href{https://doi.org/10.1088/1126-6708/2006/12/038}{\emph{JHEP}
  {\bfseries 12} (2006) 038},
  [\href{https://arxiv.org/abs/hep-ph/0605264}{{\ttfamily hep-ph/0605264}}].

\bibitem{BirrellDavies:1982}
{N. D. Birrell and P. C. Davies}, \emph{{Quantum fields in curved space}}.
\newblock {Cambridge University Press}, 1982.

\bibitem{Weinberg:1972kfs}
S.~Weinberg, \emph{{Gravitation and Cosmology}}.
\newblock John Wiley and Sons, New York, 1972.

\bibitem{Wald:106274}
R.~M. Wald, \emph{{General relativity}}.
\newblock Chicago Univ. Press, Chicago, IL, 1984.

\bibitem{Green:1987sp}
M.~B. Green, J.~H. Schwarz and E.~Witten, \emph{{SUPERSTRING THEORY. VOL. 1:
  INTRODUCTION}}.
\newblock Cambridge Monographs on Mathematical Physics. Cambridge University
  Press, 7, 1988.

\bibitem{Baumann:2007ah}
D.~Baumann, A.~Dymarsky, I.~R. Klebanov and L.~McAllister, \emph{{Towards an
  Explicit Model of D-brane Inflation}},
  \href{https://doi.org/10.1088/1475-7516/2008/01/024}{\emph{JCAP} {\bfseries
  01} (2008) 024}, [\href{https://arxiv.org/abs/0706.0360}{{\ttfamily
  0706.0360}}].

\bibitem{DeWolfe:2002nn}
O.~DeWolfe and S.~B. Giddings, \emph{{Scales and hierarchies in warped
  compactifications and brane worlds}},
  \href{https://doi.org/10.1103/PhysRevD.67.066008}{\emph{Phys. Rev. D}
  {\bfseries 67} (2003) 066008},
  [\href{https://arxiv.org/abs/hep-th/0208123}{{\ttfamily hep-th/0208123}}].

\bibitem{Kallosh:2018wme}
R.~Kallosh, A.~Linde, E.~McDonough and M.~Scalisi, \emph{{de Sitter Vacua with
  a Nilpotent Superfield}},
  \href{https://doi.org/10.1002/prop.201800068}{\emph{Fortsch. Phys.}
  {\bfseries 67} (2019) 1800068},
  [\href{https://arxiv.org/abs/1808.09428}{{\ttfamily 1808.09428}}].

\end{thebibliography}\endgroup

\end{document}